\documentclass[longauth]{aa}

\usepackage{graphicx}
\usepackage{natbib}
\usepackage{scalerel}

\usepackage[table]{xcolor}

\usepackage{txfonts}
\usepackage{glossaries}
\usepackage[pdfencoding=auto,psdextra]{hyperref}
\hypersetup{
    colorlinks=true,
    linkcolor=blue,
    filecolor=magenta,      
    urlcolor=blue,
    citecolor=blue
}
\makeatletter
\renewcommand*\aa@pageof{, page \thepage{} of \pageref*{LastPage}}
\makeatother

\usepackage[utf8]{inputenc}

\usepackage[switch, modulo]{lineno}
\usepackage{euclid}
\usepackage{multirow}
\usepackage{bm}       

\newacronym{hod}{HOD}{halo occupation distribution}
\newacronym{dic}{DIC}{deviance information criterion}
\newacronym{spt}{SPT}{standard perturbation theory}
\newacronym{eft}{EFTofLSS}{effective field theory of large-scale structure}
\newacronym{bao}{BAO}{baryon acoustic oscillation}
\newacronym{cmb}{CMB}{cosmic microwave background}
\newacronym{lss}{LSS}{large-scale structure}
\newacronym{gr}{GR}{general relativity}
\newacronym{flrw}{FLRW}{Friedmann--Lema\^{i}tre--Robertson--Walker}
\newacronym{2pcf}{2PCF}{two-point correlation function}
\newacronym{rsd}{RSDs}{redshift-space distortions}
\newacronym{cscs}{CSCS}{Swiss National Supercomputing Center}
\newacronym{fof}{FoF}{friends-of-friends}
\newacronym{pt}{PT}{perturbation theory}
\newacronym{ir}{IR}{infrared}
\newacronym{uv}{UV}{ultraviolet}
\newacronym{fob}{FoB}{figure of bias}
\newacronym{fom}{FoM}{figure of merit}
\newacronym{eds}{EdS}{Einstein--de Sitter}

\newcommand{\nbody}{\textit{N}-body }

\newcommand{\de}{\mathrm{d}}

\newcommand{\bdm}{\begin{displaymath}}
\newcommand{\edm}{\end{displaymath}}
\newcommand{\nn}{\nonumber}

\newcommand{\ie}{i.e.\,}
\newcommand{\eg}{e.g.}

\newcommand{\euler}{{\rm e}}
\newcommand{\imag}{{\rm i}}

\newcommand{\xv}{\bm{x}}

\newcommand{\kv}{\bm{k}}

\newcommand{\qv}{\bm{q}}

\newcommand{\thetav}{\bm{\theta}}

\newcommand{\kmax}{k_{\rm max}}
\newcommand{\kF}{k_{\,\rm F}}
\newcommand{\kN}{k_{\,\rm Nyq}}
\newcommand{\ks}{k_{\rm s}}
\newcommand{\kosc}{k_{\rm osc}}
\newcommand{\losc}{\ell_{\rm osc}}

\newcommand{\omegac}{\omega_{\rm c}}
\newcommand{\omegab}{\omega_{\hbox{\hglue 0.5pt}\rm b}}

\newcommand{\As}{A_{\rm s}}
\newcommand{\ns}{n_{\rm s}}
\newcommand{\bGtwo}{b_{{\cal G}_2}}
\newcommand{\bGthree}{b_{\Gamma_3}}
\newcommand{\bdtwod}{b_{\nabla^{\,2}\delta}}
\newcommand{\aPone}{\alpha_{P,1}}
\newcommand{\aPtwo}{\alpha_{P,2}}

\newcommand{\Ms}{\, h^{-1} \, M_\odot}
\newcommand{\Mpc}{\, h^{-1} \, {\rm Mpc}}

\newcommand{\cGpc}{\, h^{-3} \, {\rm Gpc}^3}
\newcommand{\kMpc}{\, h \, {\rm Mpc}^{-1}}
\newcommand{\kcMpc}{\, h^3 \, {\rm Mpc}^{-3}}
\newcommand{\Hunits}{\, {\rm km \, s^{-1} \, Mpc^{-1}}}

\newcommand{\dirac}{\delta_{\rm D}^{\,(3)}}

\newcommand{\deltainit}{\delta_{\rm L}}
\newcommand{\deltag}{\delta_{\rm g}}
\newcommand{\Plin}{P_{\rm L}}
\newcommand{\Pmm}{P_{\rm mm}}
\newcommand{\Pgg}{P_{\rm gg}}
\newcommand{\Fn}{F_n}
\newcommand{\Ponel}{P^{\,\rm 1\mbox{-}loop}}
\newcommand{\Ptwol}{P^{\,\rm 2\mbox{-}loop}}
\newcommand{\Pnw}{P_{\rm nw}}
\newcommand{\Pw}{P_{\rm w}}
\newcommand{\Pmmlo}{P_{\rm mm}^{\,{\rm IR\mbox{-}LO}}}
\newcommand{\Pmmnlo}{P_{\rm mm}^{\,{\rm IR\mbox{-}(LO+NLO)}}}
\newcommand{\Pctr}{P_{\rm ctr}}
\newcommand{\Pggtt}{P_{{\rm gg}, 22}}
\newcommand{\Pggot}{P_{{\rm gg}, 13}}
\newcommand{\Pgglo}{P_{\rm gg}^{\,\rm IR-LO}}
\newcommand{\Pggnlo}{P_{\rm gg}^{\,\rm IR-(LO+NLO)}}

\newcommand{\Peh}{P_{\rm EH}}

\newcommand{\bacco}{\texttt{baccoemu}\xspace}
\newcommand{\comet}{\texttt{COMET}\xspace}

\begin{document}

\title{\Euclid preparation}
\subtitle{XLI. Galaxy power spectrum modelling in real space}

\newcommand{\orcid}[1]{} 
\author{Euclid Collaboration: A.~Pezzotta\orcid{0000-0003-0726-2268}\thanks{\email{pezzotta@mpe.mpg.de}}\inst{\ref{aff1}}
\and C.~Moretti\orcid{0000-0003-3314-8936}\inst{\ref{aff2},\ref{aff3},\ref{aff4},\ref{aff5},\ref{aff6}}
\and M.~Zennaro\orcid{0000-0002-4458-1754}\inst{\ref{aff7}}
\and A.~Moradinezhad~Dizgah\orcid{0000-0001-8841-9989}\inst{\ref{aff8}}
\and M.~Crocce\orcid{0000-0002-9745-6228}\inst{\ref{aff9},\ref{aff10}}
\and E.~Sefusatti\orcid{0000-0003-0473-1567}\inst{\ref{aff5},\ref{aff6},\ref{aff11}}
\and I.~Ferrero\orcid{0000-0002-1295-1132}\inst{\ref{aff12}}
\and K.~Pardede\orcid{0000-0002-7728-8220}\inst{\ref{aff2},\ref{aff13},\ref{aff6},\ref{aff11},\ref{aff5}}
\and A.~Eggemeier\orcid{0000-0002-1841-8910}\inst{\ref{aff14}}
\and A.~Barreira\inst{\ref{aff15},\ref{aff16}}
\and R.~E.~Angulo\orcid{0000-0003-2953-3970}\inst{\ref{aff17},\ref{aff18}}
\and M.~Marinucci\orcid{0000-0003-1159-3756}\inst{\ref{aff19},\ref{aff20}}
\and B.~Camacho~Quevedo\orcid{0000-0002-8789-4232}\inst{\ref{aff21},\ref{aff9}}
\and S.~de~la~Torre\inst{\ref{aff22}}
\and D.~Alkhanishvili\inst{\ref{aff14}}
\and M.~Biagetti\orcid{0000-0002-5097-479X}\inst{\ref{aff23},\ref{aff6},\ref{aff2},\ref{aff5}}
\and M.-A.~Breton\inst{\ref{aff9},\ref{aff10},\ref{aff24}}
\and E.~Castorina\inst{\ref{aff25},\ref{aff26}}
\and G.~D'Amico\inst{\ref{aff19},\ref{aff27}}
\and V.~Desjacques\orcid{0000-0003-2062-8172}\inst{\ref{aff20}}
\and M.~Guidi\orcid{0000-0001-9408-1101}\inst{\ref{aff28},\ref{aff29}}
\and M.~K{\"a}rcher\orcid{0000-0001-5868-647X}\inst{\ref{aff22},\ref{aff30}}
\and A.~Oddo\inst{\ref{aff2},\ref{aff6}}
\and M.~Pellejero~Ibanez\orcid{0000-0003-4680-7275}\inst{\ref{aff17},\ref{aff4}}
\and C.~Porciani\orcid{0000-0002-7797-2508}\inst{\ref{aff14}}
\and A.~Pugno\inst{\ref{aff14}}
\and J.~Salvalaggio\orcid{0000-0002-1431-5607}\inst{\ref{aff5},\ref{aff11},\ref{aff31},\ref{aff6}}
\and E.~Sarpa\orcid{0000-0002-1256-655X}\inst{\ref{aff2},\ref{aff11}}
\and A.~Veropalumbo\orcid{0000-0003-2387-1194}\inst{\ref{aff32},\ref{aff33}}
\and Z.~Vlah\inst{\ref{aff34},\ref{aff35},\ref{aff36}}
\and A.~Amara\inst{\ref{aff37}}
\and S.~Andreon\orcid{0000-0002-2041-8784}\inst{\ref{aff32}}
\and N.~Auricchio\orcid{0000-0003-4444-8651}\inst{\ref{aff38}}
\and M.~Baldi\orcid{0000-0003-4145-1943}\inst{\ref{aff39},\ref{aff38},\ref{aff40}}
\and S.~Bardelli\orcid{0000-0002-8900-0298}\inst{\ref{aff38}}
\and R.~Bender\orcid{0000-0001-7179-0626}\inst{\ref{aff1},\ref{aff41}}
\and C.~Bodendorf\inst{\ref{aff1}}
\and D.~Bonino\inst{\ref{aff42}}
\and E.~Branchini\orcid{0000-0002-0808-6908}\inst{\ref{aff43},\ref{aff33},\ref{aff32}}
\and M.~Brescia\orcid{0000-0001-9506-5680}\inst{\ref{aff44},\ref{aff45},\ref{aff46}}
\and J.~Brinchmann\orcid{0000-0003-4359-8797}\inst{\ref{aff47}}
\and S.~Camera\orcid{0000-0003-3399-3574}\inst{\ref{aff48},\ref{aff49},\ref{aff42}}
\and V.~Capobianco\orcid{0000-0002-3309-7692}\inst{\ref{aff42}}
\and C.~Carbone\orcid{0000-0003-0125-3563}\inst{\ref{aff50}}
\and V.~F.~Cardone\inst{\ref{aff51},\ref{aff52}}
\and J.~Carretero\orcid{0000-0002-3130-0204}\inst{\ref{aff53},\ref{aff54}}
\and S.~Casas\orcid{0000-0002-4751-5138}\inst{\ref{aff55}}
\and F.~J.~Castander\orcid{0000-0001-7316-4573}\inst{\ref{aff9},\ref{aff21}}
\and M.~Castellano\orcid{0000-0001-9875-8263}\inst{\ref{aff51}}
\and S.~Cavuoti\orcid{0000-0002-3787-4196}\inst{\ref{aff45},\ref{aff46}}
\and A.~Cimatti\inst{\ref{aff56}}
\and G.~Congedo\orcid{0000-0003-2508-0046}\inst{\ref{aff4}}
\and C.~J.~Conselice\inst{\ref{aff57}}
\and L.~Conversi\orcid{0000-0002-6710-8476}\inst{\ref{aff58},\ref{aff59}}
\and Y.~Copin\orcid{0000-0002-5317-7518}\inst{\ref{aff60}}
\and L.~Corcione\orcid{0000-0002-6497-5881}\inst{\ref{aff42}}
\and F.~Courbin\orcid{0000-0003-0758-6510}\inst{\ref{aff61}}
\and H.~M.~Courtois\orcid{0000-0003-0509-1776}\inst{\ref{aff62}}
\and A.~Da~Silva\orcid{0000-0002-6385-1609}\inst{\ref{aff63},\ref{aff64}}
\and H.~Degaudenzi\orcid{0000-0002-5887-6799}\inst{\ref{aff65}}
\and A.~M.~Di~Giorgio\orcid{0000-0002-4767-2360}\inst{\ref{aff66}}
\and J.~Dinis\inst{\ref{aff64},\ref{aff63}}
\and X.~Dupac\inst{\ref{aff59}}
\and S.~Dusini\orcid{0000-0002-1128-0664}\inst{\ref{aff67}}
\and A.~Ealet\inst{\ref{aff60}}
\and M.~Farina\orcid{0000-0002-3089-7846}\inst{\ref{aff66}}
\and S.~Farrens\orcid{0000-0002-9594-9387}\inst{\ref{aff68}}
\and P.~Fosalba\orcid{0000-0002-1510-5214}\inst{\ref{aff21},\ref{aff10}}
\and M.~Frailis\orcid{0000-0002-7400-2135}\inst{\ref{aff5}}
\and E.~Franceschi\orcid{0000-0002-0585-6591}\inst{\ref{aff38}}
\and S.~Galeotta\orcid{0000-0002-3748-5115}\inst{\ref{aff5}}
\and B.~Gillis\orcid{0000-0002-4478-1270}\inst{\ref{aff4}}
\and C.~Giocoli\inst{\ref{aff38},\ref{aff69}}
\and B.~R.~Granett\orcid{0000-0003-2694-9284}\inst{\ref{aff32}}
\and A.~Grazian\orcid{0000-0002-5688-0663}\inst{\ref{aff70}}
\and F.~Grupp\inst{\ref{aff1},\ref{aff71}}
\and L.~Guzzo\orcid{0000-0001-8264-5192}\inst{\ref{aff25},\ref{aff32},\ref{aff26}}
\and S.~V.~H.~Haugan\orcid{0000-0001-9648-7260}\inst{\ref{aff12}}
\and F.~Hormuth\inst{\ref{aff72}}
\and A.~Hornstrup\orcid{0000-0002-3363-0936}\inst{\ref{aff73},\ref{aff74}}
\and K.~Jahnke\orcid{0000-0003-3804-2137}\inst{\ref{aff75}}
\and B.~Joachimi\orcid{0000-0001-7494-1303}\inst{\ref{aff76}}
\and E.~Keih\"anen\orcid{0000-0003-1804-7715}\inst{\ref{aff77}}
\and S.~Kermiche\orcid{0000-0002-0302-5735}\inst{\ref{aff78}}
\and A.~Kiessling\orcid{0000-0002-2590-1273}\inst{\ref{aff79}}
\and M.~Kilbinger\orcid{0000-0001-9513-7138}\inst{\ref{aff80}}
\and T.~Kitching\orcid{0000-0002-4061-4598}\inst{\ref{aff81}}
\and B.~Kubik\inst{\ref{aff60}}
\and M.~Kunz\orcid{0000-0002-3052-7394}\inst{\ref{aff8}}
\and H.~Kurki-Suonio\orcid{0000-0002-4618-3063}\inst{\ref{aff82},\ref{aff83}}
\and S.~Ligori\orcid{0000-0003-4172-4606}\inst{\ref{aff42}}
\and P.~B.~Lilje\orcid{0000-0003-4324-7794}\inst{\ref{aff12}}
\and V.~Lindholm\orcid{0000-0003-2317-5471}\inst{\ref{aff82},\ref{aff83}}
\and I.~Lloro\inst{\ref{aff84}}
\and E.~Maiorano\orcid{0000-0003-2593-4355}\inst{\ref{aff38}}
\and O.~Mansutti\orcid{0000-0001-5758-4658}\inst{\ref{aff5}}
\and O.~Marggraf\orcid{0000-0001-7242-3852}\inst{\ref{aff14}}
\and K.~Markovic\orcid{0000-0001-6764-073X}\inst{\ref{aff79}}
\and N.~Martinet\orcid{0000-0003-2786-7790}\inst{\ref{aff22}}
\and F.~Marulli\orcid{0000-0002-8850-0303}\inst{\ref{aff85},\ref{aff38},\ref{aff40}}
\and R.~Massey\orcid{0000-0002-6085-3780}\inst{\ref{aff86}}
\and E.~Medinaceli\orcid{0000-0002-4040-7783}\inst{\ref{aff38}}
\and Y.~Mellier\inst{\ref{aff87},\ref{aff88}}
\and M.~Meneghetti\orcid{0000-0003-1225-7084}\inst{\ref{aff38},\ref{aff40}}
\and E.~Merlin\orcid{0000-0001-6870-8900}\inst{\ref{aff51}}
\and G.~Meylan\inst{\ref{aff61}}
\and M.~Moresco\orcid{0000-0002-7616-7136}\inst{\ref{aff85},\ref{aff38}}
\and L.~Moscardini\orcid{0000-0002-3473-6716}\inst{\ref{aff85},\ref{aff38},\ref{aff40}}
\and E.~Munari\orcid{0000-0002-1751-5946}\inst{\ref{aff5}}
\and S.-M.~Niemi\inst{\ref{aff89}}
\and C.~Padilla\orcid{0000-0001-7951-0166}\inst{\ref{aff53}}
\and S.~Paltani\inst{\ref{aff65}}
\and F.~Pasian\orcid{0000-0002-4869-3227}\inst{\ref{aff5}}
\and K.~Pedersen\inst{\ref{aff90}}
\and W.~J.~Percival\orcid{0000-0002-0644-5727}\inst{\ref{aff91},\ref{aff92},\ref{aff93}}
\and V.~Pettorino\inst{\ref{aff94}}
\and S.~Pires\orcid{0000-0002-0249-2104}\inst{\ref{aff68}}
\and G.~Polenta\orcid{0000-0003-4067-9196}\inst{\ref{aff95}}
\and J.~E.~Pollack\inst{\ref{aff96},\ref{aff97}}
\and M.~Poncet\inst{\ref{aff98}}
\and L.~A.~Popa\inst{\ref{aff99}}
\and L.~Pozzetti\orcid{0000-0001-7085-0412}\inst{\ref{aff38}}
\and F.~Raison\orcid{0000-0002-7819-6918}\inst{\ref{aff1}}
\and A.~Renzi\orcid{0000-0001-9856-1970}\inst{\ref{aff100},\ref{aff67}}
\and J.~Rhodes\inst{\ref{aff79}}
\and G.~Riccio\inst{\ref{aff45}}
\and E.~Romelli\orcid{0000-0003-3069-9222}\inst{\ref{aff5}}
\and M.~Roncarelli\orcid{0000-0001-9587-7822}\inst{\ref{aff38}}
\and E.~Rossetti\inst{\ref{aff39}}
\and R.~Saglia\orcid{0000-0003-0378-7032}\inst{\ref{aff41},\ref{aff1}}
\and D.~Sapone\orcid{0000-0001-7089-4503}\inst{\ref{aff101}}
\and B.~Sartoris\inst{\ref{aff41},\ref{aff5}}
\and P.~Schneider\orcid{0000-0001-8561-2679}\inst{\ref{aff14}}
\and T.~Schrabback\orcid{0000-0002-6987-7834}\inst{\ref{aff102}}
\and A.~Secroun\orcid{0000-0003-0505-3710}\inst{\ref{aff78}}
\and G.~Seidel\orcid{0000-0003-2907-353X}\inst{\ref{aff75}}
\and M.~Seiffert\orcid{0000-0002-7536-9393}\inst{\ref{aff79}}
\and S.~Serrano\orcid{0000-0002-0211-2861}\inst{\ref{aff21},\ref{aff9},\ref{aff103}}
\and C.~Sirignano\orcid{0000-0002-0995-7146}\inst{\ref{aff100},\ref{aff67}}
\and G.~Sirri\orcid{0000-0003-2626-2853}\inst{\ref{aff40}}
\and L.~Stanco\orcid{0000-0002-9706-5104}\inst{\ref{aff67}}
\and C.~Surace\inst{\ref{aff22}}
\and P.~Tallada-Cresp\'{i}\orcid{0000-0002-1336-8328}\inst{\ref{aff104},\ref{aff54}}
\and A.~N.~Taylor\inst{\ref{aff4}}
\and I.~Tereno\inst{\ref{aff63},\ref{aff105}}
\and R.~Toledo-Moreo\orcid{0000-0002-2997-4859}\inst{\ref{aff106}}
\and F.~Torradeflot\orcid{0000-0003-1160-1517}\inst{\ref{aff54},\ref{aff104}}
\and I.~Tutusaus\orcid{0000-0002-3199-0399}\inst{\ref{aff107}}
\and E.~A.~Valentijn\inst{\ref{aff108}}
\and L.~Valenziano\orcid{0000-0002-1170-0104}\inst{\ref{aff38},\ref{aff109}}
\and T.~Vassallo\orcid{0000-0001-6512-6358}\inst{\ref{aff41},\ref{aff5}}
\and Y.~Wang\orcid{0000-0002-4749-2984}\inst{\ref{aff110}}
\and J.~Weller\orcid{0000-0002-8282-2010}\inst{\ref{aff41},\ref{aff1}}
\and G.~Zamorani\orcid{0000-0002-2318-301X}\inst{\ref{aff38}}
\and J.~Zoubian\inst{\ref{aff78}}
\and E.~Zucca\orcid{0000-0002-5845-8132}\inst{\ref{aff38}}
\and A.~Biviano\orcid{0000-0002-0857-0732}\inst{\ref{aff5},\ref{aff6}}
\and E.~Bozzo\orcid{0000-0002-8201-1525}\inst{\ref{aff65}}
\and C.~Burigana\orcid{0000-0002-3005-5796}\inst{\ref{aff111},\ref{aff109}}
\and C.~Colodro-Conde\inst{\ref{aff112}}
\and D.~Di~Ferdinando\inst{\ref{aff40}}
\and G.~Mainetti\inst{\ref{aff113}}
\and M.~Martinelli\orcid{0000-0002-6943-7732}\inst{\ref{aff51},\ref{aff52}}
\and N.~Mauri\orcid{0000-0001-8196-1548}\inst{\ref{aff56},\ref{aff40}}
\and Z.~Sakr\orcid{0000-0002-4823-3757}\inst{\ref{aff114},\ref{aff107},\ref{aff115}}
\and V.~Scottez\inst{\ref{aff87},\ref{aff116}}
\and M.~Tenti\orcid{0000-0002-4254-5901}\inst{\ref{aff40}}
\and M.~Viel\orcid{0000-0002-2642-5707}\inst{\ref{aff6},\ref{aff5},\ref{aff2},\ref{aff11},\ref{aff3}}
\and M.~Wiesmann\inst{\ref{aff12}}
\and Y.~Akrami\orcid{0000-0002-2407-7956}\inst{\ref{aff117},\ref{aff118}}
\and V.~Allevato\inst{\ref{aff45}}
\and S.~Anselmi\orcid{0000-0002-3579-9583}\inst{\ref{aff67},\ref{aff100},\ref{aff24}}
\and C.~Baccigalupi\orcid{0000-0002-8211-1630}\inst{\ref{aff2},\ref{aff5},\ref{aff11},\ref{aff6}}
\and M.~Ballardini\orcid{0000-0003-4481-3559}\inst{\ref{aff119},\ref{aff120},\ref{aff38}}
\and F.~Bernardeau\inst{\ref{aff121},\ref{aff88}}
\and A.~Blanchard\orcid{0000-0001-8555-9003}\inst{\ref{aff107}}
\and S.~Borgani\orcid{0000-0001-6151-6439}\inst{\ref{aff31},\ref{aff6},\ref{aff5},\ref{aff11}}
\and S.~Bruton\orcid{0000-0002-6503-5218}\inst{\ref{aff122}}
\and R.~Cabanac\orcid{0000-0001-6679-2600}\inst{\ref{aff107}}
\and A.~Cappi\inst{\ref{aff38},\ref{aff123}}
\and C.~S.~Carvalho\inst{\ref{aff105}}
\and G.~Castignani\orcid{0000-0001-6831-0687}\inst{\ref{aff85},\ref{aff38}}
\and T.~Castro\orcid{0000-0002-6292-3228}\inst{\ref{aff5},\ref{aff11},\ref{aff6},\ref{aff3}}
\and G.~Ca\~{n}as-Herrera\orcid{0000-0003-2796-2149}\inst{\ref{aff89},\ref{aff124}}
\and K.~C.~Chambers\orcid{0000-0001-6965-7789}\inst{\ref{aff125}}
\and S.~Contarini\orcid{0000-0002-9843-723X}\inst{\ref{aff85},\ref{aff40},\ref{aff38}}
\and A.~R.~Cooray\orcid{0000-0002-3892-0190}\inst{\ref{aff126}}
\and J.~Coupon\inst{\ref{aff65}}
\and S.~Davini\inst{\ref{aff33}}
\and G.~De~Lucia\orcid{0000-0002-6220-9104}\inst{\ref{aff5}}
\and G.~Desprez\inst{\ref{aff127}}
\and S.~Di~Domizio\orcid{0000-0003-2863-5895}\inst{\ref{aff43},\ref{aff33}}
\and H.~Dole\orcid{0000-0002-9767-3839}\inst{\ref{aff128}}
\and A.~D\'{i}az-S\'{a}nchez\orcid{0000-0003-0748-4768}\inst{\ref{aff129}}
\and J.~A.~Escartin~Vigo\inst{\ref{aff1}}
\and S.~Escoffier\orcid{0000-0002-2847-7498}\inst{\ref{aff78}}
\and P.~G.~Ferreira\inst{\ref{aff7}}
\and F.~Finelli\orcid{0000-0002-6694-3269}\inst{\ref{aff38},\ref{aff109}}
\and L.~Gabarra\orcid{0000-0002-8486-8856}\inst{\ref{aff7}}
\and K.~Ganga\orcid{0000-0001-8159-8208}\inst{\ref{aff97}}
\and J.~Garc\'ia-Bellido\orcid{0000-0002-9370-8360}\inst{\ref{aff117}}
\and F.~Giacomini\orcid{0000-0002-3129-2814}\inst{\ref{aff40}}
\and G.~Gozaliasl\orcid{0000-0002-0236-919X}\inst{\ref{aff130},\ref{aff82}}
\and A.~Hall\orcid{0000-0002-3139-8651}\inst{\ref{aff4}}
\and S.~Ili\'c\orcid{0000-0003-4285-9086}\inst{\ref{aff131},\ref{aff98},\ref{aff107}}
\and S.~Joudaki\orcid{0000-0001-8820-673X}\inst{\ref{aff132},\ref{aff91},\ref{aff92}}
\and J.~J.~E.~Kajava\orcid{0000-0002-3010-8333}\inst{\ref{aff133},\ref{aff134}}
\and V.~Kansal\inst{\ref{aff135},\ref{aff136},\ref{aff137}}
\and C.~C.~Kirkpatrick\inst{\ref{aff77}}
\and L.~Legrand\orcid{0000-0003-0610-5252}\inst{\ref{aff8}}
\and A.~Loureiro\orcid{0000-0002-4371-0876}\inst{\ref{aff138},\ref{aff139}}
\and J.~Macias-Perez\inst{\ref{aff140}}
\and M.~Magliocchetti\inst{\ref{aff66}}
\and F.~Mannucci\inst{\ref{aff141}}
\and R.~Maoli\orcid{0000-0002-6065-3025}\inst{\ref{aff142},\ref{aff51}}
\and C.~J.~A.~P.~Martins\orcid{0000-0002-4886-9261}\inst{\ref{aff143},\ref{aff47}}
\and S.~Matthew\inst{\ref{aff4}}
\and L.~Maurin\orcid{0000-0002-8406-0857}\inst{\ref{aff128}}
\and R.~B.~Metcalf\orcid{0000-0003-3167-2574}\inst{\ref{aff85},\ref{aff38}}
\and M.~Migliaccio\inst{\ref{aff144},\ref{aff145}}
\and P.~Monaco\orcid{0000-0003-2083-7564}\inst{\ref{aff31},\ref{aff5},\ref{aff11},\ref{aff6}}
\and G.~Morgante\inst{\ref{aff38}}
\and S.~Nadathur\orcid{0000-0001-9070-3102}\inst{\ref{aff132}}
\and Nicholas~A.~Walton\orcid{0000-0003-3983-8778}\inst{\ref{aff146}}
\and L.~Patrizii\inst{\ref{aff40}}
\and V.~Popa\inst{\ref{aff99}}
\and D.~Potter\orcid{0000-0002-0757-5195}\inst{\ref{aff147}}
\and A.~Pourtsidou\orcid{0000-0001-9110-5550}\inst{\ref{aff4},\ref{aff148}}
\and M.~P\"{o}ntinen\orcid{0000-0001-5442-2530}\inst{\ref{aff82}}
\and I.~Risso\orcid{0000-0003-2525-7761}\inst{\ref{aff149}}
\and P.-F.~Rocci\inst{\ref{aff128}}
\and M.~Sahl\'en\orcid{0000-0003-0973-4804}\inst{\ref{aff150}}
\and A.~G.~S\'anchez\orcid{0000-0003-1198-831X}\inst{\ref{aff1}}
\and A.~Schneider\orcid{0000-0001-7055-8104}\inst{\ref{aff147}}
\and M.~Sereno\orcid{0000-0003-0302-0325}\inst{\ref{aff38},\ref{aff40}}
\and P.~Simon\inst{\ref{aff14}}
\and A.~Spurio~Mancini\orcid{0000-0001-5698-0990}\inst{\ref{aff81}}
\and J.~Steinwagner\inst{\ref{aff1}}
\and G.~Testera\inst{\ref{aff33}}
\and R.~Teyssier\orcid{0000-0001-7689-0933}\inst{\ref{aff151}}
\and S.~Toft\orcid{0000-0003-3631-7176}\inst{\ref{aff74},\ref{aff152},\ref{aff153}}
\and S.~Tosi\orcid{0000-0002-7275-9193}\inst{\ref{aff43},\ref{aff33},\ref{aff32}}
\and A.~Troja\orcid{0000-0003-0239-4595}\inst{\ref{aff100},\ref{aff67}}
\and M.~Tucci\inst{\ref{aff65}}
\and J.~Valiviita\orcid{0000-0001-6225-3693}\inst{\ref{aff82},\ref{aff83}}
\and D.~Vergani\orcid{0000-0003-0898-2216}\inst{\ref{aff38}}
\and G.~Verza\orcid{0000-0002-1886-8348}\inst{\ref{aff154},\ref{aff155}}
\and P.~Vielzeuf\inst{\ref{aff78}}}
										   
\institute{Max Planck Institute for Extraterrestrial Physics, Giessenbachstr. 1, 85748 Garching, Germany\label{aff1}
\and
SISSA, International School for Advanced Studies, Via Bonomea 265, 34136 Trieste TS, Italy\label{aff2}
\and
ICSC - Centro Nazionale di Ricerca in High Performance Computing, Big Data e Quantum Computing, Via Magnanelli 2, Bologna, Italy\label{aff3}
\and
Institute for Astronomy, University of Edinburgh, Royal Observatory, Blackford Hill, Edinburgh EH9 3HJ, UK\label{aff4}
\and
INAF-Osservatorio Astronomico di Trieste, Via G. B. Tiepolo 11, 34143 Trieste, Italy\label{aff5}
\and
IFPU, Institute for Fundamental Physics of the Universe, via Beirut 2, 34151 Trieste, Italy\label{aff6}
\and
Department of Physics, Oxford University, Keble Road, Oxford OX1 3RH, UK\label{aff7}
\and
Universit\'e de Gen\`eve, D\'epartement de Physique Th\'eorique and Centre for Astroparticle Physics, 24 quai Ernest-Ansermet, CH-1211 Gen\`eve 4, Switzerland\label{aff8}
\and
Institute of Space Sciences (ICE, CSIC), Campus UAB, Carrer de Can Magrans, s/n, 08193 Barcelona, Spain\label{aff9}
\and
Institut de Ciencies de l'Espai (IEEC-CSIC), Campus UAB, Carrer de Can Magrans, s/n Cerdanyola del Vall\'es, 08193 Barcelona, Spain\label{aff10}
\and
INFN, Sezione di Trieste, Via Valerio 2, 34127 Trieste TS, Italy\label{aff11}
\and
Institute of Theoretical Astrophysics, University of Oslo, P.O. Box 1029 Blindern, 0315 Oslo, Norway\label{aff12}
\and
International Centre for Theoretical Physics (ICTP), Strada Costiera 11, 34151 Trieste, Italy\label{aff13}
\and
Universit\"at Bonn, Argelander-Institut f\"ur Astronomie, Auf dem H\"ugel 71, 53121 Bonn, Germany\label{aff14}
\and
Excellence Cluster ORIGINS, Boltzmannstrasse e 2, 85748 Garching, Germany\label{aff15}
\and
Ludwig-Maximilians-University, Schellingstrasse 4, 80799 Munich, Germany\label{aff16}
\and
Donostia International Physics Center (DIPC), Paseo Manuel de Lardizabal, 4, 20018, Donostia-San Sebasti\'an, Guipuzkoa, Spain\label{aff17}
\and
IKERBASQUE, Basque Foundation for Science, 48013, Bilbao, Spain\label{aff18}
\and
Dipartimento di Scienze Matematiche, Fisiche e Informatiche, Universit\`a di Parma, Viale delle Scienze 7/A 43124 Parma, Italy\label{aff19}
\and
Technion Israel Institute of Technology, Israel\label{aff20}
\and
Institut d'Estudis Espacials de Catalunya (IEEC), Carrer Gran Capit\'a 2-4, 08034 Barcelona, Spain\label{aff21}
\and
Aix-Marseille Universit\'e, CNRS, CNES, LAM, Marseille, France\label{aff22}
\and
Area Science Park, Localit\`a Padriciano 99, 34149, Trieste, Italy\label{aff23}
\and
Laboratoire Univers et Th\'eorie, Observatoire de Paris, Universit\'e PSL, Universit\'e Paris Cit\'e, CNRS, 92190 Meudon, France\label{aff24}
\and
Dipartimento di Fisica "Aldo Pontremoli", Universit\`a degli Studi di Milano, Via Celoria 16, 20133 Milano, Italy\label{aff25}
\and
INFN-Sezione di Milano, Via Celoria 16, 20133 Milano, Italy\label{aff26}
\and
INFN Gruppo Collegato di Parma, Viale delle Scienze 7/A 43124 Parma, Italy\label{aff27}
\and
Department of Mathematics and Physics, Roma Tre University, Via della Vasca Navale 84, 00146 Rome, Italy\label{aff28}
\and
INFN-Sezione di Roma Tre, Via della Vasca Navale 84, 00146, Roma, Italy\label{aff29}
\and
Aix-Marseille Universit\'e, Universit\'e de Toulon, CNRS, CPT, Marseille, France\label{aff30}
\and
Dipartimento di Fisica - Sezione di Astronomia, Universit\`a di Trieste, Via Tiepolo 11, 34131 Trieste, Italy\label{aff31}
\and
INAF-Osservatorio Astronomico di Brera, Via Brera 28, 20122 Milano, Italy\label{aff32}
\and
INFN-Sezione di Genova, Via Dodecaneso 33, 16146, Genova, Italy\label{aff33}
\and
Division of Theoretical Physics, Ru\dj er Bo\v{s}kovi\'{c} Institute, Zagreb HR-10000, Croatia\label{aff34}
\and
Kavli Institute for Cosmology Cambridge, Madingley Road, Cambridge, CB3 0HA, UK\label{aff35}
\and
DAMTP, Centre for Mathematical Sciences, Wilberforce Road, Cambridge CB3 0WA, UK\label{aff36}
\and
School of Mathematics and Physics, University of Surrey, Guildford, Surrey, GU2 7XH, UK\label{aff37}
\and
INAF-Osservatorio di Astrofisica e Scienza dello Spazio di Bologna, Via Piero Gobetti 93/3, 40129 Bologna, Italy\label{aff38}
\and
Dipartimento di Fisica e Astronomia, Universit\`a di Bologna, Via Gobetti 93/2, 40129 Bologna, Italy\label{aff39}
\and
INFN-Sezione di Bologna, Viale Berti Pichat 6/2, 40127 Bologna, Italy\label{aff40}
\and
Universit\"ats-Sternwarte M\"unchen, Fakult\"at f\"ur Physik, Ludwig-Maximilians-Universit\"at M\"unchen, Scheinerstrasse 1, 81679 M\"unchen, Germany\label{aff41}
\and
INAF-Osservatorio Astrofisico di Torino, Via Osservatorio 20, 10025 Pino Torinese (TO), Italy\label{aff42}
\and
Dipartimento di Fisica, Universit\`a di Genova, Via Dodecaneso 33, 16146, Genova, Italy\label{aff43}
\and
Department of Physics "E. Pancini", University Federico II, Via Cinthia 6, 80126, Napoli, Italy\label{aff44}
\and
INAF-Osservatorio Astronomico di Capodimonte, Via Moiariello 16, 80131 Napoli, Italy\label{aff45}
\and
INFN section of Naples, Via Cinthia 6, 80126, Napoli, Italy\label{aff46}
\and
Instituto de Astrof\'isica e Ci\^encias do Espa\c{c}o, Universidade do Porto, CAUP, Rua das Estrelas, PT4150-762 Porto, Portugal\label{aff47}
\and
Dipartimento di Fisica, Universit\`a degli Studi di Torino, Via P. Giuria 1, 10125 Torino, Italy\label{aff48}
\and
INFN-Sezione di Torino, Via P. Giuria 1, 10125 Torino, Italy\label{aff49}
\and
INAF-IASF Milano, Via Alfonso Corti 12, 20133 Milano, Italy\label{aff50}
\and
INAF-Osservatorio Astronomico di Roma, Via Frascati 33, 00078 Monteporzio Catone, Italy\label{aff51}
\and
INFN-Sezione di Roma, Piazzale Aldo Moro, 2 - c/o Dipartimento di Fisica, Edificio G. Marconi, 00185 Roma, Italy\label{aff52}
\and
Institut de F\'{i}sica d'Altes Energies (IFAE), The Barcelona Institute of Science and Technology, Campus UAB, 08193 Bellaterra (Barcelona), Spain\label{aff53}
\and
Port d'Informaci\'{o} Cient\'{i}fica, Campus UAB, C. Albareda s/n, 08193 Bellaterra (Barcelona), Spain\label{aff54}
\and
Institute for Theoretical Particle Physics and Cosmology (TTK), RWTH Aachen University, 52056 Aachen, Germany\label{aff55}
\and
Dipartimento di Fisica e Astronomia "Augusto Righi" - Alma Mater Studiorum Universit\`a di Bologna, Viale Berti Pichat 6/2, 40127 Bologna, Italy\label{aff56}
\and
Jodrell Bank Centre for Astrophysics, Department of Physics and Astronomy, University of Manchester, Oxford Road, Manchester M13 9PL, UK\label{aff57}
\and
European Space Agency/ESRIN, Largo Galileo Galilei 1, 00044 Frascati, Roma, Italy\label{aff58}
\and
ESAC/ESA, Camino Bajo del Castillo, s/n., Urb. Villafranca del Castillo, 28692 Villanueva de la Ca\~nada, Madrid, Spain\label{aff59}
\and
Universit\'e Claude Bernard Lyon 1, CNRS/IN2P3, IP2I Lyon, UMR 5822, Villeurbanne, F-69100, France\label{aff60}
\and
Institute of Physics, Laboratory of Astrophysics, Ecole Polytechnique F\'ed\'erale de Lausanne (EPFL), Observatoire de Sauverny, 1290 Versoix, Switzerland\label{aff61}
\and
UCB Lyon 1, CNRS/IN2P3, IUF, IP2I Lyon, 4 rue Enrico Fermi, 69622 Villeurbanne, France\label{aff62}
\and
Departamento de F\'isica, Faculdade de Ci\^encias, Universidade de Lisboa, Edif\'icio C8, Campo Grande, PT1749-016 Lisboa, Portugal\label{aff63}
\and
Instituto de Astrof\'isica e Ci\^encias do Espa\c{c}o, Faculdade de Ci\^encias, Universidade de Lisboa, Campo Grande, 1749-016 Lisboa, Portugal\label{aff64}
\and
Department of Astronomy, University of Geneva, ch. d'Ecogia 16, 1290 Versoix, Switzerland\label{aff65}
\and
INAF-Istituto di Astrofisica e Planetologia Spaziali, via del Fosso del Cavaliere, 100, 00100 Roma, Italy\label{aff66}
\and
INFN-Padova, Via Marzolo 8, 35131 Padova, Italy\label{aff67}
\and
Universit\'e Paris-Saclay, Universit\'e Paris Cit\'e, CEA, CNRS, AIM, 91191, Gif-sur-Yvette, France\label{aff68}
\and
Istituto Nazionale di Fisica Nucleare, Sezione di Bologna, Via Irnerio 46, 40126 Bologna, Italy\label{aff69}
\and
INAF-Osservatorio Astronomico di Padova, Via dell'Osservatorio 5, 35122 Padova, Italy\label{aff70}
\and
University Observatory, Faculty of Physics, Ludwig-Maximilians-Universit{\"a}t, Scheinerstr. 1, 81679 Munich, Germany\label{aff71}
\and
von Hoerner \& Sulger GmbH, Schlo{\ss}Platz 8, 68723 Schwetzingen, Germany\label{aff72}
\and
Technical University of Denmark, Elektrovej 327, 2800 Kgs. Lyngby, Denmark\label{aff73}
\and
Cosmic Dawn Center (DAWN), Denmark\label{aff74}
\and
Max-Planck-Institut f\"ur Astronomie, K\"onigstuhl 17, 69117 Heidelberg, Germany\label{aff75}
\and
Department of Physics and Astronomy, University College London, Gower Street, London WC1E 6BT, UK\label{aff76}
\and
Department of Physics and Helsinki Institute of Physics, Gustaf H\"allstr\"omin katu 2, 00014 University of Helsinki, Finland\label{aff77}
\and
Aix-Marseille Universit\'e, CNRS/IN2P3, CPPM, Marseille, France\label{aff78}
\and
Jet Propulsion Laboratory, California Institute of Technology, 4800 Oak Grove Drive, Pasadena, CA, 91109, USA\label{aff79}
\and
AIM, CEA, CNRS, Universit\'{e} Paris-Saclay, Universit\'{e} de Paris, 91191 Gif-sur-Yvette, France\label{aff80}
\and
Mullard Space Science Laboratory, University College London, Holmbury St Mary, Dorking, Surrey RH5 6NT, UK\label{aff81}
\and
Department of Physics, P.O. Box 64, 00014 University of Helsinki, Finland\label{aff82}
\and
Helsinki Institute of Physics, Gustaf H{\"a}llstr{\"o}min katu 2, University of Helsinki, Helsinki, Finland\label{aff83}
\and
NOVA optical infrared instrumentation group at ASTRON, Oude Hoogeveensedijk 4, 7991PD, Dwingeloo, The Netherlands\label{aff84}
\and
Dipartimento di Fisica e Astronomia "Augusto Righi" - Alma Mater Studiorum Universit\`a di Bologna, via Piero Gobetti 93/2, 40129 Bologna, Italy\label{aff85}
\and
Department of Physics, Institute for Computational Cosmology, Durham University, South Road, DH1 3LE, UK\label{aff86}
\and
Institut d'Astrophysique de Paris, 98bis Boulevard Arago, 75014, Paris, France\label{aff87}
\and
Institut d'Astrophysique de Paris, UMR 7095, CNRS, and Sorbonne Universit\'e, 98 bis boulevard Arago, 75014 Paris, France\label{aff88}
\and
European Space Agency/ESTEC, Keplerlaan 1, 2201 AZ Noordwijk, The Netherlands\label{aff89}
\and
Department of Physics and Astronomy, University of Aarhus, Ny Munkegade 120, DK-8000 Aarhus C, Denmark\label{aff90}
\and
Centre for Astrophysics, University of Waterloo, Waterloo, Ontario N2L 3G1, Canada\label{aff91}
\and
Department of Physics and Astronomy, University of Waterloo, Waterloo, Ontario N2L 3G1, Canada\label{aff92}
\and
Perimeter Institute for Theoretical Physics, Waterloo, Ontario N2L 2Y5, Canada\label{aff93}
\and
Universit\'e Paris-Saclay, Universit\'e Paris Cit\'e, CEA, CNRS, Astrophysique, Instrumentation et Mod\'elisation Paris-Saclay, 91191 Gif-sur-Yvette, France\label{aff94}
\and
Space Science Data Center, Italian Space Agency, via del Politecnico snc, 00133 Roma, Italy\label{aff95}
\and
CEA Saclay, DFR/IRFU, Service d'Astrophysique, Bat. 709, 91191 Gif-sur-Yvette, France\label{aff96}
\and
Universit\'e Paris Cit\'e, CNRS, Astroparticule et Cosmologie, 75013 Paris, France\label{aff97}
\and
Centre National d'Etudes Spatiales -- Centre spatial de Toulouse, 18 avenue Edouard Belin, 31401 Toulouse Cedex 9, France\label{aff98}
\and
Institute of Space Science, Str. Atomistilor, nr. 409 M\u{a}gurele, Ilfov, 077125, Romania\label{aff99}
\and
Dipartimento di Fisica e Astronomia "G. Galilei", Universit\`a di Padova, Via Marzolo 8, 35131 Padova, Italy\label{aff100}
\and
Departamento de F\'isica, FCFM, Universidad de Chile, Blanco Encalada 2008, Santiago, Chile\label{aff101}
\and
Universit\"at Innsbruck, Institut f\"ur Astro- und Teilchenphysik, Technikerstr. 25/8, 6020 Innsbruck, Austria\label{aff102}
\and
Satlantis, University Science Park, Sede Bld 48940, Leioa-Bilbao, Spain\label{aff103}
\and
Centro de Investigaciones Energ\'eticas, Medioambientales y Tecnol\'ogicas (CIEMAT), Avenida Complutense 40, 28040 Madrid, Spain\label{aff104}
\and
Instituto de Astrof\'isica e Ci\^encias do Espa\c{c}o, Faculdade de Ci\^encias, Universidade de Lisboa, Tapada da Ajuda, 1349-018 Lisboa, Portugal\label{aff105}
\and
Universidad Polit\'ecnica de Cartagena, Departamento de Electr\'onica y Tecnolog\'ia de Computadoras,  Plaza del Hospital 1, 30202 Cartagena, Spain\label{aff106}
\and
Institut de Recherche en Astrophysique et Plan\'etologie (IRAP), Universit\'e de Toulouse, CNRS, UPS, CNES, 14 Av. Edouard Belin, 31400 Toulouse, France\label{aff107}
\and
Kapteyn Astronomical Institute, University of Groningen, PO Box 800, 9700 AV Groningen, The Netherlands\label{aff108}
\and
INFN-Bologna, Via Irnerio 46, 40126 Bologna, Italy\label{aff109}
\and
Infrared Processing and Analysis Center, California Institute of Technology, Pasadena, CA 91125, USA\label{aff110}
\and
INAF, Istituto di Radioastronomia, Via Piero Gobetti 101, 40129 Bologna, Italy\label{aff111}
\and
Instituto de Astrof\'isica de Canarias, Calle V\'ia L\'actea s/n, 38204, San Crist\'obal de La Laguna, Tenerife, Spain\label{aff112}
\and
Centre de Calcul de l'IN2P3/CNRS, 21 avenue Pierre de Coubertin 69627 Villeurbanne Cedex, France\label{aff113}
\and
Institut f\"ur Theoretische Physik, University of Heidelberg, Philosophenweg 16, 69120 Heidelberg, Germany\label{aff114}
\and
Universit\'e St Joseph; Faculty of Sciences, Beirut, Lebanon\label{aff115}
\and
Junia, EPA department, 41 Bd Vauban, 59800 Lille, France\label{aff116}
\and
Instituto de F\'isica Te\'orica UAM-CSIC, Campus de Cantoblanco, 28049 Madrid, Spain\label{aff117}
\and
CERCA/ISO, Department of Physics, Case Western Reserve University, 10900 Euclid Avenue, Cleveland, OH 44106, USA\label{aff118}
\and
Dipartimento di Fisica e Scienze della Terra, Universit\`a degli Studi di Ferrara, Via Giuseppe Saragat 1, 44122 Ferrara, Italy\label{aff119}
\and
Istituto Nazionale di Fisica Nucleare, Sezione di Ferrara, Via Giuseppe Saragat 1, 44122 Ferrara, Italy\label{aff120}
\and
Institut de Physique Th\'eorique, CEA, CNRS, Universit\'e Paris-Saclay 91191 Gif-sur-Yvette Cedex, France\label{aff121}
\and
Minnesota Institute for Astrophysics, University of Minnesota, 116 Church St SE, Minneapolis, MN 55455, USA\label{aff122}
\and
Universit\'e C\^{o}te d'Azur, Observatoire de la C\^{o}te d'Azur, CNRS, Laboratoire Lagrange, Bd de l'Observatoire, CS 34229, 06304 Nice cedex 4, France\label{aff123}
\and
Institute Lorentz, Leiden University, PO Box 9506, Leiden 2300 RA, The Netherlands\label{aff124}
\and
Institute for Astronomy, University of Hawaii, 2680 Woodlawn Drive, Honolulu, HI 96822, USA\label{aff125}
\and
Department of Physics \& Astronomy, University of California Irvine, Irvine CA 92697, USA\label{aff126}
\and
Department of Astronomy \& Physics and Institute for Computational Astrophysics, Saint Mary's University, 923 Robie Street, Halifax, Nova Scotia, B3H 3C3, Canada\label{aff127}
\and
Universit\'e Paris-Saclay, CNRS, Institut d'astrophysique spatiale, 91405, Orsay, France\label{aff128}
\and
Departamento F\'isica Aplicada, Universidad Polit\'ecnica de Cartagena, Campus Muralla del Mar, 30202 Cartagena, Murcia, Spain\label{aff129}
\and
Department of Computer Science, Aalto University, PO Box 15400, Espoo, FI-00 076, Finland\label{aff130}
\and
Universit\'e Paris-Saclay, CNRS/IN2P3, IJCLab, 91405 Orsay, France\label{aff131}
\and
Institute of Cosmology and Gravitation, University of Portsmouth, Portsmouth PO1 3FX, UK\label{aff132}
\and
Department of Physics and Astronomy, Vesilinnantie 5, 20014 University of Turku, Finland\label{aff133}
\and
Serco for European Space Agency (ESA), Camino bajo del Castillo, s/n, Urbanizacion Villafranca del Castillo, Villanueva de la Ca\~nada, 28692 Madrid, Spain\label{aff134}
\and
ARC Centre of Excellence for Dark Matter Particle Physics, Melbourne, Australia\label{aff135}
\and
Centre for Astrophysics \& Supercomputing, Swinburne University of Technology, Victoria 3122, Australia\label{aff136}
\and
W.M. Keck Observatory, 65-1120 Mamalahoa Hwy, Kamuela, HI, USA\label{aff137}
\and
Oskar Klein Centre for Cosmoparticle Physics, Department of Physics, Stockholm University, Stockholm, SE-106 91, Sweden\label{aff138}
\and
Astrophysics Group, Blackett Laboratory, Imperial College London, London SW7 2AZ, UK\label{aff139}
\and
Univ. Grenoble Alpes, CNRS, Grenoble INP, LPSC-IN2P3, 53, Avenue des Martyrs, 38000, Grenoble, France\label{aff140}
\and
INAF-Osservatorio Astrofisico di Arcetri, Largo E. Fermi 5, 50125, Firenze, Italy\label{aff141}
\and
Dipartimento di Fisica, Sapienza Universit\`a di Roma, Piazzale Aldo Moro 2, 00185 Roma, Italy\label{aff142}
\and
Centro de Astrof\'{\i}sica da Universidade do Porto, Rua das Estrelas, 4150-762 Porto, Portugal\label{aff143}
\and
Dipartimento di Fisica, Universit\`a di Roma Tor Vergata, Via della Ricerca Scientifica 1, Roma, Italy\label{aff144}
\and
INFN, Sezione di Roma 2, Via della Ricerca Scientifica 1, Roma, Italy\label{aff145}
\and
Institute of Astronomy, University of Cambridge, Madingley Road, Cambridge CB3 0HA, UK\label{aff146}
\and
Department of Astrophysics, University of Zurich, Winterthurerstrasse 190, 8057 Zurich, Switzerland\label{aff147}
\and
Higgs Centre for Theoretical Physics, School of Physics and Astronomy, The University of Edinburgh, Edinburgh EH9 3FD, UK\label{aff148}
\and
Dipartimento di Fisica, Universit\`a degli studi di Genova, and INFN-Sezione di Genova, via Dodecaneso 33, 16146, Genova, Italy\label{aff149}
\and
Theoretical astrophysics, Department of Physics and Astronomy, Uppsala University, Box 515, 751 20 Uppsala, Sweden\label{aff150}
\and
Department of Astrophysical Sciences, Peyton Hall, Princeton University, Princeton, NJ 08544, USA\label{aff151}
\and
Niels Bohr Institute, University of Copenhagen, Jagtvej 128, 2200 Copenhagen, Denmark\label{aff152}
\and
Cosmic Dawn Center (DAWN)\label{aff153}
\and
Center for Cosmology and Particle Physics, Department of Physics, New York University, New York, NY 10003, USA\label{aff154}
\and
Center for Computational Astrophysics, Flatiron Institute, 162 5th Avenue, 10010, New York, NY, USA\label{aff155}}    

\abstract{
We investigate the accuracy of the perturbative galaxy bias expansion in view of the forthcoming analysis of the \Euclid spectroscopic galaxy samples. We compare the performance of a Eulerian galaxy bias expansion using state-of-the-art prescriptions from the \gls{eft} with a hybrid approach based on Lagrangian perturbation theory and high-resolution simulations. These models are benchmarked against comoving snapshots of the flagship I N-body simulation at $z=(0.9,1.2,1.5,1.8)$, which have been populated with H$\alpha$ galaxies leading to catalogues of millions of objects within a volume of about $58\cGpc$. Our analysis suggests that both models can be used to provide a robust inference of the parameters $(h, \omegac)$ in the redshift range under consideration, with comparable constraining power. We additionally determine the range of validity of the \gls{eft} model in terms of scale cuts and model degrees of freedom. From these tests, it emerges that the standard third-order Eulerian bias expansion ---which  includes local and non-local bias parameters, a matter counter term, and a correction to the shot-noise contribution--- can accurately describe the full shape of the real-space galaxy power spectrum up to the maximum wavenumber of $\kmax=0.45\kMpc$, and with a measurement precision of well below the percentage level. Fixing either of the tidal bias parameters to physically motivated relations still leads to unbiased cosmological constraints, and helps in reducing the severity of projection effects due to the large dimensionality of the model. We finally show how we repeated our analysis assuming a volume that matches the expected footprint of Euclid, but without considering observational effects, such as purity and completeness, showing that we can get constraints on the combination $(h,\omegac)$ that are consistent with the fiducial values to better than the 68\% confidence interval over this range of scales and redshifts.
}

%
%

%
%
\titlerunning{Galaxy power spectrum modelling in real space}
\authorrunning{A. Pezzotta et al.}
   
\maketitle


\section{Introduction}
\label{sec:intro}

The large-scale distribution of galaxies is an extremely important source of cosmological information from the low-redshift Universe, complementing observations of the \gls{cmb}, such as those made using the \WMAP \citep[WMAP; ][]{Hinshaw2013} and \Planck \citep{planck2020}. In the course of the past two decades, observations of the \gls{lss} from spectroscopic galaxy redshift surveys, such as the 2dF Galaxy Redshift Survey \citep[2dFGRS; ][]{Colless2001}, the 6dF Galaxy Survey \citep[6dFGS; ][]{Jones2009}, the VIMOS VLT Deep Survey \citep[VVDS; ][]{LeFevre2013}, the Sloan Digital Sky Survey \citep[SDSS; ][]{York2000}, the WiggleZ Dark Energy Survey \citep[WiggleZ; ][]{Drinkwater2010}, the VIMOS Public Extragalactic Redshift Survey \citep[VIPERS; ][]{Guzzo2014}, the Galaxy And Mass Assembly \citep[GAMA; ][]{Driver2011}, and the Baryon Oscillation Spectroscopic Survey \citep[BOSS; ][]{Dawson2012} and its extension \citep[eBOSS; ][]{Dawson2016}, have provided a wealth of information on how gravitational instability shapes the large-scale matter distribution and on the relation between matter and galaxy density perturbations. At the same time, such observations have stood as a  testing ground for what has ultimately emerged as the standard cosmological model.

In the next decade, this picture is going to be significantly enriched by analyses performed by Stage-IV spectroscopic surveys, such as the Dark Energy Spectroscopic Instrument \citep[DESI; ][]{DESI2016} and \Euclid \citep{Euclid2011}, which are going to explore a still relatively uncharted epoch at $1\lesssim z\lesssim2$, when the Universe was only about half of its current age. In particular, \Euclid is going to collect the redshift of millions of H$\alpha$-emitting galaxies across a total sky surface of $15\,000\deg^2$, therefore increasing the statistical constraining power on the cosmological parameters to an unprecedented level for spectroscopic analyses in the low-redshift Universe. It comes with no surprise that the increase in statistical significance of the observations must necessarily be accompanied by an equivalent increase in the accuracy of the theoretical recipes used to analyse the data in order to keep systematic errors in the theory at a fraction of the statistical error budget. This becomes even more relevant in terms of the range of validity of the considered models, whose reach must be properly benchmarked against realistic mock samples.

The standard cosmological probe for galaxy clustering is the galaxy  \gls{2pcf}, or its Fourier transform, the galaxy power spectrum. Both statistics quantify the excess probability of finding galaxy pairs at a given separation with respect to the case of a purely random (Poissonian) distribution. These observables have been extensively used by recent experiments to place constraints on the cosmological parameters, either focusing on specific features such as \gls{bao} and \gls{rsd} in the so-called template-fitting approach \citep{Peacock2001, Tegmark2006, Guzzo2008, Blake2011, Reid2012, Beutler2012, Contreras2013, Howlett2015, Beutler2016, Okumura2016, Alam2017, Pezzotta2017, GilMarin2018, Hou2018, Wang2018, Zhao2018}, or in full-shape analyses \citep{Sanchez2014, Sanchez2016, Grieb2017, IvanovEtal2020, Troster2020, DAmicoEtal2020, Semenaite2022, Chen2022, Philcox2022, Carrilho2023, Semenaite2023, Moretti2023}.

For both kinds of approaches, the inference of cosmological information is made intrinsically more difficult by the presence of three separate effects that build on linear theory predictions. These are the non-linear gravitational evolution of the dark matter density field \citep{Bernardeau2002, BaumannEtal2012, Carrasco2012}, the relationship between the galaxy $\deltag$ and the matter $\delta$ density fields, known as thr `galaxy bias' (\citealp{kaiser:1984, bbks}; see \citealp{Desjacques2018} for a recent review), and finally the apparently anisotropic pattern in the distribution of galaxies due to the effect of their peculiar velocities on the observed redshift  \citep[\gls{rsd},][]{Kaiser1987, hamilton:1992, fisher:1995, Scoccimarro1999, Scoccimarro2004b, Taruya2010, Senatore2014, Perko2016}. Each of these effects needs to be carefully modelled in order to infer accurate cosmological constraints from the full shape of the galaxy power spectrum/\gls{2pcf}. This goal can be achieved in different ways: using numerical methods, such as \nbody simulations \citep[\eg][]{Kuhlen2012, Schneider2016, Springel2021}, adopting analytical approaches based on a perturbative solution to the equations governing the evolution of the matter and galaxy density fields \citep[\eg][]{FryGaz1993, Bernardeau2002,  McDonald2009, Carrasco2012, Assassi2014, Senatore2015, Desjacques2018}, or resorting to hybrid methods that combine the previous two methodologies \citep[\eg][]{Knabenhans_2021, Angulo2021, ZennaroEtal2021, Pellejero-IbanezEtal2022}.

In terms of galaxy bias, it has become standard practice to adopt a perturbative expansion of the galaxy density field using Eulerian coordinates, which can be expressed as a sum of partial derivatives of the gravitational and velocity divergence potentials, each one weighted by a corresponding free parameter to be fitted against measurements. When considering the one-loop galaxy power spectrum, the list includes the linear bias, $b_1$, expressed in terms of the dark matter density field in the large-scale limit, as $\deltag=b_1\delta$ \citep{kaiser:1984, bbks, cole/kaiser:1989, nusser/davis:1994, mo/white:1996, sheth/tormen:1999}, plus the next-to-leading-order correction obtained from a spherically symmetric gravitational collapse via a power-law Taylor expansion $\deltag=\sum_n b_n\,\delta^n/n!$ of the matter density field \citep{szalay:1988, Coles1993, FryGaz1993, scoccimarro/etal:2001, Smith2007, manera/etal:2010, Desjacques2009, frusciante/sheth:2012, schmidt/etal:2013}. This is further supplemented by the presence of non-local contributions generated by the cosmic tidal field \citep{bouchet/etal:1992, Catelan1998, Catelan2000, McDonald2009}, which have been proven to be essential for a correct description of the clustering of dark matter halos \citep{Manera2011, Roth2011} and to secure consistency between the results from the analysis of two- and three-point correlation measurements \citep{Pollack2012, Pollack2014}. These extra corrections were first detected in \nbody simulations in \cite{Chan2012} and \cite{Baldauf2012}, and have since then become a standard ingredient in the bias expansion. Additionally, the latter also takes into account the effects of short-range non-localities during the processes of galaxy formation, which lead to the presence of higher-than-second-order derivatives of the gravitational potential. At leading order in the power spectrum, higher derivatives appear with a term scaling as $\nabla^2\delta$ \citep{bbks, Matsubara1999, Desjacques2008, Desjacques2009}. Finally, the dependence on short-wavelength modes is included via an additional stochastic field $\varepsilon_{\rm g}(\xv)$ \citep{Dekel1999, Sheth1999, Taruya1999, Matsubara1999, bonoli/pen:2009, hamaus/etal:2010, Schmidt2016, Ginzburg2017},\footnote{In order for $\varepsilon_{\rm g}(\xv)$ to be completely uncorrelated from large-scale fluctuations, the hypothesis of primordial Gaussianity must hold true. On the contrary, $\varepsilon_{\rm g}(\xv)$ cannot be treated as a purely stochastic contribution.} which is responsible for a shot-noise contribution to the power spectrum. This correction deviates from the predictions of a purely Poissonian distribution, and at the same time can introduce a scale dependence due to the physical scale at which two objects can be mistaken for a single one, similarly to the exclusion effect for dark matter halos \citep{Scherrer1998, Sheth1999, cooray/sheth:2002, Smith2007, baldauf/etal:2013, baldauf/etal:2016}.

The high dimensionality of the parameter space for the model described above can be reduced by employing a set of physically motivated relations expressing a few higher-order bias parameters in terms of lower-order ones. A typical assumption is the conserved evolution of tracers (coevolution), which, from a local-in-matter-density expansion at the moment of formation, leads to the well-known local Lagrangian relations \citep{Chan2012, Baldauf2012, Eggemeier2019}. The latter have been adopted in the literature as a fairly conservative trade-off between sampling the whole set of bias parameters and fixing some of the model degrees of freedom, most notably in the analysis of the BOSS DR12 data release \citep{Sanchez2016, Grieb2017} to improve the statistical constraints on the cosmological parameters obtained from the anisotropic \gls{2pcf} and power spectrum. 

The standard bias expansion has been the subject of several tests in the literature, together with a validation of the coevolution relations mentioned in the previous paragraph. As an example, \cite{Saito2014} checked the consistency between the bias parameters fitted from the halo power spectrum and bispectrum (the Fourier transform of the three-point function) using a sample of measurements in different mass bins and at different redshifts, revealing an agreement between the two sets of bias measurements up to $k\sim0.1\kMpc$. The use of an irreducible bias basis, and also properly including a higher-derivative correction, was tested in \cite{Angulo2015b}, who showed that with this approach it is possible to extend the validity of the one-loop galaxy bias expansion up to $k\sim0.3\kMpc$ even at $z=0$. More recently, \cite{EggScoCro2011} analysed the accuracy of this expansion at fixed cosmology using simulated \gls{hod} catalogues built to mimic the clustering properties of the SDSS Main Galaxy Sample \citep{Strauss2002} and of the BOSS CMASS and LOWZ samples \citep{Eisenstein2011, Dawson2012, Reid2015}. The authors focused on the necessity to introduce both a higher-derivative term and a scale-dependent correction to shot noise while analysing the auto galaxy and cross galaxy-matter power spectrum. Findings from this study indicate that the standard one-loop bias expansion can be broken on scales $k\sim0.2\kMpc$ unless higher-order stochastic corrections are taken into consideration. \cite{PezCroEgg2108} and \cite{EggemeierEtal2021} extended this analysis to include a determination of the cosmological parameters, and explored the additional constraining power coming from the one-loop galaxy bispectrum. These works show how fixing the quadratic tidal bias as a function of the linear bias provides accurate results up to $k\sim0.35\kMpc$ for the galaxy power spectrum. A similar analysis was carried out by \cite{Oddo2021}, who assessed the constraining power of the galaxy bispectrum on the cosmological parameters, displaying a consistency up to $k\sim0.3\kMpc$ for the one-loop power spectrum and $k\sim0.09\kMpc$ for the tree-level bispectrum. Equivalent analyses in redshift space \citep{Markovic2019, Bose2020, delaBella2020, Gualdi2021, Rizzo2022, NicolaEtal2023arXiv}, or in terms of field-level comparisons \citep{SchmittfullEtal2019}, have also appeared in recent years, leading to compatible scenarios.

On a partially different side, numerical simulations \citep[see \eg][for a review]{Kuhlen2012} have proven to be an optimal way to reproduce the evolution of the matter and galaxy density field deep into the non-linear regime, and their use in analyses of galaxy redshift surveys has therefore multiplied in recent years thanks to a large number of different suites, such as\, DEMNUni \citep{Demnuni2015}, UNIT \citep{Chuang2019}, Quijote \citep{QuijoteSims2019}, Uchuu \citep{Ishiyama2021}, and AbacusSummit \citep{Maksimova2021, Yuan2022} simulations. In quantitative terms, state-of-art \nbody simulations can achieve an accuracy of  better than $2\%$  on the shape of the non-linear matter power spectrum down to scales of $k\sim10\kMpc$ \citep{Schneider2016, Springel2021, Angulo2021}, but unfortunately their application as a tool to infer cosmological information from real observations is limited by their extreme computational cost. However, in recent years, different methods have been proposed with the goal of increasing their range of applicability. This ranges from methods meant to speed up their production \citep[\eg][]{Monaco2002, Tassev2013, Izard2016} to ones designed to find an optimal interpolation strategy among a limited pool of high-resolution simulations \citep{Heitmann2014, Liu2018, Nishimichi2019, DeRose2019, Giblin2019, Wibking2019, Winther2019, Knabenhans2019, Knabenhans_2021}. Among this second category of methods, we highlight \bacco \citep{Angulo2021}, an emulator for the non-linear matter power spectrum that was recently extended to also include biased tracers in real \citep{ZennaroEtal2021} and redshift space \citep{Pellejero-IbanezEtal2022}, assuming a hybrid Lagrangian bias model, with the individual terms of the expansion directly emulated from high-resolution simulations, and a cosmology-rescaling technique meant to reduce the number of simulations required to train the emulator.

In this paper, we compare different models and test different scale cuts and bias relations on a sample of synthetic galaxy catalogues tailored to reproduce ---to the best of our knowledge--- the clustering signal of the H$\alpha$ sample that will be targeted by \Euclid. As we are interested in the relative performance of different theory models, we do not consider the presence of observational systematic uncertainties  in
this analysis. For example, effects such as purity and completeness of the sample will induce variations in the comoving number density considered in this work. At the same time, the sample purity, which is determined by the presence of line and noise interlopers, will also modify the overall shape of the measured $n$-point statistics. All of these effects are going to be investigated by a dedicated group in the Euclid Consortium, while a specific analysis on theory model selection with a more realistic analysis (including survey mask, selection effects, and combining multiple redshift bins) is going to be developed in a future paper (Euclid Collaboration: Moretti et al., in prep.).

Our goal is to test the range of validity of the one-loop galaxy bias expansion, which we quantify by means of three different performance metrics \citep{Osato2019}: the figure of bias, quantifying the accuracy of the model in terms of the recovery of the model parameters; the goodness of fit, measuring how well the best-fit model compares to the input data vectors; and the figure of merit, quantifying the statistical power of the model in constraining the cosmological parameters. These metrics are computed for each of the configurations we test, as a function of the maximum wave mode $k_\mathrm{max}$ included in the fit, exploring different bias relations meant to reduce the dimensionality of the parameter space. As we limit our attention to the real-space galaxy power spectrum alone, we focus on the recovery of the dimensionless Hubble parameter $h$, which is defined in terms of the Hubble constant $H_0$ as $H_0=100\,h\,\Hunits$, and of the cold dark matter density parameter $\omegac\equiv\Omega_{\rm c}h^2$, where $\Omega_{\rm c}$ is the corresponding fractional density parameter. At the same time, we avoid sampling the scalar amplitude $\As$, as this would lead to a strong degeneracy with the linear bias parameter $b_1$. This degeneracy can be partially broken if only considering the additional constraining power from higher-order statistics, cross-correlation with the matter density field, or in a multi-tracer analysis, or by considering the apparently anisotropic clustering amplitude when also including \gls{rsd}.

This work is the first installment in a series of \Euclid preparation papers meant to validate the theoretical framework used to analyse the full shape of two- and three-point clustering measurements from the final data sample. Here, we focus on the real-space galaxy power spectrum, while the corresponding three-point equivalent for the real-space galaxy bispectrum is going to be presented in Euclid Collaboration: Eggemeier et al. (in prep.). Both of these papers describe tests conducted in real space, that is, using the true comoving positions of galaxies inside the box instead of the positions  displaced because of \gls{rsd}. While this choice excludes one of the main observational probes of galaxy clustering, such analyses can provide an important testing ground for the model of galaxy bias. This includes calibration of optimal scale cuts for the model,\footnote{The reader should bear in mind that these scale cuts are determined only from the performance of galaxy bias. More realistic scale cuts, also including the effect of, for example, \gls{rsd} will be provided in one of the next entries of the series (Euclid Collaboration: Camacho et al., in prep.).} as well as testing different ways to reduce the dimensionality of the parameter space, such as using the coevolution relations. In addition, real-space analyses can become relevant in the context of modelling 3$\times$2 point statistics (photometric galaxy clustering, weak gravitational lensing, and galaxy--galaxy lensing), such as in the analysis performed by the Dark Energy Survey \citep[see \eg][for cosmological inference that requires a proper modelling of photometric galaxy bias]{Pandey2022, Porredon2022}. On the other hand, future installments of this series will focus on the modelling of the redshift-space equivalents of the statistics adopted in these works.

This article is structured in the following way. In Sect.~\ref{sec:data}
we present the simulated galaxy samples and the power-spectrum measurements and covariances used throughout the paper. In Sect.~\ref{sec:theory} we describe the theoretical models that we employed for the analysis, while in Sect.~\ref{sec:fitting_procedure} we describe the fitting procedure and the performance metrics used to quantify the goodness of the models for different configurations as a function of the maximum mode included in the fit. Finally, in Sect.~\ref{sec:results} we present the results of the analysis, and we draw our conclusions in Sect.~\ref{sec:conclusions}.


\section{Data}
\label{sec:data}

\subsection{\Euclid simulations}
\label{sec:simulations}

\begin{table}
\caption{Fiducial parameters of the flat $\Lambda$CDM cosmological model of the Flagship I simulation. From left to right, the list includes the value of the Hubble parameter $h$, the density parameter of cold dark matter $\omegac$, and baryons $\omegab$, the total neutrino mass $M_\nu$, the rms density fluctuations inside a sphere of radius $8\Mpc$, $\sigma_8$, and the scalar index $\ns$.} 
  \centering
   \renewcommand{\arraystretch}{1.3}
  \begin{tabular}{|c|c|c|c|c|c|}
    \hline
     \rowcolor{blue!5}
    $h$ & $\omegac$ & $\omegab$ & $M_\nu\,[\rm{eV}]$ & $\sigma_8$ & $\ns$ \\ 
    \hline
    $0.67$ & $0.121203$ & $0.0219961$ & $0$ &  $0.83$ & $0.97$ \\
    \hline
  \end{tabular}
  \label{tab:flagship_cosmology}
\end{table}

In order to determine the performance of the selected theoretical models, we first need a set of simulated data samples spanning the same redshift range that will be observed by \Euclid, and for which the input cosmology is known a priori.

In the following we make use of four comoving outputs, selected to cover the redshift range $0.9<z<1.8$ of the Flagship I simulation.\footnote{The roman numeral `I' is meant to differentiate the simulation adopted in this work from its more recent version, \ie Flagship II. The latter has been upgraded with respect to its predecessor in a number of way, such as by displaying a 2.5 times larger mass resolution, accounting for relativistic effects, and including massive neutrinos. However, because of the unavailability of halo comoving snapshots at the time the analysis presented in this paper first started, we decided to employ the older version of the Flagship for this work.} The latter has been carried out on the supercomputer Piz Daint, which is hosted by the \gls{cscs}, using the \texttt{PKDGRAV3} algorithm \citep{Potter2017}, and consists of a record-setting \nbody run with two trillions dark matter particles moving under the effect of gravity within a box of size $L=3780\Mpc$. The mass resolution of the simulation ($m_{\rm p} \sim 2.398 \times 10^9 \Ms$\,) allows us to marginally resolve halos with a typical mass $M_{\rm h}$ of few $10^{10}\Ms$\,, which host the majority of the $\mathrm{H}\alpha$ emission line galaxies that are going to be targeted by \Euclid. The nominal flat $\Lambda$CDM cosmology adopted to run the simulation as stated in \citet{Potter2017} differs from the fiducial cosmology assumed in this paper in the value of the spectral index ($n_s=0.96$ vs. $n_s=0.97$). This choice has been motivated since, during the course of this study, we observed subtle yet significant inconsistencies between
our models and the measurement in the Flagship I simulation. After contacting the
team responsible for running the simulation, they confirmed that the nominal parameters of the
simulation were previously wrongly communicated, and that they are in agreement with the ones we have identified. The latter are obtained by performing cosmological fits to high-resolution dark matter power spectrum measurements at various redshifts, and are summarized in Table \ref{tab:flagship_cosmology}. This procedure is detailed in Appendix~\ref{app:matter_power_spectrum}.

Each comoving snapshot has been populated with galaxies by firstly generating a \gls{fof} halo catalogue with a linking length $b=0.2$ and a minimum halo mass corresponding to ten dark matter particles,\footnote{Despite this small number, we only select halos hosting H$\alpha$ emitters with a minimum mass corresponding to few tens of matter particles, based on the redshift of the considered snapshot.} where the halo mass $M_{\rm h}$ is defined as the sum of all the identified particles. Subsequently, halos have been populated with galaxies using a \gls{hod} algorithm to match the abundance and clustering of the H$\alpha$ samples implemented in the main Flagship I lightcone catalog. \footnote{The particle lightcone has been built on the fly when running the N-body code at the \gls{cscs}, and features a full-sky distribution of dark matter particles in the redshift range $0<z<2.3$. This has been later populated with \texttt{Rockstar} halos \citep{BehWecWu2013} and galaxies using \gls{hod} and abundance matching techniques to reproduce the expected number density and luminosity profile of the H$\alpha$ models described in \cite{Pozzetti2016}.} The latter, in turn, is meant to reproduce the number density and clustering properties corresponding to two different $\mathrm{H}\alpha$ profiles, labelled as Model 1 and Model 3 in \cite{Pozzetti2016}. These samples are defined by different templates for the evolution of the luminosity function $\phi(L,z)$, from the use of a standard Schechter parametrization for Model 1 \citep{Schechter1976}, to the direct fit to real observations for Model 3. The net effect in terms of number density is that the Model 1 sample has almost twice as many objects as Model 3, which is more conservative in the selection of H$\alpha$ emitters, as shown in Fig.~4 of \cite{Pozzetti2016}.\footnote{We do not consider the additional Model 2 in this analysis since the total number density of this sample is of the same order of the one of the Model 1 sample, which already provides an optimistic number count.} A more detailed description of the Flagship H$\alpha$ lightcone and the pipeline for its construction will be provided in  Euclid Collaboration: Castander et al., in prep.


\begin{figure}
    \centering
    \includegraphics[width=\columnwidth]{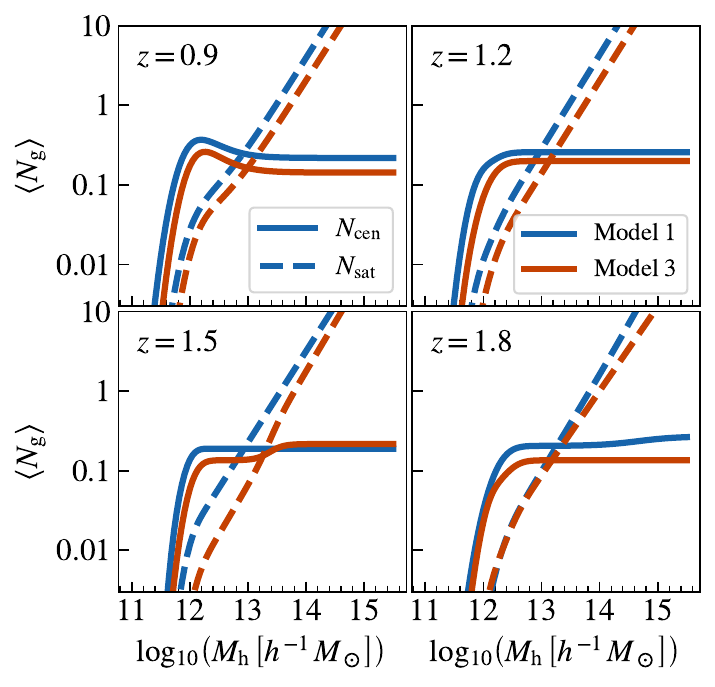}
    \caption{Halo occupation dsitribution profiles of the eight H$\alpha$ samples employed in this analysis. Individual panels show the profiles at different redshift (increasing from the top left to the bottom right panel) for both Model 1 (blue) and 3 (orange). Solid and dashed lines identify the average number of central and satellite galaxies, respectively, in halos of mass $M_{\rm h}$.}
    \label{fig:hod}
\end{figure}

In terms of the comoving snapshots, the \gls{hod} we implemented consists of a 8-parameter model, where the mean occupation numbers of central and satellite galaxies are defined as
\be
\begin{split}
    \ave{N_{\rm cen}}(M_{\rm h}) = \,&\frac{1}{2}f_{\rm cen}^{\,\rm max}\brackets{1+{{\rm erf}\paren{\frac{\log M_{\rm h}-\log M_{\rm min}}{\sigma_{\log M}}}}} \\
    & \times\left[1-\frac{1-f_{\rm cen}^{\rm min}/f_{\rm cen}^{\rm max}}{1+10^{\frac{2}{k}\left(\log M_{\rm h}-\log M_{\rm drop}\right)}}\right]\,,
\end{split}
\ee
\be
    \ave{N_{\rm sat}}(M_{\rm h}) = N_{\rm cen} \paren{\frac{M_{\rm h}}{M_1}}^\alpha.
\ee
Here, $M_{\rm min}$ is the typical minimum mass of halos hosting a central galaxy, $\sigma_{\log M}$ is the dispersion around $M_{\rm min}$, and $f_{\rm cen}^{\,\rm max}$ is the amplitude of the central galaxy occupation. We include mass-dependency of $\ave{N_{\rm cen}}$ above the transition scale $M_{\rm min}$ using three additional parameters, $M_{\rm drop}$, $k$, and $f_{\rm cen}^{\rm min}$. Finally, the mean occupancy of satellite galaxies is regulated by $M_1$, which is a simple normalisation factor, and $\alpha$, which corresponds to the slope of the power law distribution of satellites. In order to determine the distribution of galaxies inside halos, we employ a standard NFW profile \citep{NFW}.

In Figure~\ref{fig:hod} we show the fitted \gls{hod} profiles as a function of the host halo mass $M_{\rm h}$ for the eight samples: four redshifts times two different models. In all panels, both centrals and satellites profiles are shown, marked respectively with continuous and dashed lines. The mean occupation of central galaxies $f_{\rm cen}^{\,\rm max}$ inside dark matter halos does not converge to one, even for the most massive halos selected by the halo finder. This is a consequence of having selected a subsample (H$\alpha$ in this case) from the whole population present in the ligthcone. The typical expection value for the occupancy of central galaxies for $M_{\rm h}>M_{\rm min}$ slightly varies with the different samples, but is typically close to 0.2. This follows from the property of $\mathrm{H}\alpha$ emitters to be relatively young, blue, and star-forming galaxies, whereas, in massive halos, environmental effects such as galaxy-galaxy interactions, ram pressure stripping, and AGN feedback can suppress star formation in galaxies, effectively reducing the likelihood of finding actively star-forming central galaxies.

\begin{table}
\caption{Specifications for the \gls{hod} galaxy samples used in this analysis. The table lists the total number of objects $N_\mathrm{g}$, the mean comoving number density $\bar{n}$ of the sample, and the scale $k_{\rm sn}$ at which Poissonian shot-
noise becomes the leading contribution in the galaxy power spectrum. Following our convention on the normalization of the power spectrum, the latter is simply defined as the inverse of the mean number density. All the considered samples share the same volume, which coincides with the one of the Flagship I comoving outputs, \ie $(3780\Mpc)^3$. The last columns shows the volume factor $\eta$ between the full-box volume and the one of a \Euclid-like shell, as defined in Eq. (\ref{eq: eta_volume}).}
  \centering
  \renewcommand{\arraystretch}{1.3}
  \begin{tabular}{|c|c|r|c|c|c|}
    \hline
    \rowcolor{blue!5}
    & &  & $\bar{n}$ & $k_{\rm sn}$ & \\
    \rowcolor{blue!5}
     \multirow{-2}{*}{$z$}& \multirow{-2}{*}{HOD} & \multirow{-2}{*}{$N_{\rm g}$\hspace{18pt}} & \scalebox{0.88}{$\left[\kcMpc\right]$} & \scalebox{0.88}{$\left[\kMpc\right]$} & \multirow{-2}{*}{$\eta$}\\[3pt]
    \hline
    \multirow{2}{*}{0.9} & 1 & 201\,816\,513 & 0.0037 & 0.64 & \multirow{2}{*}{6.67}\\
    \cline{2-5}
                             & 3 & 110\,321\,755 & 0.0020 & 0.51 & \\
    \hline
    \multirow{2}{*}{1.2} & 1 & 108\,057\,141 & 0.0020 & 0.56 & \multirow{2}{*}{5.88}\\
    \cline{2-5}
                             & 3 & 55\,563\,490 & 0.0010 & 0.39 & \\
    \hline
    \multirow{2}{*}{1.5} & 1 & 69\,132\,138 & 0.0013 & 0.45 & \multirow{2}{*}{5.26}\\
    \cline{2-5}
                             & 3 & 31\,613\,213 & 0.0006 & 0.26 &  \\
    \hline
    \multirow{2}{*}{1.8} & 1 & 24\,553\,758 & 0.0005 & 0.26 & \multirow{2}{*}{3.33}\\
    \cline{2-5}
                             & 3 & 16\,926\,864 & 0.0003 & 0.22 &  \\
    \hline
  \end{tabular}
  \label{tab:hod_samples}
\end{table}

The total number of galaxies for each sample, their number density, and the scale $k_{\rm sn}$ at which the Poissonian shot-noise $P_{\rm sn}\equiv1/\bar{n}$ becomes the dominant contribution in the data vectors, are listed in Table \ref{tab:hod_samples}. A warning to be made is that, to determine the parameters for the \gls{hod} models, we selected galaxies from the lightcone assuming a significantly faint H$\alpha$ flux limit, corresponding to $f_{\,{\rm H}\alpha}=2\times10^{-16}\,{\rm erg}\,{\rm cm}^{-2}\,{\rm s}^{-1}$ \citep{Scaramella2022}, without assuming any realistic observational effect, such as target incompleteness, purity of the sample, and the impact of the angular footprint and radial selection function (Euclid Collaboration: Granett et al., in prep., Euclid Collaboration: Monaco et al., in prep.). This results in a sample with higher number density \citep[see \eg][for a recent forecast of H$\alpha$ emitters from HST]{Bagley2020}, with measured galaxy power spectra that are less affected by shot-noise. At the same time, the lack of line and noise interlopers allows us to neglect any extra contribution (Euclid Collaboration: Risso et al., in prep, Euclid Collaboration: Lee et al., in prep.) to the model galaxy power spectrum presented in Sect.~\ref{sec:theory}. As a consequence, given the high precision assumed to validate the theory models, we believe that our tests should provide a conservative estimate of their range of validity. We leave to future \Euclid analyses a more dedicated study of the impact of observational systematics.

\subsection{Measurements and covariances}

\begin{figure}
    \centering
    \includegraphics[width=\columnwidth]{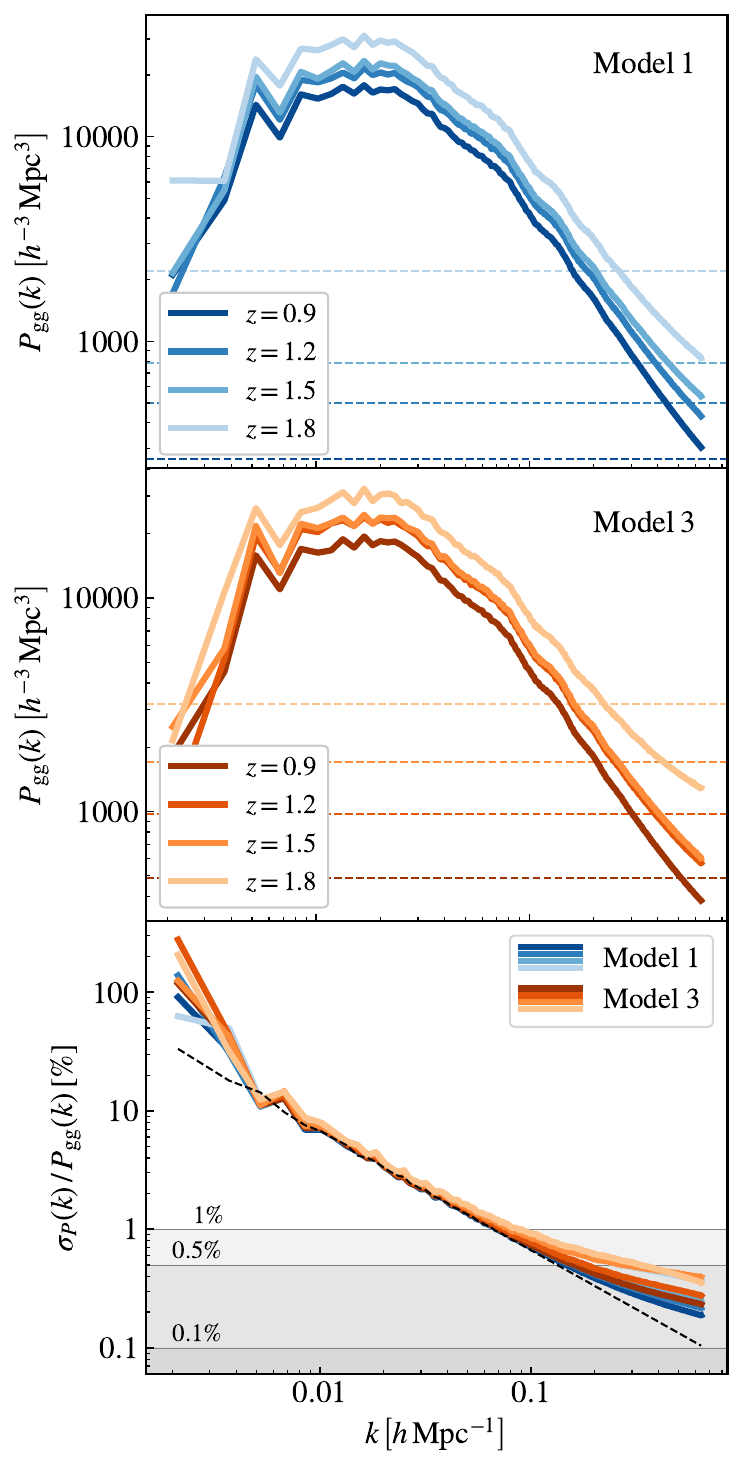}
    \caption{Galaxy power spectrum measurements and uncertainties obtained from the Flagship I comoving snapshots. \emph{Top}: Measurements of the Model 1 \gls{hod} samples. The colour gradient identifies the different redshifts of the samples, as shown in the legend. Dashed horizontal lines correspond to the amplitude of the Poisson shot-noise term $P^{\,\rm sn}$ -- obtained as the inverse of the number density specified in the last column of Table \ref{tab:hod_samples} -- for the different redshifts. \emph{Centre}: Same but for the Model 3 \gls{hod} samples. \emph{Bottom}: Error-to-measurement ratios, assuming a Gaussian covariance matrix as in Eq. (\ref{eq:gaussian_cov}). The coloured solid lines are obtained using the Poisson noise-subtracted power spectra, while the dashed black line highlights the linear relationship from Eq. (\ref{eq:gaussian_cov}), \ie $2/N_k$. Grey bands mark the $1\%$, $0.5\%$, and $0.1\%$ limit.}
    \label{fig:pk_measurements}
\end{figure}

For each of the samples described above, we measured the real-space galaxy power spectrum $\Pgg(k)$ using the publicly available \texttt{PowerI4} code.\footnote{Available at \url{https://github.com/sefusatti/PowerI4}. For this purpose, we do not use the official \Euclid code, LE3-PK-GC, since our only need is to measure the power spectrum from periodic boxes, without including also radial and angular selection effects that can be properly included using the official code.} The latter provides the functionality to compute the power spectrum from a particle distribution within a regular cubic box, using a variety of particle assignment schemes. For this analysis, we made use of a fourth-order interpolation scheme, otherwise known as piecewise cubic spline \citep[PCS; see][for the exact form of the kernel]{Sefusatti+2016}, coupled with an interlacing method to reduce the  aliasing contribution at high wave modes $k$.

We measured the power spectrum in the range defined by $\left[\kF,\kN\right]$, where $\kF=2\pi/L\sim0.0017\kMpc$ is the fundamental frequency in a box of linear size $L\,$, and $\kN=\pi \, N_{\rm grid}/L$ is the Nyquist frequency corresponding to a density grid of linear size $N_\mathrm{grid}$. We choose a grid resolution of $N_\mathrm{grid}=1024$ for the three dimensions of the box, to obtain measurements of the power spectrum up to a maximum wave mode of $0.8\kMpc$, and we sampled the $k$ range using a linear binning with step $\Delta k=\kF$.\footnote{For a limited number of the configurations presented in later sections, we carried out consistency checks with a different linear binning, namely $\Delta k=2\kF$ and $\Delta k=3\kF$, showing how the final constraints do not depend significantly on this choice.} The top and central panels of Fig.~\ref{fig:pk_measurements} show the power spectrum measurements for the Model 1 and 3 \gls{hod} samples respectively, with the redshift evolution over the available simulation snapshots marked by different lines in each panel. Differently from the evolution of the matter power spectrum, the galaxy power spectrum features an increasingly lower amplitude at lower redshifts. This can be explained by a larger linear galaxy bias at high redshift that overcomes the growth of matter fluctuations.

Since only a single realization is available for each of the \gls{hod} samples, we estimate the error covariance matrices associated to the data vectors using an analytical prediction in the Gaussian approximation, as explained in \cite{Grieb2016}. This implies that the variance $\sigma_P^{\,2}(k)$ associated to each $k$ bin is independent from the value of the galaxy power spectrum at different modes, and can be written as
\be
    \sigma_P^{\,2}\,(k)=\frac{2}{N_k}\,\Pgg^{\,2}(k)\,,
    \label{eq:gaussian_cov}
\ee
where $N_k$ identifies the number of independent wave modes falling in the bin $\left[k-\Delta k/2,k+\Delta k/2\right]$, while $\Pgg(k)$ is the theoretical non-linear galaxy power spectrum including shot-noise contributions. The latter has been obtained from a preliminary fit of the full non-linear model to the data vector of each sample, assuming, in a first iteration, an approximate but reasonable evaluation of the covariance itself. We expect the Gaussian approximation to be sufficient to our goals. The only additional contribution due to the galaxy trispectrum, here neglected, while noticeable at the relevant scales \citep{ScoccimarroZaldarriagaHui1999, Sefusatti2006, BlotEtal2015, BlotEtal2016, Bertolini2016a, Wadekar2020} does not lead to significant differences ($\lesssim 10\%$) on parameters constraints in the mildly non-linear regime \citep{Blot2019, WadekarIvanovScoccimarro2020}. This is also supported by the goal of this analysis, which is testing the relative performance of different theory models rather than providing absolute values for the parameter uncertainties. However, we highlight how the Gaussian approximation is not expected to deliver completely realistic error bars (to the level of accuracy mentioned above), and that future analyses, both on simulated and real data will be integrated with a more complex model also including non-linear corrections (either analytical or from N-body simulations).

The bottom panel of Fig.~\ref{fig:pk_measurements} shows the standard deviation normalised by the corresponding galaxy power spectrum, where the latter is shot-noise-subtracted to highlight the different level of noise in our samples. For this reason, the eight cases exhibit a deviation from the linear relation in Eq. \ref{eq:gaussian_cov} (shown with a black dashed line) at small ($k\gtrsim0.1\kMpc$) scales, where the shot-noise correction starts to become dominant with respect to the power spectrum signal. It should be noted how the relative error over this range of scales is well below the 1\% level. Similarly, at large scales ($k\lesssim0.005\kMpc$) we find a departure from the linear relationship due to the small amplitude of $\Pgg$, which can be clearly observed from the top and middle panel of Fig. \ref{fig:pk_measurements}. Finally, we note that this range of scales is also partially dominated by cosmic variance, due to the use of a single realisation of the Flagship simulation. 

\subsection{Volume rescaling}
\label{sec:vol_resc}

The main goal of our analysis is to carry out stringent tests to determine the range of validity of the standard one-loop galaxy bias model on the redshift range that will be explored by \Euclid. For this we make use of a volume $V_{\rm box}$ corresponding to the full-box size of the Flagship comoving outputs, which is significantly larger than the volume that will be covered by \Euclid. At the same time, we are interested in assessing the constraining power of the real-space galaxy power spectrum using a reference volume close to the one of an expected redshift bin of the full \Euclid volume, $V_{\rm shell}$. With this purpose in mind, we define new covariance matrices for the different samples presented in the previous sections, with an overall amplitude rescaled by the ratio between the volume of the Flagship comoving outputs and that of the \Euclid-like shells,
\be
        \eta=\frac{V_{\rm box}}{V_{\rm shell}}\,,
 \label{eq: eta_volume}
\ee
such that the rescaled covariance $C_{\rm shell}$ can be expressed\,\footnote{This rescaling is not valid in general, but can be performed when working under the assumption of a diagonal covariance matrix \citep[see \eg][for a similar rescaling to partially match the signal-to-noise ratio of different galaxy and halo samples]{EggScoCro2011}. We note that the approach adopted in this analysis bears some limitations, since the amplitude of shot-noise is also rescaled, leading to slightly larger data uncertainties. However, this effect should partially account for the fact that we assume only a Gaussian recipe to predict the covariance matrix, thus underestimating the error.} in terms of the original full-box covariance $C_{\rm box}$ as
\be
        C_{\rm shell} = \eta \, C_{\rm box}\,.
\ee

We follow \cite{EuclidForecast2019} and assume four non-overlapping redshift shells, centered at $z=(0.9,1.2,1.5,1.8)$, and with a depth of $\Delta z=(0.2,0.2,0.2,0.3)$, respectively, over a total projected area of $15\,000$ square degrees. With these values, we derive volume factors $\eta$ for each of the considered redshift bins, shown in the last column of Table \ref{tab:hod_samples}. We note that the mean values of the four redshift shells used in \cite{EuclidForecast2019} do not match perfectly the redshifts of the four comoving snapshots used in this work. However, this is only marginally relevant, since we do not carry out a proper comparison to the Fisher forecasts obtained in that analysis. In fact, this will be a more suited aspect of investigation when considering the same observables, that is, the Legendre multipoles of the anisotropic galaxy power spectrum, and especially when considering more realistic number densities, as pointed out in Sect.~\ref{sec:simulations}. 

A proper comparison between the results obtained using the full-box volume and the rescaled ones is presented in Sect.~\ref{sec:results_diff_volumes}. In addition to the \Euclid-like shells, we consider three additional volume rescalings, by dividing the range between $V_\mathrm{box}$ and $V_\mathrm{shell}$ into four evenly sized intervals. This leads to a total  of five different sets of covariances, based on the volumes defined above.


\section{Theoretical model}
\label{sec:theory}

In this section we describe the theoretical framework of \gls{pt}, which is essential to understand the evolution of post-inflationary fluctuations in the matter density field $\delta$ into the current large-scale distribution of galaxies via gravitational instability. This description is expected to be accurate only down to the mildly non-linear regime, where the amplitude of the density contrast $\delta$ is small enough to be perturbatively expanded. In the strong non-linear regime we expect this model to fail, as gravitational collapse leads to the formation of bound structures beyond the regime of validity of perturbative approaches.

For convenience, in the rest of this article we use the following notation for the integration over the infinite volume of a loop variable $\qv$,
\be
    \int_{\qv} \equiv \int \frac{\de^3 q}{(2\pi)^3}\,,
    \label{eq:int_loop_variable}
\ee
and adopt the following convention for the direct and inverse Fourier transform of the density contrast,
\begin{align}
        \delta(\kv) & \equiv (2\pi)^3 \int_{\xv} \euler^{-\imag\kv\cdot\xv}\delta(\xv)\,,\\
        \delta(\xv) & \equiv \int_{\kv} \euler^{\,\imag\kv\cdot\xv}\delta(\kv)\,.
        \label{eq:fourier_to_conf_transform}
\end{align}
The three-dimensional Dirac function is represented with the standard notation $\dirac$. Finally, the power spectrum $P_\mathrm{XX}(k,z)$ of any component, matter or biased tracer, is defined as the auto-correlation of the corresponding density field $\delta_{\rm X}$, such that
\be
    \ave{\delta_\mathrm{X}(\kv)\,\delta_\mathrm{X}(\kv')} \equiv (2\pi)^3\, P_\mathrm{XX}(k)\, \dirac\left(\kv+\kv'\right)\,,
    \label{eq:def_power_spectrum}
\ee
where the presence of the Dirac function and the independence of the power spectrum from the orientation of the wave mode $\kv$ reflect the underlying assumption of homogeneity and isotropy.

\subsection{Eulerian framework and effective field theory}
\label{sec:theory_eft}

\subsubsection{Modelling of the non-linear matter power spectrum}
\label{sec:modelling_of_the_non-linear_matter_power_spectrum}

We begin by summarising the most relevant outcomes of standard perturbation theory \citep[SPT; see \eg][for a review on the subject]{Bernardeau2002}. Its main assumption is that on large scales the dynamics of dark matter can be approximated as that of a perfectly pressureless fluid, with negligible effects from particle shell-crossing in multi-streaming regions. Under the so-called \gls{eds} approximation, we can write the matter density contrast using a perturbative expansion,
\be
    \delta=\delta^{\,(1)}+\delta^{\,(2)}+\delta^{\,(3)}+\ldots\,,
    \label{eq:delta_expansion}
\ee
where at each order $n$ the individual contribution $\delta^{\,(n)}$ is a function of the linear density contrast $\deltainit$,\footnote{In details, $\deltainit$ represents the initial density contrast linearly extrapolated to the redshift under consideration.}
\be
    \begin{split}
        \delta^{\,(n)}(\kv) = \int_{\qv_1\ldots\,\qv_n} & \dirac(\kv-\qv_{1\ldots\,n}) \: \Fn\,(\qv_1,\,\ldots,\,\qv_n) \\
        & \times \deltainit(\qv_1)\,\ldots\,\deltainit(\qv_n)\,.    
   \end{split}
    \label{eq:delta_n}
\ee
Here $\qv_{1\ldots\,n}\equiv\qv_1+\ldots+\qv_n\,$, and $\Fn$ is the $n$-th order symmetrised PT kernel describing the non-linear interaction among fluctuations at different wave modes $\qv_1,\,\ldots,\,\qv_n$. The vanishing argument of $\dirac$ reflects the translational invariance of the equations of motion in a spatially homogeneous universe.

Similarly, the non-linear matter power spectrum $\Pmm(k)$ can be expanded by combining Eqs. (\ref{eq:def_power_spectrum}) and (\ref{eq:delta_expansion}), leading to
\be
    \Pmm(k) = \Plin(k) + \Ponel(k) + \Ptwol(k) + \ldots\,,
    \label{eq:pmm_expansion}
\ee
where $\Plin\sim\ave{\deltainit^{\,2}}$ corresponds to the linear matter power spectrum, and at one-loop the only non-vanishing contributions are
\be
    \begin{split}
        \Ponel(k) = \: & P_{22}(k) + P_{13}(k) \\
        = \; & 2 \int_{\qv} F_2^{\;2}\paren{\kv-\qv,\qv} \, \Plin\paren{\abs{\kv-\qv}} \, \Plin(q) \\
        & + 6 \, \Plin(k) \int_{\qv} F_3\paren{\qv,-\qv,\kv} \, \Plin(q)\,.
    \end{split}
    \label{eq:pmm_oneloop}
\ee
For the sake of completeness, we report the expanded expressions for the second- and third-order symmetrised kernels, $F_2(\qv_1,\qv_2)$ and $F_3(\qv_1,\qv_2,\qv_3)$ in Appendix \ref{app:formulas}.

The one-loop model in SPT, however, fails to accurately describe the non-linear damping of the acoustic oscillations due to bulk flow displacements \citep{Eisenstein2007, Crocce2008, Baldauf-ir-res2015}. At first order, this effect can be reproduced in the theoretical model for $\Pmm(k)$ by a proper resummation of all \gls{ir} modes $q<k$, that is, of comoving separations larger than the one under consideration  (see \citealp{Crocce2006b} and \citealp{Crocce2008} for a description of the \gls{bao} smearing in the context of renormalised perturbation theory). 

A more standard procedure to include these corrections is based on the split of the linear power spectrum $\Plin$ as the sum of a smooth $\Pnw$ and wiggly $\Pw$ component \citep{Seo2008, Baldauf-ir-res2015, Blas2016a}, that is 
\be
    \Plin(k)=\Pnw(k)+\Pw(k)\,.
    \label{eq:plin=pnw+pw}
\ee
At leading order, it is possible to estimate the amplitude of the damping factor making use of the Zeldovich approximation \citep{Zeldovich1970}. This leads to an expression for the leading-order, \gls{ir}-resummed power spectrum,
\be
    \Pmmlo(k)=\Pnw(k)+\euler^{-k^2\Sigma^2}\Pw(k)\,,
    \label{eq:Pmmlo}
\ee
where $\Sigma^2$, representing the variance of the relative displacement field \citep{Eisenstein2007}, is defined as
\be
    \Sigma^2=\frac{1}{6\pi^2}\int_0^{\,k_{\rm s}}\Pnw(q)\,\brackets{1-j_0\paren{\frac{q}{\kosc}}+2j_2\paren{\frac{q}{\kosc}}} \,\de q\,.
    \label{eq:sigma2}
\ee
Here $j_n$ is the $n$-th order spherical Bessel function of the first kind, $\kosc=1/\losc$ is the wavelength corresponding to the BAO scale $\losc=110\,\Mpc$,\footnote{The value of $\losc$ should be varied as a function of the cosmological parameters. However, we cross-checked that for the relatively small parameter space explored in this analysis, its value do not deviate significantly from the one of a \Planck-like cosmology.} and $\ks=0.2\,\kMpc$ is the \gls{uv} integration limit.\footnote{Despite the correct integration range being scale-dependent, as it accounts for all wave modes $q<k$, we fix the \gls{uv} limit, similarly to what is done in \cite{IvanovEtal2020}, as it can be shown that the integrand of Eq. \eqref{eq:sigma2} is not providing significant contributions at $q>0.2\,\kMpc$.}

The next-to-leading order correction can be written by using the leading order term of Eq. \eqref{eq:Pmmlo} inside the expression for the one-loop corrections of Eq. \eqref{eq:pmm_oneloop}. This leads to the final formulation for the non-linear IR-resummed power spectrum \citep{Baldauf-ir-res2015, Blas2016a},
\be
    \begin{split}
        \Pmmnlo(k)=\,&\Pnw(k)+\paren{1+k^2\Sigma^2}\,\euler^{-k^2\Sigma^2}\Pw(k)\\
        &+\Ponel\brackets{\Pnw+\euler^{-k^2\Sigma^2}\Pw}(k)\,,
    \end{split}
    \label{eq:Pmmnlo}
\ee
where the square brackets of the last term mean that the evaluation of the one-loop correction is carried out using the leading order \gls{ir}-resummed power spectrum in place of the linear one.

Another partial failure of the model, which is equally shared by any recipe based on perturbative methods, is that its range of validity is limited to quasi-linear scales, where the assumption of a pressureless fluid is still justified. However, on scales approaching the non-linear scale $k_{\rm NL}$,\footnote{This is typically defined as the scale at which the dimensionless matter power spectrum, $$\Delta^2(k)\equiv\frac{k^3P(k)}{2\pi^2}\,,$$ becomes unity, that is, $\Delta^2(k_{\rm NL})\equiv1$\,.} dark matter particles experience shell-crossing, effectively introducing a non-zero pressure, which under more realistic conditions is further enhanced by the presence of baryonic processes, such as galaxy formation, ISM cooling, and AGN and supernovae feedback. These effects can be described in terms of a non-trivial stress-energy tensor which, at leading order, results in an additional contribution to the matter power spectrum \citep{PueblasScoccimarro2009, Carrasco2012, BaumannEtal2012},
\be
    \Pctr(k)=-2 \, c_\mathrm{s}^2 \, k^2\Pmmlo(k)\,,
    \label{eq:Pctr}
\ee
usually denoted as counterterm in the \gls{eft} framework. 
Here, the parameter $c_{\rm s}$ can be interpreted as an effective speed of sound \citep{BaumannEtal2012, Carrasco2014b, Baldauf2015b}, reflecting the influence of short-wavelength perturbations,
but accounts as well for the complex physics behind galaxy formation (when considering biased tracers of the matter density field).

Summarising, we can write the final expression for the model of the non-linear matter power spectrum as
\be
    \Pmm(k)=\Pmmnlo(k)+\Pctr(k)\,,
    \label{eq:Pmm_final}
\ee
which contains one free parameter, $c_\mathrm{s}$, which must be treated as a nuisance parameter to be fitted against real or, in our case, simulated measurements.

\subsubsection{Modelling of the non-linear galaxy power spectrum}
\label{sec:modelling_of_the_non-linear_galaxy_power_spectrum}

The general perturbative expansion of the galaxy density field $\deltag$ is based on the sum of all the individual operators that are a function of properties of the environment in which galaxies reside, such as the underlying matter density field and the large-scale tidal field. More precisely, this sum includes all those operators that are sourced by second derivatives of the gravitational potential $\Phi$ and the velocity potential $\Phi_{v}$ \citep[see][for a detailed review on the subject]{Desjacques2018}.

If we restrict our model to the one-loop prediction for the power spectrum, the relation between $\deltag$ and $\delta$ can be described considering only terms up to third order in the perturbations. In detail, this relation can be written as
\be
    \begin{split}
        \deltag(\xv)=\;&b_1\,\delta(\xv)+\bdtwod\,\nabla^{\,2}\delta(\xv)+\varepsilon_\mathrm{g}(\xv) \\
        &+\frac{b_2}{2}\,\delta^{\,2}(\xv)+\bGtwo\,\mathcal{G}_2\left(\Phi_{v}\,|\,\xv\right)+\bGthree\,\Gamma_3(\xv)+\ldots\,,
    \end{split}
    \label{eq:deltag_expansion}
\ee
where each operator is multiplied by a free bias parameter that determines its overall amplitude.\footnote{We note that this set of bias parameters needs to be renormalised before one can write the expression for the galaxy power spectrum \citep{McDonald2006, Assassi2014}. This procedure is meant to remove the dependence on the cutoff scale used to define the galaxy density field,  and to cancel the effect of higher-order bias parameters on large scales.} The different terms in Eq. \eqref{eq:deltag_expansion} can be summarised as follows.
\bi
    \item[(i)] At leading order, the shot-noise-corrected galaxy density field can be expressed using a linear and local relation in $\delta$. This relation is characterised by a linear bias parameter, $b_1$, which simply rescales the underlying matter density contrast by a constant factor \citep{kaiser:1984}.
     \item[(ii)] The effect of short-range non-localities during the process of galaxy formation is characterised by the presence of higher derivatives of the gravitational potential \citep{bbks, McDonald2009, Desjacques2009}. At leading order, the only non-zero term scales with the Laplacian of the matter density field, $\nabla^{\,2}\delta$, and has an amplitude regulated by the free parameter $\bdtwod$. The formation of structures involves the collapse of matter from a finite region of space, which for dark matter halos is well approximated by their Lagrangian radius $R$. Since the estimation of the corresponding radius for a given galaxy sample can be cumbersome, here we absorb the value of $R$ inside the definition of $\bdtwod$.
     \item[(iii)] The impact of short-scale fluctuations on the galaxy density field at larger separations is determined by an additional stochastic field, $\varepsilon_{\rm g}$, which, under the assumption of Gaussian initial conditions, is completely uncorrelated from large-scale perturbations. If galaxies are randomly distributed, the stochastic contribution to the galaxy power spectrum is purely represented by the Poisson limit, $1/\bar{n}$, with $\bar{n}$ corresponding to the mean number density of the selected sample.
    \item[(iv)] Moving to mildly non-linear scales, higher-order correlations of the density field appear \citep{Coles1993, FryGaz1993}, starting with a term proportional to $\delta^{\,2}$, characterised by a quadratic local bias $b_2$. This factor is expected from a spherically symmetric gravitational collapse, in which higher powers of $\delta$ become more relevant at progressively smaller scales. The third power of the matter density field is not included in Eq. (\ref{eq:deltag_expansion}) since its effect on the one-loop galaxy power spectrum is an extra contribution to the large-scale limit, which can be absorbed in the renormalization of the linear bias.
    \item[(v)] Even starting with a purely local-in-matter-density bias expansion at the time of formation, non-linear evolution is responsible for the generation of large-scale tidal fields \citep{Chan2012, Baldauf2012}. At leading order, the corrections given by the tidal stress tensor are represented by a non-local quadratic bias, $\bGtwo$, and by the second-order Galileon operator, $\mathcal{G}_2$, defined as
    \be
        \mathcal{G}_2\paren{\Phi\,|\,\xv} \equiv \brackets{\nabla_{ij}\,\Phi(\xv)}^{\,2}-\brackets{\nabla^{\,2}\,\Phi(\xv)}^{\,2}.
        \label{eq:G2_operator_conf}
    \ee
    In Fourier space, Eq. \eqref{eq:G2_operator_conf} can be written as
    \be
        \mathcal{G}_2(\kv)=\int_{\qv}S(\qv,\kv-\qv)\,\delta(\qv)\,\delta(\kv-\qv) \, ,
        \label{eq:G2_operator_fourier}
    \ee
    where
    \be
        S(\kv_1,\kv_2)\equiv \frac{\paren{\kv_1\cdot\kv_2}^{\,2}}{k_1^{\,2}\,k_2^{\,2}}-1
        \label{eq:S_operator}
    \ee
    is the Fourier-space kernel corresponding to the second-order Galileon operator $\mathcal{G}_2$.
    \item[(vi)] The next-to-leading-order correction to the tidal field can be obtained considering terms up to second-order in the potential of the displacement field \citep{Chan2012}. This contribution is represented by an additional non-local cubic bias, $\bGthree$, and by the operator
    \be
        \Gamma_3(\xv)\equiv\mathcal{G}_2\paren{\Phi\,|\,\xv}-\mathcal{G}_2\paren{\Phi_{v}\,|\,\xv},
        \label{eq:Gamma3_operator}
    \ee
    whose net effect inside Eq. \eqref{eq:deltag_expansion} is to include terms up to third order in perturbations of $\delta$.
\ei

All the terms giving a non-zero contribution to the one-loop galaxy power spectrum are listed in Eq. \eqref{eq:deltag_expansion}. The complete expression for $\Pgg$ then reads
\be
    \Pgg(k)=\Pgg^{\,\rm tree}(k) + \Pgg^{\,\rm1\mbox{-}loop}(k) + \Pgg^{\,\rm ctr}(k) + \Pgg^{\,\rm noise}(k) \, ,
    \label{eq:Pgg}
\ee
where the individual contributions can be written as
\be
    \Pgg^{\,\rm tree}(k)=b_1^{\,2} \, \Plin(k)\,,
    \label{eq:Pgg_tree}
\ee
\be
    \begin{split}
        \Pgg^{\,\rm 1\mbox{-}loop}(k) = \: &\Pggtt(k)+\Pggot(k) \\
        =\:& 2\int_{\qv}K_2^{\,2}\paren{\qv,\kv-\qv} \, \Plin\paren{\abs{\kv-\qv}} \, \Plin(q)  \\
        & + 6 \, b_1 \, \Plin(k) \int_{\qv} K_3\paren{\qv,-\qv,\kv} \, \Plin(q) \, ,
    \end{split}
    \label{eq:Pgg_oneloop}
\ee
\be
    \begin{split}
        \Pgg^{\,\rm ctr}(k)\;&=-2 \, b_1\paren{b_1\,c_{\rm s}^{\,2}+\bdtwod} \, k^2 \, \Plin(k) \\
        &\equiv -2 \, c_0 \, k^2 \, \Plin(k)\,,
    \end{split}
    \label{eq:Pgg_ctr}
\ee
\be
    \Pgg^{\,\rm noise}(k)=\frac{1}{\bar{n}}\paren{1+\aPone+\aPtwo \, k^2}\,.
    \label{eq:Pgg_noise}
\ee
For the sake of completeness, a complete list of the individual one-loop corrections can be found in Appendix \ref{app:formulas}. In the previous expressions, $K_2$ and $K_3$ are the symmetrised mode-coupling kernels for a generic biased tracer of the matter density field that follows the parametrization given in Eq. \eqref{eq:deltag_expansion}. In detail, they read
\be
    K_2(\kv_1,\kv_2)=b_1 \, F_2(\kv_1,\kv_2)+\frac{1}{2}b_2+\bGtwo \, S(\kv_1,\kv_2)\,,
    \label{eq:K2_kernel}
\ee
and
\be
    \begin{split}
        K_3(\kv_1,\kv_2,\kv_3)=\;&b_1 \, F_3(\kv_1,\kv_2,\kv_3) + b_2 \, F_2(\kv_1,\kv_2) \\
        & + 2 \, \bGtwo \, S(\kv_1,\kv_{23}) \, F_2(\kv_2,\kv_3) \\
        & + 2 \, \bGthree \, S(\kv_1,\kv_{23})\brackets{F_2(\kv_2,\kv_3)-G_2(\kv_2,\kv_3)}\,,
    \end{split}
    \label{eq:K3_kernel}
\ee
where $G_2(\kv_1,\kv_2)$ is the standard one-loop kernel for the non-linear evolution of the velocity divergence field, and Eq. \eqref{eq:K3_kernel} has to be symmetrised with respect to its arguments $\paren{\kv_1,\kv_2,\kv_3}$.

Inside Eq. \eqref{eq:Pgg_noise}\,, $\aPone$ is a free nuisance parameter that accounts for deviations from a purely Poissonian shot-noise.\footnote{This is expected since there is a physical separation under which two galaxies cannot simultaneously form, similarly to the exclusion effect for dark matter halos. The observed shot-noise can be either super- (signature of high-satellite star-forming galaxies) or sub-Poissonian (mostly typical of red central galaxies in massive halos), depending on the considered galaxy sample.} In addition, it is also required as a way to reabsorb the otherwise non-zero low-$k$ limit of one of the individual one-loop contributions, as explained in Appendix \ref{app:formulas}. Similarly, $\aPtwo$ parametrises the next-to-leading order correction, which scale as $k^2$. 

Since the leading-order higher-derivative correction is completely degenerate with the matter counterterm, as they are both proportional to the combination $k^2\,\Plin(k)$, we define a new more suited parameter,
\be
    c_0 \equiv b_1 \paren{b_1 \, c_{\rm s}^{\,2} + \bdtwod},
    \label{eq:def_counterterm}
\ee
to avoid the presence of unnecessary degeneracies between the parameters of the model.

In the previous scheme we have deliberately omitted the resummation of infrared modes, but, similarly to the case discussed in Sect.~\ref{sec:modelling_of_the_non-linear_matter_power_spectrum}, galaxy two-point clustering also has to be corrected for the effect of large-scale bulk motions. For this reason, we write the relations for the leading- and next-to-leading order \gls{ir}-resummed galaxy power spectra (mimicking Eqs. \ref{eq:Pmmlo} and \ref{eq:Pmmnlo}) as
\be
    \Pgglo(k) = b_1^{\,2}\brackets{\Pnw(k) + \euler^{-k^2\,\Sigma^2}\Pw(k)}+\frac{1}{\bar{n}}\paren{1+\aPone},
    \label{eq:Pgglo}
\ee
\be
    \begin{split}
        \Pggnlo(k) = \;& b_1^{\,2}\,\brackets{\Pnw(k) + \paren{1+k^2\Sigma^2}\euler^{-k^2\Sigma^2}\Pw(k)} \\
        & +\Pgg^{\,\rm 1\mbox{-}loop}\brackets{\Pnw+\euler^{-k^2\Sigma^2}\Pw}(k) \\
        & +\Pgg^{\,\rm ctr}\brackets{\Pnw+\euler^{-k^2\Sigma^2}\Pw}(k) +\Pgg^{\,\rm noise}(k)\,,
    \end{split}
    \label{eq:Pggnlo}
\ee
where, once again, the square brackets of the second and third terms in Eq. \eqref{eq:Pggnlo} reflect how the evaluation of the one-loop and counterterm contributions is carried out sourcing the leading order \gls{ir}-resummed matter power spectrum, $\Pmmlo$, in place of the linear power spectrum, $\Plin(k)$.

\subsubsection{Coevolution relations}
\label{sec:bias_rel}

A significant fraction of the bias parameters that have been introduced in this section enters in the expression for $\Pgg(k)$ only at higher-order, as clearly pointed out by the presence of only the linear bias $b_1$ in the expression for the leading-order galaxy power spectrum Eq. \eqref{eq:Pgglo}. This is significantly different from higher-order correlators of the galaxy density field, such as the galaxy bispectrum, for which both the local and non-local quadratic biases, $b_2$ and $\bGtwo$, appear also in the expression for the leading-order term, and can therefore be constrained with much better accuracy \citep{Oddo2021, EggemeierEtal2021}.

Given the poor constraining power of the galaxy power spectrum alone, it has become standard practice in real-data analyses to fix some of them to some physically motivated values or relations. This is important not only to obtain a larger constraining power for the remaining parameters, but also to ensure that none of them experiences strong degeneracies such as the one exhibited by the $\paren{\bGtwo,\bGthree}$ pair (see Appendix \ref{app:formulas}). In this work, we test two different relations, which are briefly summarised in the next paragraphs.

As already explained in Sect.~\ref{sec:modelling_of_the_non-linear_galaxy_power_spectrum}, even starting with a purely
local-in-matter-density expression, $\deltag(\delta)$, at the time of formation, non-linear gravitational evolution is responsible for the generation of a large-scale tidal field \citep{Fry1996, Chan2012}. This means that, even expressing the initial galaxy density field assuming only a spherically symmetric gravitational collapse -- and thus with only local bias parameters $b_n\neq0$ -- tidal contributions appear at later times because of gravitational evolution, leading to the presence of non-negligible tidal biases. Assuming that the total number of objects is conserved in time, it is possible to find a relation between the late-time non-local parameters and lower-order bias parameters, such that
\begin{align}
    & \bGtwo^{\rm coev}=-\frac{2}{7}(b_1-1)+b_{\mathcal{G}_2}^{\,\mathcal{L}}\,,\label{eq:bGtwo_coev} \\
    & \bGthree^{\rm coev}=-\frac{1}{6}(b_1-1)-\frac{5}{2}\bGtwo+b_{\Gamma_3}^{\,\mathcal{L}}\,,
    \label{eq:bGthree_coev}
\end{align}
where the bias parameters with a superscript $\mathcal{L}$ stand for the corresponding Lagrangian quantities, that is, at the time of formation. The previous relations are commonly referred to as coevolution, or local Lagrangian relations when setting to zero the Lagrangian bias, and have been extensively used in most real-data analyses to fix one or both non-local parameters \citep{Feldman2001, GilMarin2015, Sanchez2016, Grieb2017}. However, recent results \citep{LazSch1809,AbiBal1807} have indicated that measurements from numerical simulations seem to suggest lower values for $\bGtwo$ with respect to its local Lagrangian relation.

An alternative approach for fixing $\bGtwo$, found to be more accurate when compared to results from \nbody simulations and derived using the excursion-set formalism, has been proposed by \cite{SheChaSco1304}. In this case, it is possible to express $\bGtwo$ as a quadratic form in terms of the linear bias $b_1$, such that 
\be
    \bGtwo^\mathrm{ex-set}=0.524-0.547\,b_1 + 0.046\,b_1^{\,2}\,.
    \label{eq:bG2_exset}
\ee
Such expressions are based on theoretical considerations on halo bias that only take into account the halo mass. As a consequence, they neglect potentially important effects, such as assembly bias \citep[see \eg][for an analysis carried out, respectively, on dark matter halos and galaxies from hydrodinamical simulations]{Lazeyras2021, Barreira_2021}. This means that their applicability to a real-data analysis must be carefully assessed \citep[see][for recent applications]{EggScoCro2011,PezCroEgg2108}. Nonetheless, their use in this analysis is well justified, since we focus on \gls{hod} samples for which the assignment of a galaxy into a host halo is only determined by the mass of the latter.

In Sect.~\ref{sec:results} we carry out tests to determine whether the previously defined relations can be employed to analyse clustering measurements adopting \Euclid requirements.

\subsection{Hybrid Lagrangian bias expansion model}
\label{sec:bacco}

In the previous sections, the relationship between the galaxy and the matter density field has been described through an Eulerian-based framework. However, this is not the only description of the galaxy power spectrum in the quasi-linear regime. Other approaches are possible, often based to various degrees on results from numerical simulations. We consider here the so-called `hybrid Lagrangian' models. They draw from Lagrangian perturbation theory for the bias expression connecting galaxy and matter overdensities, but rely on simulations to capture the development of non-linearities when converting Lagrangian quantities to the observable Eulerian quantities.

The Lagrangian bias expansion describes the clustering of biased tracers in terms of a superposition of Lagrangian operators advected to Eulerian coordinates. It was first developed at one-loop in \gls{pt} by \cite{Matsubara2008}, while \cite{ModiChenWhite2020} proposed to combine the perturbative approach on bias with measurements of the advected operators from \nbody simulations. This hybrid approach potentially allows us to push the bias expansion formalism to smaller scales with respect to purely perturbative approaches \citep[see \eg][for recent applications of the model to DES Y1 data and numerical simulations]{Hadzhiyska2021, DeRose2023}. Recent implementations include the works of \cite{KokronEtal2021} and \cite{ZennaroEtal2021} in real space, and of \cite{Pellejero-IbanezEtal2022} in redshift space. In the present work we consider the implementation in the code \bacco \citep{ZennaroEtal2021}.\footnote{\url{https://bacco.dipc.org/emulator.html}} It describes the Eulerian galaxy overdensity in terms of a second-order expansion of the Lagrangian galaxy density field $\deltag(\qv)$ where $\qv$ is the Lagrangian position corresponding to the Eulerian position $\xv=\qv+\boldsymbol{\Psi}(\qv)$ with $\boldsymbol{\Psi}(\qv)$ being the displacement field-connecting initial and final positions. This means that the Eulerian overdensity is given by
\be
1+\deltag(\xv) = \int_{\qv} \, w(\qv) \, \dirac\big(\xv - \qv - \boldsymbol{\Psi}(\qv)\big)\,,
\ee
where $w(\qv)$ expresses the weighting function that transforms the matter field into the galaxy field,
\be
        \begin{split}
                w(\qv)=\;&1+b_1^{\,{\cal L}}\,\delta(\qv)+b_2^{\,{\cal L}}\,\paren{\delta^{\,2}(\qv)-\ave{\delta^{\,2}}}\\
                &+b_{s^{\,2}}^{\,{\cal L}}\,\brackets{s^{\,2}(\qv)-\ave{s^{\,2}}}+b_{\nabla^2\delta}^{\,{\cal L}}\,\nabla^2\delta(\qv)\,.
        \end{split}
 \label{eq:lagrangian_exp}
\ee
Here the total list of operators built on the matter density field $\delta(\qv)$ consists of $\mathcal{O}=\{1, \delta, \delta^{\,2}, s^{\,2}, \nabla^2 \delta\}$, and the individual entries correspond to the fully non-linear matter distribution, not weighted $(1)$ and weighted $(\delta)$ by the linear overdensity field, the squared linear overdensity field $\delta^{\,2}$, the squared traceless tidal field $s^{\,2}$,\footnote{In this expansion $s_{ij}(\qv)=\partial_i\,\partial_j\,\Phi(\qv)-\delta^{\rm K}_{ij}\,\delta(\qv)$, where $\delta^{\rm K}_{ij}$ is the Kronecker delta function. This definition matches the one of the second-order Galileon operator $\mathcal{G}_2$, as in Eq. \eqref{eq:G2_operator_conf}, with the only difference being that the two operators are defined in Lagrangian and Eulerian space, respectively. This is different from the parametrisation adopted in \eg\, \cite{Desjacques2018}, where the quadratic tidal operator is defined as $K_{ij} =\partial_i\,\partial_j\,\Phi(\kv)-\frac{1}{3}\delta^{\rm K}_{ij}\,\delta(\kv)$.} and the Laplacian of the linear overdensity field $\nabla^2 \delta$, respectively. Note that unlike the Eulerian bias basis presented before, the expansion in Eq. (\ref{eq:lagrangian_exp}) does not include the next-to-leading-order correction to the tidal field, captured by the operator $\Gamma_3$. \footnote{We note that there is nothing preventing a complete expansion up to third order in $\delta$ even in Lagrangian space. While the presence of this contribution may be partially relevant in terms of field level or higher-order statistics, such as the galaxy bispectrum, accuracy checks carried out by the \bacco team have led to the conclusion that neglecting the cubic operator is a robust assumption for the analysis of the galaxy power spectrum.} This implies that the two bases are only equivalent under the assumption of coevolution for the Eulerian parameter $\bGthree$ (see Eq. \ref{eq:bGthree_coev}).

The final model depends on four free parameters, the linear bias $b_{1}^{\,{\cal L}}$, the local quadratic bias $b_{2}^{\,{\cal L}}$, the tidal quadratic bias $b_{s^2}^{\,{\cal L}}$, and the higher-derivative bias $b_{\nabla^2\delta}^{\,{\cal L}}$, to which we add the extra free parameter $\aPone$ to account (at first order) for non-Poissonian shot-noise, in the same way as it is done in the Eulerian \gls{pt} model. We use a different notation for the quadratic tidal bias, since the definition of the tidal field operator is slightly different from the one presented in Sect.~\ref{sec:theory_eft}. The same is true for the Laplacian bias, which in this case only models higher-derivative corrections, but could also (partially) absorb unmodelled non-local effects coming from higher orders, extra physics, such as baryonic effects, or the smoothing of the density field performed in Lagrangian space.

The galaxy power spectrum can then be expressed as
\be
    P_{\rm gg}(k) = \sum_{i,j} b_i^{\,{\cal L}}\, b_j^{\,{\cal L}}\, P_{ij}\,(k) + \dfrac{1+\aPone}{\bar{n}_{\rm g}}\,,
    \label{eq:Pgg_lag}
\ee
where $P_{ij}(k)$ are the 15 cross-spectra of the five previously defined advected operators. To compute the $P_{ij}$ terms, \bacco has been trained with high-resolution $P_{ij}$ measurements from 800 combinations of cosmologies and redshifts, obtained applying the cosmology-rescaling technique to four main \nbody simulations \citep{AnguloWhite2010, ZennaroEtal2019, ContrerasEtal2020}.

As a final remark, we note that, even if it is possible to find a relation between the Lagrangian and Eulerian bias parameters, the two sets do not exactly correspond to the same physical quantities. This happens because in the purely perturbative Eulerian framework they properly represent the response of galaxy formation to large-scale perturbations, whereas in the hybrid Lagrangian one this physical meaning is lost due to the extrapolation of the individual operators to the non-linear regime (since the individual operators of the bias expansion are measured from N-body simulations and thus contain higher-order contributions).


\section{Model selection and fitting procedure}
\label{sec:fitting_procedure}

In this section we describe the methodology used to determine the best combination between different models, scale cuts, and bias configurations. In addition we list the details of the fitting procedure and the priors of the selected parameter spaces.

\subsection{Performance metrics}
\label{sec:perf_metrics}

In the context of model selection, the most relevant aspects to take into consideration are the range of validity of a given model and the precision and accuracy of the constraints on the parameters of interest. The procedure that we adopt is based on the selection of the maximum wave mode $\kmax$ up to which the model is still capable of providing a good description of the data vectors, while still recovering the correct input parameters. This can be quantified by means of three different performance metrics \citep[employed in \eg][]{Osato2019, EggScoCro2011, PezCroEgg2108, EggemeierEtal2021}, which are described in the next subsections.

\subsubsection{Figure of bias}
\label{sec:figure_of_bias}

One of the main requirement that the theoretical model has to satisfy is that its fit to the data return unbiased model parameters. The parameters controlling bias, shot-noise, and counterterms can be effectively treated as free nuisance parameters, to be marginalised over after sampling the joint posterior distribution. The set of parameters of interest is therefore restricted to the cosmological parameters, in our case $\thetav\equiv\left\{h,\,\omegac\right\}$.

We quantify the unbiasedness of the model in the recovery of $\thetav$ in terms of the \gls{fob} defined as
\be
    {\rm FoB}(\thetav) \equiv \brackets{\big(\ave{\thetav}-\thetav_{\,\rm fid}\big)^{\intercal}\,S^{-1}(\thetav)\,\big(\ave{\thetav}-\thetav_{\,\rm fid}\big)}^\frac{1}{2},
    \label{eq:fob}
\ee
where $\ave{\thetav}$ and $\thetav_{\,\rm fid}$ represent the mean of the posterior distribution of the selected parameters and their fiducial values, respectively, and $S(\thetav)$ is a square matrix containing the auto- and cross-covariance among all the entries of the vector $\thetav$.\footnote{This means that, for the case we are considering, where $\thetav=\left\{h,\,\omegac\right\}$, $S$ is a $2\times2$ matrix containing the variance of $h$ and $\omegac$ on its diagonal, and the cross-covariance between them on the off-diagonal entries.} The meaning of Eq. \eqref{eq:fob} is straightforward: we are quantifying the deviation of the posterior distribution from the fiducial values of the corresponding parameters, and expressing this information in terms of the intrinsic error of those parameters. In the case where $\thetav$ consists of only one parameter, the \gls{fob} simply expresses how far the posterior is from the fiducial value in units of the standard deviation of the parameter, with the $68\%$ and $95\%$ percentiles corresponding to values of \gls{fob} of 1 and 2, respectively. Note that when considering more than one parameter these values change, as they need to be computed by directly integrating a multivariate normal distribution with the corresponding number of dimensions. For $n=2$, we evaluate that the new thresholds for the $68\%$ and $95\%$ percentiles are $1.52$ and $2.49$, respectively.

\subsubsection{Goodness of fit}
\label{sec:chi2}

The goodness of fit quantifies the consistency of the theoretical model $P^{\,\rm th}$ with the input data vector $P^{\,\rm data}$. We consider the standard $\chi^2$ test, corresponding to
\be
        \chi^2(\thetav)=\sum_{i=1}^{N_{\rm bins}}\sum_{j=i}^{N_{\rm bins}}\brackets{P^{\,\rm th}_i\,(\thetav)-P^{\,\rm data}_i\,} \,C^{-1}_{ij}(\thetav)\,\brackets{P^{\,\rm th}_j\,(\thetav)-P^{\,\rm data}_j\,}\,.
\ee
This results in a distribution of $\chi^2$ values across the sampled parameter space. Instead of picking the $\chi^2$ corresponding to the maximum-likelihood position, whose estimation from the sampled posterior distribution is subject to noise, \footnote{An optimal research of the maximum-likelihood position could be carried out employing a $\chi^2$ minimiser, or in the context of a more frequentist approach based on a profile likelihood, which we do not perform in this work.} we compute the posterior-averaged value, $\ave{\,\chi^2}$, from a weighted average over all sampled parameter combinations, which is instead a more stable quantity (see Appendix \ref{app:chi2} for a comparison between the two approaches). The posterior-averaged $\chi^2$ is then compared to the predictions from the $68\%$ and $95\%$ percentiles of the $\chi^2$ distribution with the corresponding number of degrees of freedom. The latter is simply defined as $N_{\rm{dof}}=N_{\rm{bins}}-N_{\rm{pars}}$, where $N_{\rm{bins}}$ is the total number of independent wave mode bins up to the selected $\kmax$, and $N_{\rm{pars}}$ is the total number of free parameters of the model. 

\subsubsection{Figure of merit}

Finally, each configuration of the model -- that is a given scale cut and bias assumptions -- is inspected to determine its statistical power in constraining the parameters $\thetav$. For this purpose, similarly to what is done for the figure of bias, we define a \gls{fom} for a given set of model parameters $\thetav$ as \citep{Wang2008}
\be
{\rm FoM}(\thetav) = \brackets{\det\Big(S(\thetav)\Big)}^{\,-{1}/{2} } ,
\ee
where $S(\thetav)$ is once again the covariance matrix of the parameters $\thetav$, and $\det(S)$ its determinant. The meaning of this quantity can be more clearly understood assuming a flat posterior distribution with null correlation between the entries of $\thetav$. In this case, $\det(S)$ represents the volume of the hyper-rectangle over which the posterior distribution of $\thetav$ is distributed. Similarly, for non-zero parameter correlations, $\det(S)$ represents the hyper-volume contained in the hyper-surface defined by the covariance matrix $S$. Therefore, a high value of the \gls{fom} corresponds to a more statistically significant constraint of the model parameters. 

In order to visualise how much can be gained by pushing the model to higher $\kmax$ values, in the next section we plot the \gls{fom} of each individual configuration normalised to that of a reference case, corresponding to the configuration with the \gls{eft} model at $\kmax=0.1\kMpc$ with all nuisance parameters sampled as free parameters.

\subsection{Fitting procedure}

In order to properly sample the posterior distribution we need to compute the galaxy power spectrum and the likelihood for a large number of points in parameter space. To achieve convergence while keeping the number of evaluations as low as possible, an efficient sampling algorithm is needed. 

All the results presented in this work have been obtained using a nested sampling approach \citep{Skilling2006}, which differs from a standard Metropolis--Hastings \citep{Metropolis1953, Hastings1970} Markov-chain sampler in a number of ways. The main difference is that, using nested sampling, the whole hyper-dimensional parameter space is explored within the specified priors by means of a given number of live points, which are subsequently modified to track the posterior distribution of the parameters according to the value of the evidence. In this analysis we make use of the public code \texttt{PyMultiNest} \citep{Buchner2014} with a total number of 1800 live points, after having checked that the output posterior distribution has properly converged with this number. Further details, together with a comparison of different samplers, are presented in Appendix \ref{sec:samplers}.

We adopt the approach of a full-shape analysis. This means that we directly sample the cosmological parameter space, with a model galaxy power spectrum that is generated at each step. A single evaluation of the theory models presented in Sect.~\ref{sec:theory} can take up to few seconds, since it combines a call to the Boltzmann solver to obtain linear theory predictions, and a call to the routines responsible for computing the non-linear corrections. Since the typical number of model evaluations for a single Markov chain can reach order of $\mathcal{O}\,(10^6)$, the final running time necessary to obtain a converged posterior distribution can take up to several days.

In order to speed up the model evaluation, we make use of the publicly available \comet package \citep{comet}\footnote{\url{https://pypi.org/project/comet-emu/}} to emulate the \gls{eft} model, providing an evaluation of the full one-loop prediction in about $\mathcal{O}\,(10\,{\rm ms})$. The code has been validated against a set of 1500 theory data vectors in a range of redshifts that covers the one we explore in this analysis, showing an averaged $0.1\%$ systematic error for the final $\Pgg(k)$ model, and it is therefore suited to be used for this analysis.\footnote{Further validation tests have been carried out against other codes owned by the authors of this paper, as shown in Appendix \ref{app:modelling_challenge}.} The evaluation of the hybrid Lagrangian-bias-based model is instead carried out using the public emulator \bacco, as mentioned in Sect.~\ref{sec:bacco}.

In all the cases, we assume a Gaussian likelihood function defined as
\begin{align}
        -2&\ln\mathcal{L}(\thetav)\nn\\
 &=\sum_{i=1}^{N_{\rm bins}}\sum_{j=i}^{N_{\rm bins}}\brackets{P^{\,\rm th}_i(\thetav)-P^{\,\rm data}_i\,}\,C^{\,-1}_{ij}(\thetav)\,\brackets{P^{\,\rm th}_j(\thetav)-P^{\,\rm data}_j\,}\,,
    \end{align}
which is computed at each point in parameter space explored by the sampler, and whose value is used to determine whether to assign to the current point one of the live points. The final output, which is saved to external files ready to be post-processed, consists of a list of points in parameter space together with the corresponding value of the log-likelihood.

\subsection{Parameter priors}

\begin{table}
\caption{List of model parameters, split into cosmological and nuisance ones, with the latter further divided into the two bias models described in Sect.~\ref{sec:theory}. The nuisance parameters consist of bias parameters, EFTofLSS counterterm, and shot-noise terms. For each parameter, the imposed prior is specified in the last column of the table. The letter $\mathcal{U}$ stands for a uniform distribution, with edges identified by the first and second element of the pair, respectively.}
  \renewcommand{\arraystretch}{1.3}
  \centering
  \begin{tabular}{|c|c|c|}
    \hline
    \rowcolor{blue!5}
     & Parameter & Prior\\
     \hline
     \rowcolor{blue!5}
    \multicolumn{3}{|c|}{\bf Cosmology} \\
    \hline
    \multirow{2}{*}{} & $h$ & \small{$\mathcal{U}\,[0.55,0.85]$}\\
    \cline{2-3}
    & $\omega_\mathrm{c}$ & \small{$\mathcal{U}\,[0.08,0.16]$} \\
    \hline
    \rowcolor{blue!5}
    \multicolumn{3}{|c|}{{\bf Eulerian bias} expansion} \\
    \hline
    \multirow{4}{*}{Bias} & $b_1$ & \small{$\mathcal{U}\,[0.25,4]$}\\
    \cline{2-3}
    & $b_2$ & \small{$\mathcal{U}\,[-10,10]$}\\
    \cline{2-3}
    & $\bGtwo$ & \small{$\mathcal{U}\,[-4,4]$ or fixed to Eq. \eqref{eq:bG2_exset}}\\
    \cline{2-3}
    & $\bGthree$ & \small{$\mathcal{U}\,[-8,8]$ or fixed to Eq. \eqref{eq:bGthree_coev}}\\
    \hline
    Counterterm & $c_0\,\big[({\rm Mpc}/h)^2\big]$ & \small{$\mathcal{U}\,[-100,100]$}\\    
    \hline
    \multirow{2}{*}{Shot-noise} & $\aPone$ & \small{$\mathcal{U}\,[-1,2]$} \\
    \cline{2-3}
    & $\aPtwo\,\big[({\rm Mpc}/h)^2\big]$ & \small{$\mathcal{U}\,[-5,5]$ or fixed to 0}\\
    \hline
    \rowcolor{blue!5}
    \multicolumn{3}{|c|}{{\bf Hybrid Lagrangian bias} expansion}\\
    \hline
    \multirow{4}{*}{Bias} & $b_1^{\,{\cal L}}$ & \small{$\mathcal{U}\,[-1,3]$} \\
    \cline{2-3}
    & $b_2^{\,{\cal L}}$ & \small{$\mathcal{U}\,[-3,3]$} \\
    \cline{2-3}
    & $b_{s^2}^{\,{\cal L}}$ & \small{$\mathcal{U}\,[-10,10]$} \\
    \cline{2-3}
    & $b_{\nabla^2 \delta}^{\,{\cal L}}\,\big[({\rm Mpc}/h)^2\big]$ & \small{$\mathcal{U}\,[-10,10]$} \\
    \hline
    Shot-noise & $\aPone$ & \small{$\mathcal{U}\,[-1,1]$}\\
    \hline
  \end{tabular}
  \label{tab:priors}
\end{table}

\begin{figure*}
         \centering
    \includegraphics[width=2\columnwidth]{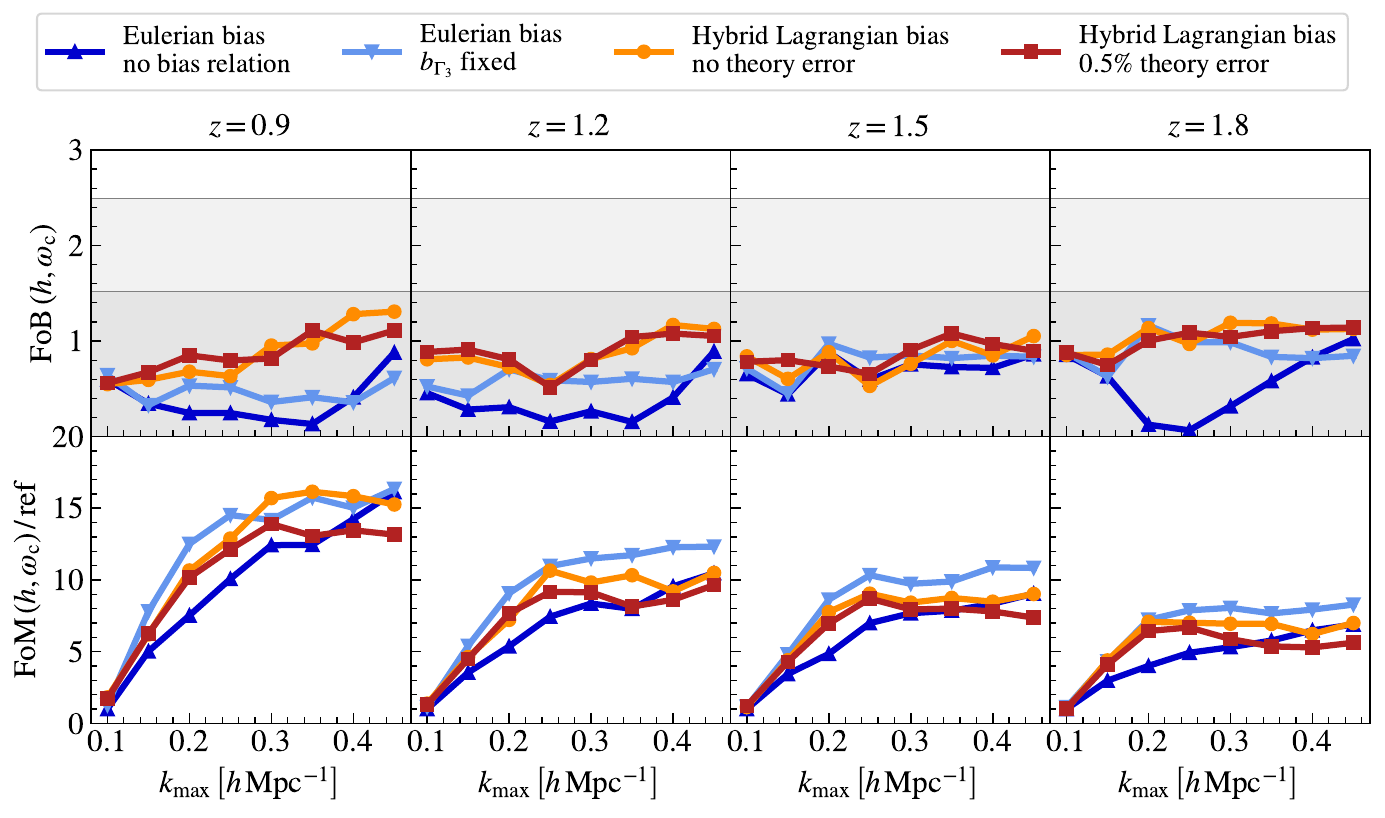}
    \caption{Performance metrics (\gls{fob} in the top row and \gls{fom} in the bottom row) extracted from the Model 3 \gls{hod} samples as a function of the maximum wave mode $\kmax$ of the fit, assuming the rescaled covariance matrices matching the four \Euclid spectroscopic redshift bins described in Sect.~\ref{sec:vol_resc}. Different curves correspond to different models, as described in the legend. The \gls{fom} panels are normalised in units of the reference \gls{fom}, corresponding to the one of the EFT model with all parameters free at $\kmax=0.1\kMpc$. The grey bands in the \gls{fob} panels represent the $68\%$ and $95\%$ percentiles of the corresponding \gls{fob} distribution, as explained in Sect.~\ref{sec:figure_of_bias}.}
    \label{fig:compare_baccoemu_scaled}
\end{figure*}

Our parameter space consists of both cosmological and nuisance parameters. Sampled cosmological parameters comprise the Hubble parameter $h$ and the cold dark matter density parameter $\omegac$. The latter can be constrained only through the full-shape of the galaxy power spectrum, especially via the position of the matter-radiation equality $k_{\rm eq}$, since geometric distance information is lost due to the fact that we conduct our analysis using real-space coordinates. For the same reason, $h$ can be artificially constrained because we fix all the other parameters affecting the amplitude of the matter power spectrum \citep{Sanchez2022}. \footnote{In practice, expressing the galaxy power spectrum in $\kMpc$ units makes possible to constrain evolution parameters $\paren{h,\As,w_0,w_{\rm a},\omega_{\rm K},\ldots}$ even when they are varied together in the same fit of the galaxy power spectrum data vector. This happens because with this set of units it is possible to break the degeneracy experienced by the evolution parameters that is otherwise present when expressing the data vector in ${\rm Mpc}^{-1}$ units \citep{Sanchez2022}.} We keep fixed the baryon density parameter $\omegab$ and the scalar spectral index $\ns$, since galaxy clustering measurements on their own are not able to constrain them with the same level of precision of \gls{cmb} data. At the same time, since in real space the primordial scalar amplitude $\As$ is strongly degenerate with the linear bias parameter $b_1$, at least on sufficiently large scales,\footnote{While the linear galaxy power spectrum depends on the combination $b_1^2\As$, the non-linear corrections depend on a different combination of the linear bias and the scalar amplitude, so that they can in principle break the degeneracy. However, since loop corrections are subdominant with respect to the amplitude of the linear galaxy power spectrum, we find that a strong degeneracy is still present, even when including mildly non-linear scales in the fits.} we keep $A_\mathrm{s}$ fixed to its fiducial value, along with the rest of the cosmological parameters, to the values shown in Table \ref{tab:flagship_cosmology}. 

The nuisance parameters are split into two sets, depending on the considered model. The parameters of the Eulerian bias expansion are composed of a mixture of bias parameters, $\curly{b_1,b_2,\bGtwo,\bGthree}$, counterterms, $\curly{c_0}$, and shot-noise parameters, $\curly{\aPone,\aPtwo}$. All of them enter in the final expression for the galaxy power spectrum as shown in Sect.~\ref{sec:theory_eft}. When testing the bias relations presented in Sect.~\ref{sec:bias_rel}, the parameters subject to the bias relations are not sampled over, but computed at each step in the chain as a function of the lower-order bias parameters. The scale-dependent noise parameter $\aPtwo$ is kept fixed to 0 for the majority of the runs we carry out, except for the ones presented in Sect.~\ref{sec:scale-dep_shot}, where we explicitly test the constraining power of the \gls{eft} model on this parameter in the range of redshifts that we are considering.

For the hybrid Lagrangian model we sample over a different set of bias parameters, $\curly{b_1^{\,{\cal L}}, b_2^{\,{\cal L}}, b_{s^2}^{\,{\cal L}}, b_{\nabla^2 \delta}^{\,{\cal L}}}$, and shot-noise, $\curly{\aPone}$. In this case we do not consider relations among bias parameters, but every run will assume the full set. 

When not mentioned otherwise, we adopt a completely agnostic approach, setting an uninformative flat prior for all the parameters, as shown in Table \ref{tab:priors}. The size of the prior for the two cosmological parameters and for most of the nuisance parameters has been selected to prevent the posterior distribution from becoming dominated by the imposed prior.


\section{Results}
\label{sec:results}

\begin{figure*}
         \centering
    \includegraphics[width=2\columnwidth]{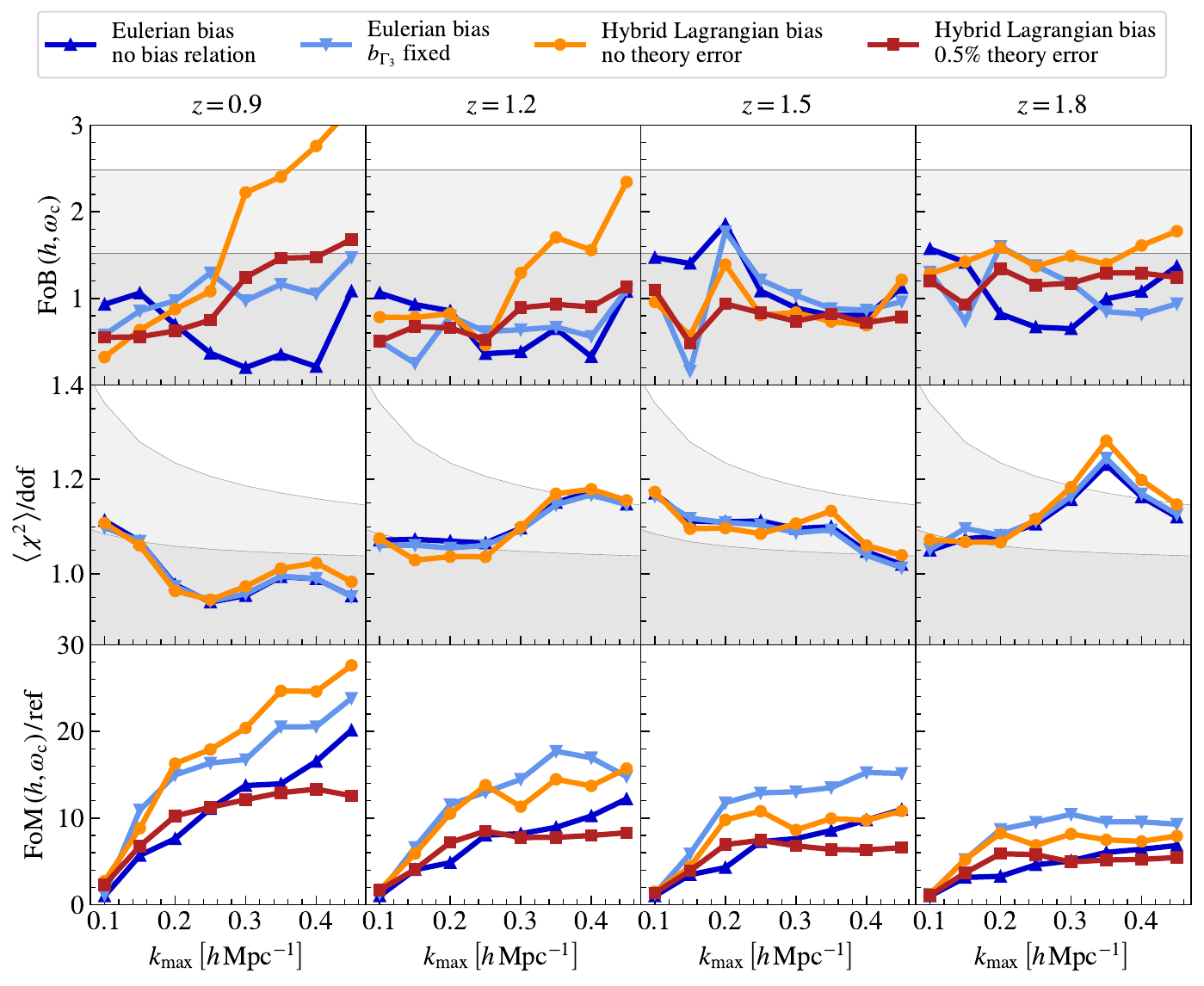}
    \caption{Same as Fig.~\ref{fig:compare_baccoemu_scaled} but assuming a covariance matrix corresponding to the full simulation volume. The additional middle row corresponds to the averaged $\chi^2$ normalised by the total number of degrees of freedom. Similarly to the \gls{fob} panels, the grey shaded bands in the $\chi^2$ panels mark the 68th and 95th percentile of the corresponding $\chi^2$ distribution with the corresponding number of degrees of freedom.}
    \label{fig:compare_baccoemu_full}
\end{figure*}

In this section we present the results obtained by fitting the data samples presented in Sect.~\ref{sec:data} with the two theoretical models described in Sect.~\ref{sec:theory}. 
We start off with a comparison between the performance of these models in Sect.~\ref{sec:eulerian_vs_lagrangian}, and leave to later sections a more detailed description of the model selection carried out for the \gls{eft} model in terms of scale cuts and bias relations. For compactness, we call this model simply the EFT model.

\subsection{Performance of Eulerian and hybrid Lagrangian bias expansion}
\label{sec:eulerian_vs_lagrangian}

In this section we carry out a comparison between the Eulerian expansion and the hybrid Lagrangian bias approach implemented in \bacco in terms of the three performance metrics previously defined in Sect.~\ref{sec:perf_metrics}. For this goal, we focus on fitting the galaxy power spectra of the Model 3 \gls{hod} sample at the redshifts of the four comoving snapshots, using both the rescaled \Euclid-like covariances, and the ones from the full simulation box. For each case, we run multiple chains to assess the stability of the results as a function of the maximum wave mode $\kmax$. The latter is selected in the range $\brackets{0.10,0.45}\kMpc$ using a linear spacing of $\Delta k=0.05\kMpc$, for a total of eight different cases.

We select two different configurations of the EFT model. The first one corresponds to the case in which all the nuisance parameters $\curly{b_1, b_2, \bGtwo, \bGthree, c_0, \aPone}$ are free to vary, with the only exception of the scale-dependent shot-noise parameter $\aPtwo$ which we set to zero.  In the second one we additionally fix the cubic tidal bias $\bGthree$ to its coevolution relation (Eq. \ref{eq:bGthree_coev}). The latter case is chosen in order to provide an alternative model based on the Eulerian expansion of Sect.~\ref{sec:theory_eft}, with the same assumptions on galaxy bias as in \bacco (see discussion in Sect.~\ref{sec:bacco}). In addition, this is one of the best configurations when considering the performance metrics on the combination $\curly{h,\omegac}$, as we properly validate in Sect.~\ref{sec:results_free_cosmo}.

As for \bacco, we leave all bias parameters $\curly{b_1^{\,{\cal L}}, b_2^{\,{\cal L}}, b_{s^2}^{\,{\cal L}},b_{\nabla^2 \delta}^{\,{\cal L}}}$ free to vary, with the addition of the parameter controlling the amplitude of the non-Poissonian stochastic noise, $\aPone$. Since \bacco is an emulator based on \nbody simulations, it is affected by two sources of noise: first, the emulation error, that is the noise introduced by the accuracy of the trained neural network itself; second, the training set error, that is the inaccuracies already present in the data used for training. The former is a scale-dependent quantity, which becomes progressively larger at small scales and caps at a maximum 0.5\% of the galaxy power spectrum signal at $k \sim 0.7\kMpc$ for $\Lambda$CDM cosmologies well within the allowed parameter space; it can get to the order of $\mathcal{O}\,(1\%)$ of the power spectrum signal for cosmologies closer to the limits of the emulator parameter space \citep{ZennaroEtal2021}. On the other hand, the intrinsic error of the training set is induced by the cosmology-rescaling technique employed during its construction; it once again depends on scale, and is subpercent in the case of $\Lambda$CDM cosmologies, but could reach percent levels when also massive neutrinos and dynamical dark energy are considered \citep{ContrerasEtal2020,ZennaroEtal2021}. To account for these combined effects, we consider two cases for the chains run with \bacco. In the first one we employ the same covariance matrix used to analyse the data galaxy power spectra as in the EFT chains, while in the second one we add in quadrature a theory error corresponding to 0.5\% of the galaxy power spectrum signal.\footnote{The assumption of choosing an extra contribution of 0.5\% of the power spectrum is well justified by the fact that we are only exploring a $\Lambda$CDM parameter space that is completely contained within the prior range of the emulator.}

In Fig.~\ref{fig:compare_baccoemu_scaled} we show the performance metrics (\gls{fob}, \gls{fom}) in the case of the realistic \Euclid-like volume. Note that we do not show the averaged reduced $\chi^2$ in this case, since the rescaled covariance matrix does not describe the fluctuations in the data vector, and therefore the collection of $\sum(P^{\,\rm th}-P^{\,\rm data})^2/\sigma^{\,2}$ values deviates from a $\chi^2$ distribution. At all redshifts the fits obtained with both models display a \gls{fob} within the 68\% confidence interval up to $\kmax = 0.45\kMpc$, with only a partial preference for the EFT framework when considering the value of the \gls{fob}, which is anyway consistent to $1\sigma$ for all the cases.

In the figure, the \gls{fom} is normalised by the value obtained with the EFT configuration at $\kmax=0.1\kMpc$, to show relative gains. As expected, the \gls{fom} of the EFT model is larger with a fixed $\bGthree$, because of the smaller number of free parameters. Similarly, we note that the combined constraining power on $(h,\omegac)$ of \bacco is degraded when including theory errors in the data covariance matrix.

When comparing the two different models, we note that \bacco without including theory errors (orange line) reaches its maximum \gls{fom} value, comparable to the maximum value achieved by the Eulerian model (blue line), already at a lower $\kmax$, of about $0.25$-$0.3\kMpc$. This is a consequence of the extra parameter, $\bGthree$, present in the Eulerian bias model. As further evidence, the Eulerian bias model display higher \gls{fom} values when considering a fixed $\bGthree$ (light-blue line), in particular on scales $\kmax<0.2\kMpc$. Above this threshold, we note that the EFT configuration features a slightly larger \gls{fom} than the one of the hybrid model, with the only exception of the $z=0.9$ snapshot, for which the two curves have a similar amplitude at all scales (light blue vs orange).

Except for the main EFT case, including smaller scales does not seem to increase the \gls{fom} beyond a scale of about $0.3\kMpc$. Since on these scales the theory error associated to \bacco is of similar magnitude as the data covariance, we note that, including the extra $0.5\%$ contribution (red line), the \gls{fom} starts flattening at a slightly lower $\kmax\sim0.25\kMpc$. This is mostly noticeable for the $z=0.9$ snapshot, for which shot-noise becomes the dominant contribution at a much larger $\kmax$.

A plot similar to the one in Fig.~\ref{fig:compare_baccoemu_scaled} is shown in Fig.~\ref{fig:compare_baccoemu_full}, this time considering the covariance matrix corresponding to the full simulation volume of about 54$\cGpc$. Since now the covariance matrix correctly represents the statistical fluctuations in the data vectors, we additionally show the $\chi^2$ averaged over the chain and normalised to the numbers of degrees of freedom. In this case, it is clear that not accounting for the theory error of \bacco can lead to a bias in the cosmological parameters, most notably at low redshift. On the contrary, including the reference $0.5\%$ theory error is enough to recover unbiased results, with the sole exception of the case at $\kmax = 0.45\kMpc$ and $z=0.9$.

The EFT model also returns unbiased measurements, with some spurious configurations outside the $1\sigma$ confidence interval for low $\kmax$ values at $z=1.5$. The main reason for this effect is likely imputable to the presence of projection effects when marginalising the posterior distribution in the $\curly{h,\omega_{\rm c} }$ plane, as we explain later in Sect.~\ref{sec:results_free_cosmo}. The averaged $\chi^2$ behaves in a consistent way between the two models, displaying an amplitude that is constantly lower than the 95th percentile of the corresponding $\chi^2$ distribution for both sets of curves, with the only exception of the largest $\kmax$ values of the $z=1.8$ snapshot.

In terms of goodness of fit, we note that, due to our choice of reporting the posterior-averaged $\chi^2$ value instead of the maximum-likelihood value, the normalised $\chi^2$ can start off with values larger than 1 at low $\kmax$ values. This is mostly caused by the non-gaussianity of the sampled posterior distribution when the data vectors cannot properly constrain the whole set of sampled parameters. In this case the averaged posterior can increase the value of the $\chi^2$ and making it appear artificially larger. In Appendix \ref{app:chi2} we include an example using the maximum-likelihood $\chi^2$, showing how in this case the goodness of fit typically assumes values consistent with 1 on those scales. 

While the constraining power of \bacco is in this case limited by the theory error being of similar order as the statistical error of the synthetic data considered, it is highly competitive with the Eulerian approach on scales that are free from this limitation, at $\kmax\lesssim0.2\kMpc$. On the one hand, the full-volume test considered here leads to very conservative results: the errors associated with the full volume of the Flagship simulation are roughly a factor 2 smaller than the scaled errors considered in this work, and these scaled errors for Model 3, in turn, are roughly another factor 2 smaller than the errors expected assuming the volumes and number densities for typical redshift bins of the spectroscopic sample described in the forecasts of \cite{EuclidForecast2019}. In addition, all covariance matrices are computed in the Gaussian approximation, which might underestimate the amplitude of the errors at small scales. On the other hand, these results provide a motivation to further reduce the noise associated with emulators -- for example through larger training sets, and employing Zeldovich control variates \citep{ChartierEtal2021, KokronEtal2022}. This is key for the design of the next generation of emulators. In fact, even if the configuration without theory errors shows the limitation of currently available codes, the corresponding \gls{fom} curve highlights the potential gain achievable with a more accurate version of the emulator.

\subsection{Testing the EFT model: Fixed cosmology}
\label{sec:results_fixed_cosmo}

\begin{table}
\caption{Marginalised mean values of the linear bias $b_1$ and the shot-noise parameter $\aPone$ measured using the two-parameter model for the ratio $\Pgg/\Pmm$ presented in Eq. \eqref{eq:two_pars_model}. Fits are carried out only considering scales up to $k_\mathrm{max}=0.08\kMpc$.}
  \label{tab:fiducial_b1_aP1}
  \renewcommand{\arraystretch}{1.3}
  \centering
  \begin{tabular}{|c|c|c|c|}
    \hline
    \rowcolor{blue!5}
    & HOD &  &  \\
    \rowcolor{blue!5}
    \multirow{-2}{*}{Redshift} & Model & \multirow{-2}{*}{$b_1$} &  \multirow{-2}{*}{$\aPone\left[\frac{1}{\bar{n}}\right]$}\\
    \hline
    \multirow{2}{*}{$z=0.9$} & 1 & $1.350\pm0.004$ & $0.220\pm0.220$\\
    \cline{2-4}
                             & 3 & $1.395\pm0.003$ & $0.253\pm0.079$\\
    \hline
    \multirow{2}{*}{$z=1.2$} & 1 & $1.661\pm0.006$ & $0.424\pm0.152$\\
    \cline{2-4}
                             & 3 & $1.751\pm0.004$ & $0.289\pm0.057$\\
    \hline
    \multirow{2}{*}{$z=1.5$} & 1 & $1.977\pm0.007$ & $0.386\pm0.104$\\
    \cline{2-4}
                             & 3 & $2.030\pm0.005$ & $0.219\pm0.032$\\
    \hline
    \multirow{2}{*}{$z=1.8$} & 1 & $2.474\pm0.007$ & $0.257\pm0.039$\\
    \cline{2-4}
                             & 3 & $2.486\pm0.005$ & $0.346\pm0.018$\\
    \hline
  \end{tabular}
\end{table}

In the rest of this section we focus on testing the range of validity of the EFT model using different scale cuts, bias relations, and reference volumes. This test is limited only to the EFT model because, as shown in the previous subsection, we cannot run fits using \bacco without accounting for the extra contribution from theory errors in the covariance matrix, especially when considering the extremely large precision of the full-box covariance. In order to assess the level of accuracy of the EFT model and determine its range of validity, we first carry out fits at fixed cosmology assuming the full volume of the simulation box. In this way, we focus exclusively on the performance of one-loop galaxy bias prediction with highly precise measurements, testing which scale cuts and bias relations lead to the best agreement between the theory model and the input data vectors.

The validation of the model includes an accuracy test consisting in recovering fiducial values for the linear bias $b_1$ and the shot-noise parameter $\aPone$ determined from the large-scale limit of the ratio between the measurements of the galaxy and of the matter power spectrum. This will reduce the effect of cosmic variance on the linear bias estimate (the cross galaxy-matter power spectrum is unfortunately not available). At large scales, we can assume a simple two-parameter, linear model given by
\be
    \Pgg(k) = b_1^2\,\Pmm(k) + \frac{1+\aPone}{\bar{n}}\,,
    \label{eq:two_pars_model}
\ee
to be fit on scales $\kmax<0.08\kMpc$. As a reference, the marginalised mean posterior values of both $b_1$ and $\aPone$ are listed in Table \ref{tab:fiducial_b1_aP1}.

\begin{figure*}
    \centering
    \includegraphics[width=2\columnwidth]{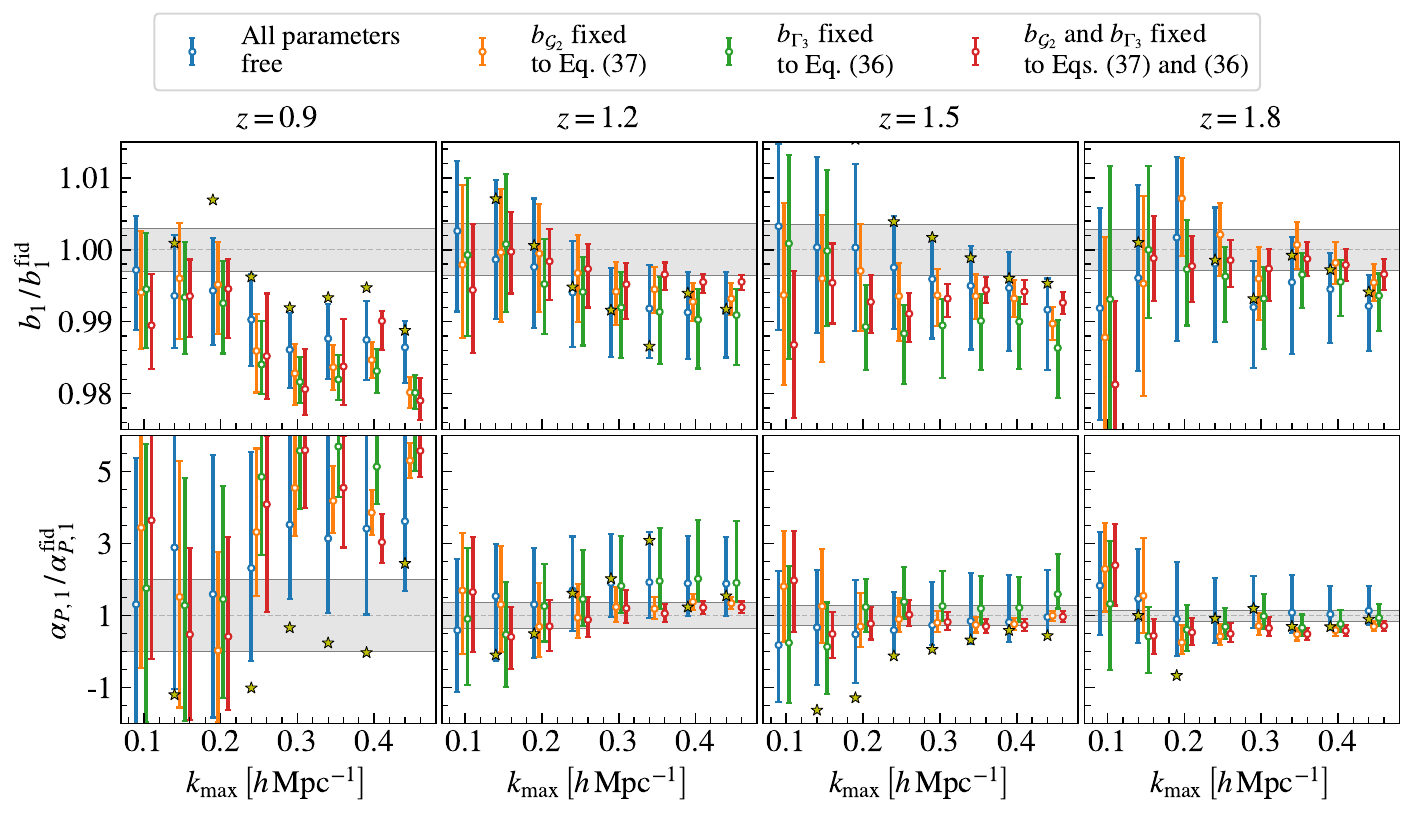}
    \includegraphics[width=2\columnwidth]{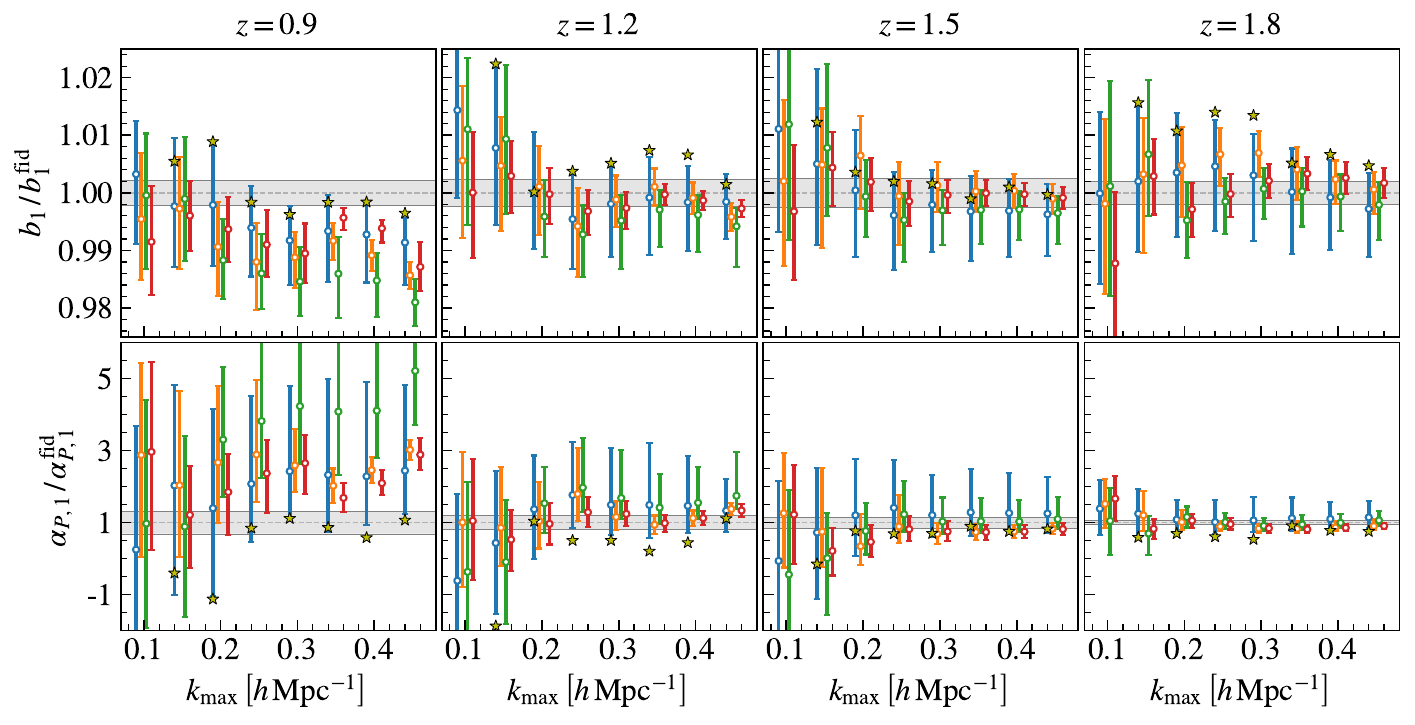}
    \caption{Comparison between the marginalised constraints on the linear bias parameter $b_1$ and the shot-noise parameter $\aPone$ obtained at fixed cosmology, and the fiducial values listed in Table~\ref{tab:fiducial_b1_aP1} obtained using only the large scale-limit of Eq. (\ref{eq:two_pars_model}). The first two and last two rows show results for the Model 1 and Model 3 \gls{hod} samples, respectively. In both cases, the upper panels show constraints on the linear bias $b_1$, while the bottom ones show constraints on the constant shot-noise parameter $\aPone$. Different colours correspond to different assumptions on the total number of free bias parameters, as shown in the legend. Star symbols highlight the position of the maximum-likelihood for the case with all bias parameters free to vary. Dashed grey lines and shaded bands mark the fiducial value and $1\sigma$ confidence interval from Table \ref{tab:fiducial_b1_aP1}.}
    \label{fig:marg_1d_fixed_cosmo}
\end{figure*}

\begin{figure*}
    \centering
    \includegraphics[width=2\columnwidth]{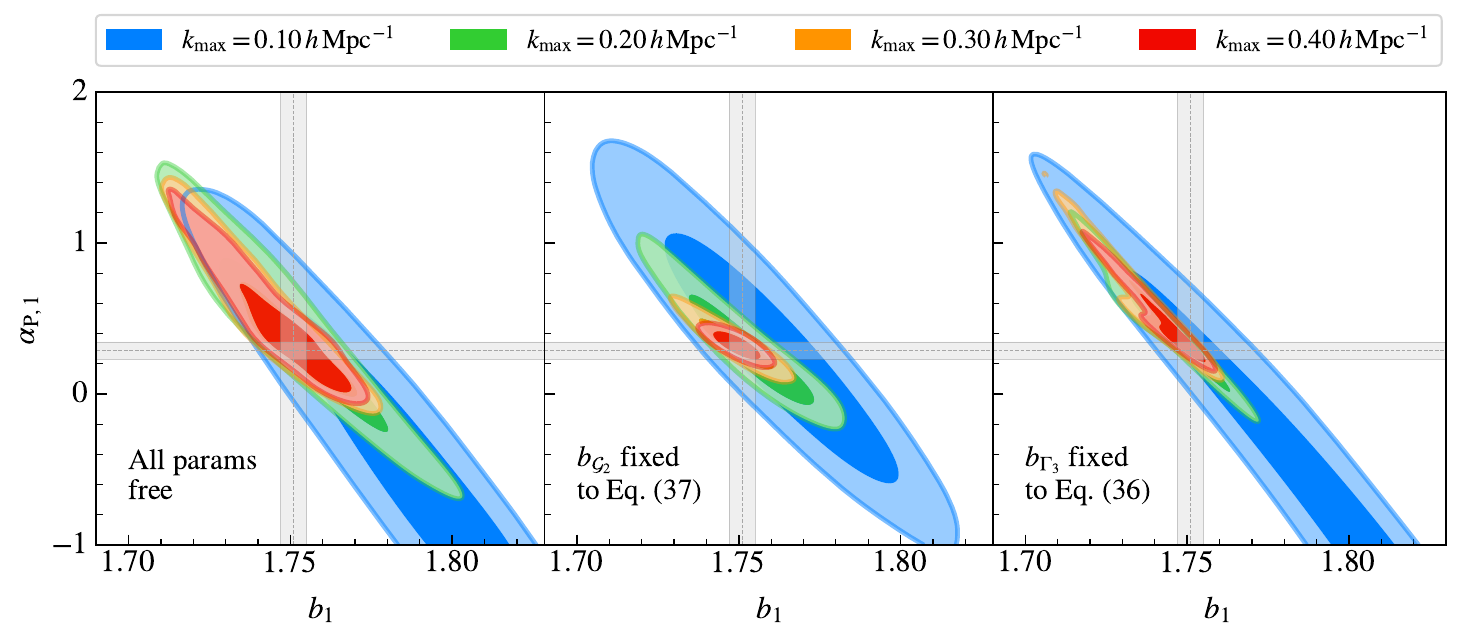}
    \caption{Marginalised 2D constraints on the pair $\paren{b_1,\aPone}$ for the case corresponding to the Model 3 \gls{hod} sample at $z=1.2$. The three different panels correspond to the case in which all the parameters are left free (left panel), $\bGtwo$ is fixed to the excursion-set relation as in Eq. \eqref{eq:bG2_exset} (middle panel), and $\bGthree$ is fixed to the coevolution relation as in Eq. \eqref{eq:bGthree_coev} (right panel). Different colours represent different $\kmax$ values, as listed in the legend. Fiducial $68\%$ confidence intervals for both $b_1$ and $\aPone$ are shown with grey bands.}
        \label{fig:2d_contours}
\end{figure*}

In Fig.~\ref{fig:marg_1d_fixed_cosmo} we show the marginalised constraints obtained fitting the full model of Eq. \eqref{eq:Pggnlo} to the eight data vectors, against the fiducial values of $\curly{b_1, \aPone}$ obtained from the large-scale limit as in Eq. \eqref{eq:two_pars_model}. We test four different model configurations, which differ by the total number of bias parameters that are kept fixed to the relations presented in Sect.~\ref{sec:bias_rel}.
\bi
    \item[(i)] All nuisance parameters are left free to vary while sampling the posterior distribution, for a total of six free parameters -- linear bias $b_1$, local quadratic bias $b_2$, non-local quadratic bias $\bGtwo$, non-local cubic bias $\bGthree$, matter counterterm and higher-derivative bias $c_0$, and constant shot-noise parameter $\aPone$\,.
    \item[(ii)] $\bGtwo$ is fixed to the excursion-set-based relation defined in Eq. \eqref{eq:bG2_exset}, for a total number of five parameters.
    \item[(iii)] $\bGthree$ is fixed to the coevolution relation defined in Eq. \eqref{eq:bGthree_coev}, for a total number of five parameters.
    \item[(iv)] Both $\bGtwo$ and $\bGthree$ are fixed to the relations assumed in (ii) and (iii), respectively, for a total number of four parameters.
\ei
In all these cases, we keep the scale-dependent shot-noise parameter $\aPtwo$ fixed to zero (we test the validity of this assumption in Sect.~\ref{sec:scale-dep_shot}).

Overall, the configuration with the largest number of free parameters -- case (i) in the previous list, shown with blue points and error bars in Fig.~\ref{fig:marg_1d_fixed_cosmo} -- is capable of capturing the correct amplitude of both $b_1$ and $\aPone$ for the majority of the tested $\kmax$ values and redshifts, showing a mild running of the one-dimensional marginalised values that becomes relevant only for the lowest redshift snapshot we consider, on scales $\kmax>0.2\kMpc$. The same effect is partially present for the $z=1.2$ snapshot, although less significant: as a matter of fact, the marginalised constraints are consistent with their fiducial values at better than $2\sigma$. Rather than only considering the mean posterior distribution, it is instructive to also plot the maximum-likelihood point in the parameter space under consideration. We estimate this quantity using as a proxy the point in the sampled posterior distribution that maximises the likelihood, even though the latter is partially affected by a certain degree of stochasticity. In this case (star symbols in Fig.~\ref{fig:marg_1d_fixed_cosmo}) we observe a shift towards the fiducial values, even if not for all configurations. A discrepancy between the maximum-likelihood point and the mean of the marginalised posterior is a clear hint at the presence of projection effects, also known as prior volume effects, due to the high dimensionality of the parameter space and to non-linear degeneracies among the model parameters.

Fixing either $\bGtwo$ -- case (ii), orange points and error bars -- or $\bGthree$ -- case (iii), green points and error bars -- does not significantly help in terms of accuracy of the marginalised constraints, with systematic deviations that can still become larger than the $1\sigma$ confidence interval. However we find that, while fixing $\bGthree$ typically results in similar constraining power on both $b_1$ and $\aPone$, imposing a relation on $\bGtwo$ leads to definitely tighter posteriors. This is the result of breaking the strong degeneracy between the two non-local bias parameters, $\bGtwo$ and $\bGthree$, and at the same time the one between the quadratic biases, $b_2$ and $\bGtwo$, leaving the remaining parameters to be more tightly constrained (a clear example of these is displayed in the right panel of Fig.~\ref{fig:comp_samplers}). The same clearly happens when combining the two previous relations -- case (iv), red points and error bars -- since with this setup we completely break the degeneracies in the considered parameter space. However, in this case we observe a deviation from the fiducial values of $b_1$ which can reach more than $2\sigma$ for some of the configurations, in particular at low redshift, hinting at a departure from the assumption of conserved evolution.

The effect of the strong $b_2$-$\bGtwo$-$\bGthree$ degeneracy can be observed in a more direct way by inspection of the 2d marginalised constraints in the $b_1$-$\aPone$ subspace. In Fig.~\ref{fig:2d_contours} we show such posterior distributions, taking as a reference the Model 3 sample at $z=1.2$. The different panels correspond to different bias relations, from left to right: the case with all the parameters free to vary, with $\bGtwo$ fixed to the excursion-set-based relation, and with $\bGthree$ fixed to the coevolution relation. A first consideration to make is that there is a non-trivial degeneracy between the two parameters, for which projection effects might bias the 1d constraints without necessarily meaning that the hyper-dimensional posterior distribution does not cover the fiducial values of the parameters. Secondly, we can observe how the case with fixed $\bGtwo$ gives the tightest constraints for both parameters, with an increase in the merit of the constraints that is directly related to the maximum scale adopted in the fit, up to $\kmax=0.4\kMpc$. Once more, this trend can be easily explained by the effective breaking of the degeneracy among the higher-order bias parameters.

For a limited number of cases, even when fixing one or more degrees of freedom, we find that the final posterior distribution can still appear multi-modal, leading to enlarged constraints when marginalising over the remaining parameters. For this reason, some of the chains where both tidal bias parameters $\bGtwo$ and $\bGthree$ are fixed feature marginalised constraints that are larger than the ones with one additional degree of freedom. This is clearly noticeable for the largest $\kmax$ bin of the Model 3 sample at $z=0.9$.

\begin{figure}
    \centering
    \includegraphics[width=\columnwidth]{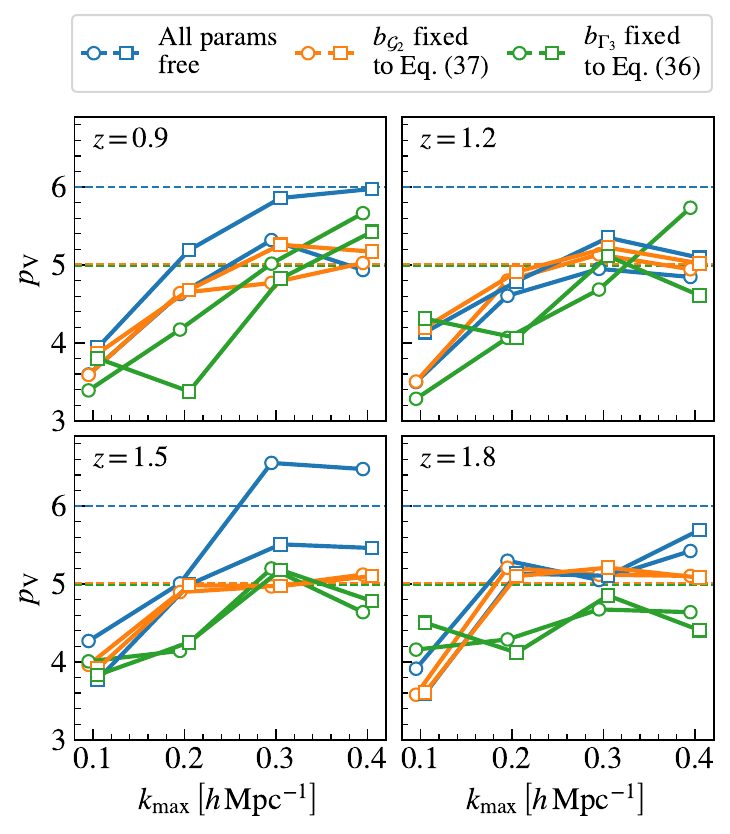}
    \caption{Number of effective parameters that can be properly constrained by the input data vectors of the Model 1 (circles) and 3 (squares) \gls{hod} samples, respectively, as a function of the maximum wave mode included in the analysis and for different configurations of the bias model, as listed in the legend. Different panels correspond to different redshifts, as shown in the corresponding top left corner.}
    \label{fig:pv}
\end{figure}

The self-consistency of the different models in terms of the number of model parameters can be assessed using an additional statistics. In this context, we are interested in determining the total number of parameters that can be effectively constrained by the data vectors. A commonly employed statistics is represented by the $p_{\rm V}$ value, defined as \citep{Gelman2014}
\be
    p_{\rm V}=\frac{1}{2}\ave{\paren{\,\chi^{\,2}-\ave{\,\chi^{\,2}}}^2}\,,
\ee
that is, the variance of the corresponding $\chi^{\,2}$ distribution. This number indirectly tracks the presence of degeneracies among the model parameters, and only converges to the total number of free parameters for a normal distribution. In order for a theory model to effectively constrain a given number of parameters, the $p_V$ value is expected to reach that same value, and can therefore be used as a proxy for the self-consistency of different model configurations. In Fig.~\ref{fig:pv} we show this value as measured from both sets of \gls{hod} samples and for different values of $\kmax$. In practice, we observe that the model with all parameters free never reaches the expected value of $p_V=6$, even for the largest value of $\kmax$, with the exception of a couple of configurations. This shows that a six-parameter model is most likely resulting in overfitting. On the contrary, fixing one of the two tidal biases makes the $p_V$ reach the expected limit above some $\kmax$, with a transition that typically happens sooner for the case with fixed $\bGtwo$. This reinforces the conclusion that this configuration is preferred with respect to the others under consideration.

We note that these results may be partially affected by the presence of cosmic variance in the data vectors. For this reason, in Appendix \ref{app:sample_variance} we explicitly assess the impact of this extra contribution, using both a smooth and a noisy realization of the data, generated using the theory code. This test shows that most of the residuals observed in Fig.~\ref{fig:marg_1d_fixed_cosmo} can be explained by sample variance affecting our data vectors.

\begin{figure}
        \centering
        \includegraphics[width=\columnwidth]{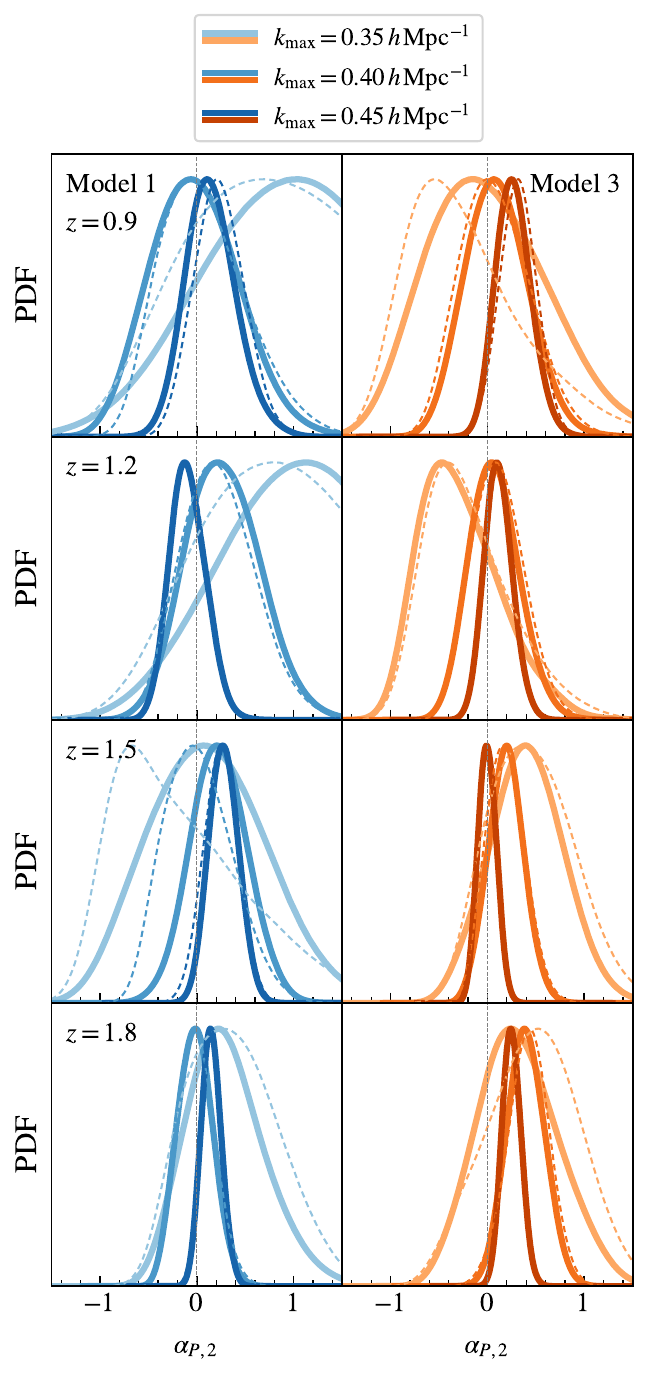}
        \caption{Marginalised 1D constraints on the scale-dependent shot-noise parameter $\aPtwo$ in the fits with fixed cosmological parameters. Different rows correspond to different redshifts (top to bottom, low to high redshift), while different columns correspond to different \gls{hod} samples. Within each panel, the colour gradient marks different values of $\kmax$, as detailed in the legend. Solid/dashed lines correspond to the configurations with all the nuisance parameters free to vary, and with $\bGtwo$ fixed to the excursion-set relation as in Eq. \eqref{eq:bG2_exset}, respectively.}
        \label{fig:marg_aP2}
\end{figure}

\subsection{Constraints on scale-dependent shot noise}
\label{sec:scale-dep_shot}

So far, the stochastic field $\varepsilon_{\rm g}$ entering the expression for the galaxy density field in Eq. \eqref{eq:deltag_expansion} has been assumed responsibile only for a constant offset from Poissonian predictions, via the parameter $\aPone$. An immediate check on the performance of the one-loop galaxy bias expansion can be carried out by further extending the model parameter space to also include the next-to-leading order correction to the stochastic field $\varepsilon_{\rm g}$. As already mentioned in Sect.~\ref{sec:theory}, this leads to the presence of an additional $k^2$-dependent term in the galaxy power spectrum, whose amplitude is regulated by the extra parameter $\aPtwo$.

Figure~\ref{fig:marg_aP2} shows the marginalised one-dimensional constraints on $\aPtwo$, for both \gls{hod} samples, Model 1 on the left and Model 3 on the right, respectively. Since the large-scale limit of the galaxy power spectrum does not have enough constraining power on $\aPtwo$, we only consider values of $\kmax$ above $0.35\kMpc$.\footnote{For some of the samples, this range of scales is already dominated by the Poissonian shot-noise contribution, as can be observed from the top and middle panel of Fig.~\ref{fig:pk_measurements}.} We never observe a statistically significant detection of the $\aPtwo$ parameter, with the majority of the marginalised constraints being consistent with $\aPtwo=0$ well within the $2\sigma$ confidence interval. The only configurations for which this does not happen are the ones at high redshifts, specifically when considering high values of $\kmax$, since these are the configurations for which the parameter $\aPtwo$ is constrained with the highest precision. Performing the same test with one of the tidal bias fixed\,\footnote{We choose to fix $\bGtwo$, motivated by the results of Sect.~\ref{sec:results_fixed_cosmo} that suggested this is the configuration less affected by degeneracies between the model parameters} does not lead to significantly different conclusions. This seems to suggest that the $\aPtwo$ parameter might have a more important role over a range of redshifts where the shot-noise correction is more relevant.

We postpone further tests to a next installment of this series of papers, since a more careful check should be carried out adopting samples with different number densities -- as this value determines the range of scales where the transition from signal to noise takes place -- in particular considering values that would represent in a more reliable way the expected H$\alpha$ galaxy distribution detected by \Euclid. At the same time, an important test should be carried out using the redshift-space galaxy power spectrum (Euclid Collaboration: Camacho et al., in prep.), for which extra $k^2$-dependent noise corrections are required, as a function of the orientation with respect to the line of sight \citep{Philcox2022, Carrilho2023, Moretti2023}. Finally, we note that the findings of this analysis are in line with the conclusions from \cite{PezCroEgg2108}, for which, in terms of constraints on the cosmological parameters $\curly{h,\omegac}$, a clear detection of scale-dependent stochastic parameters happens only when considering the combined information from the galaxy-galaxy and galaxy-matter power spectra.

\subsection{Testing the EFT model: results on cosmological parameters}
\label{sec:results_free_cosmo}

\begin{figure*}
    \centering
    \includegraphics[width=2\columnwidth]{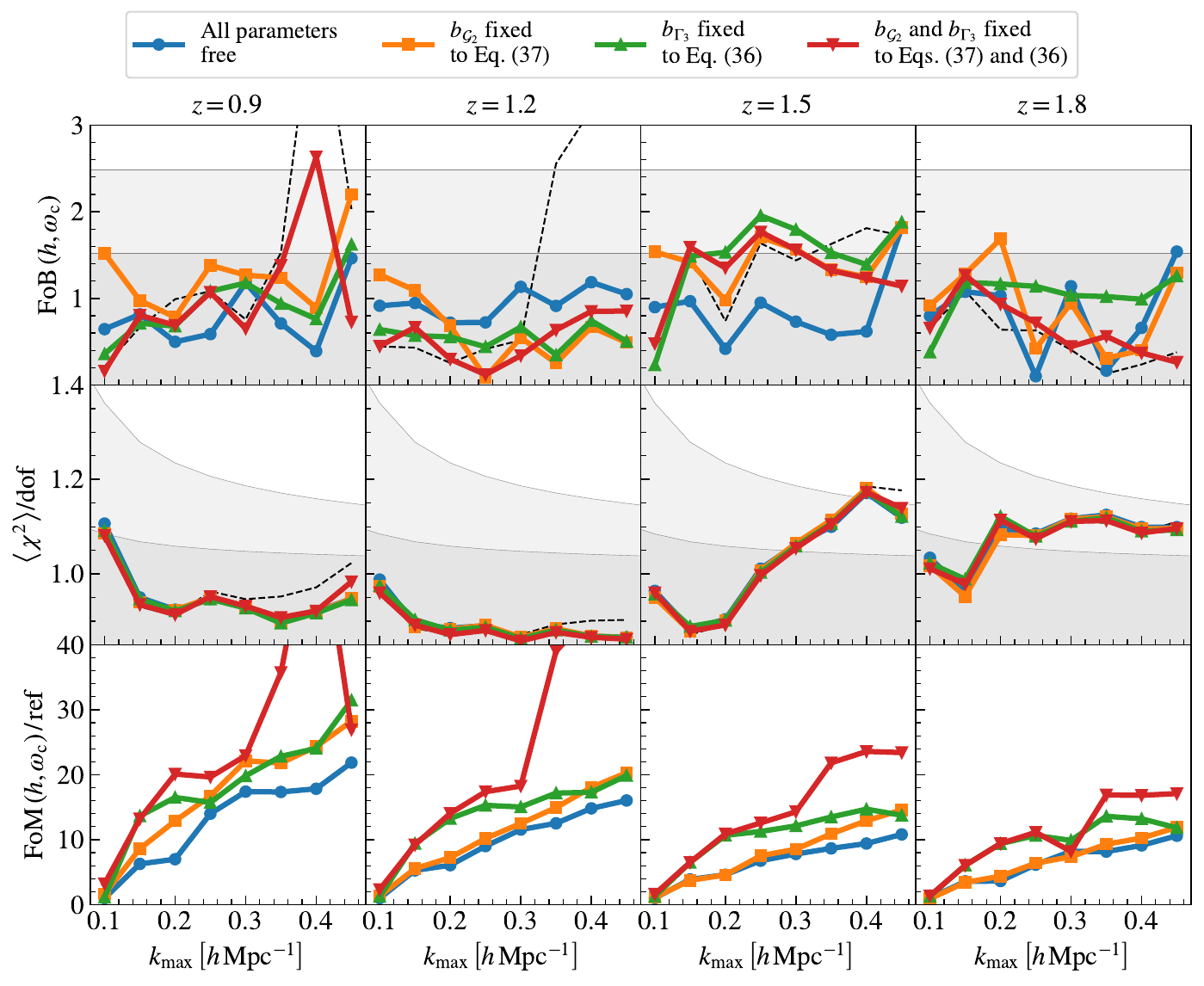}
    \caption{Performance metrics of the Model 1 \gls{hod} samples for the various configurations defined in Sect.~\ref{sec:results_fixed_cosmo} as a function of the maximum wave mode $\kmax$, and for the four different redshifts of the samples. The metrics shown are figure of bias (top), goodness of fit
(middle), and figure of merit (bottom). Different colours correspond to different model configurations, as listed in the legend. The black dashed line shows as a reference the case in which both tidal bias parameters, $\bGtwo$ and $\bGthree$, are set to 0. The grey bands in the \gls{fob} and $\chi^2$ panels represent the $68\%$ and $95\%$ percentiles of the corresponding distributions. The \gls{fom} panels show the figure of merit normalised to the one of the standard run ---with all bias parameters free to vary--- at $\kmax=0.1\kMpc$.}
    \label{fig:pm_mod1}
\end{figure*}

\begin{figure*}
    \centering
    \includegraphics[width=2\columnwidth]{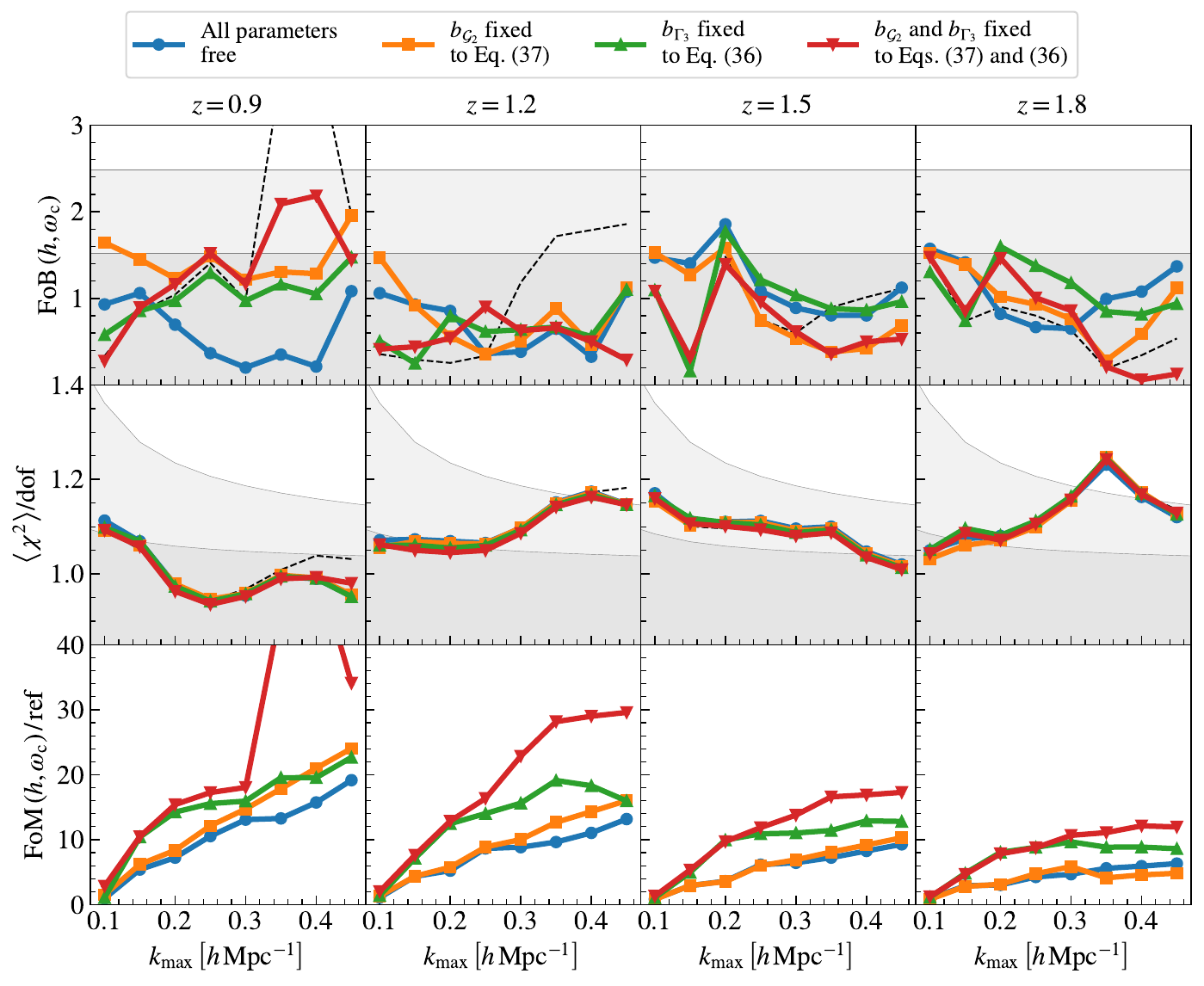}
    \caption{Same as in Fig.~\ref{fig:pm_mod1}, but for the Model 3 \gls{hod} samples.}
    \label{fig:pm_mod3}
\end{figure*}

\begin{figure*}
    \centering
    \includegraphics[width=2\columnwidth]{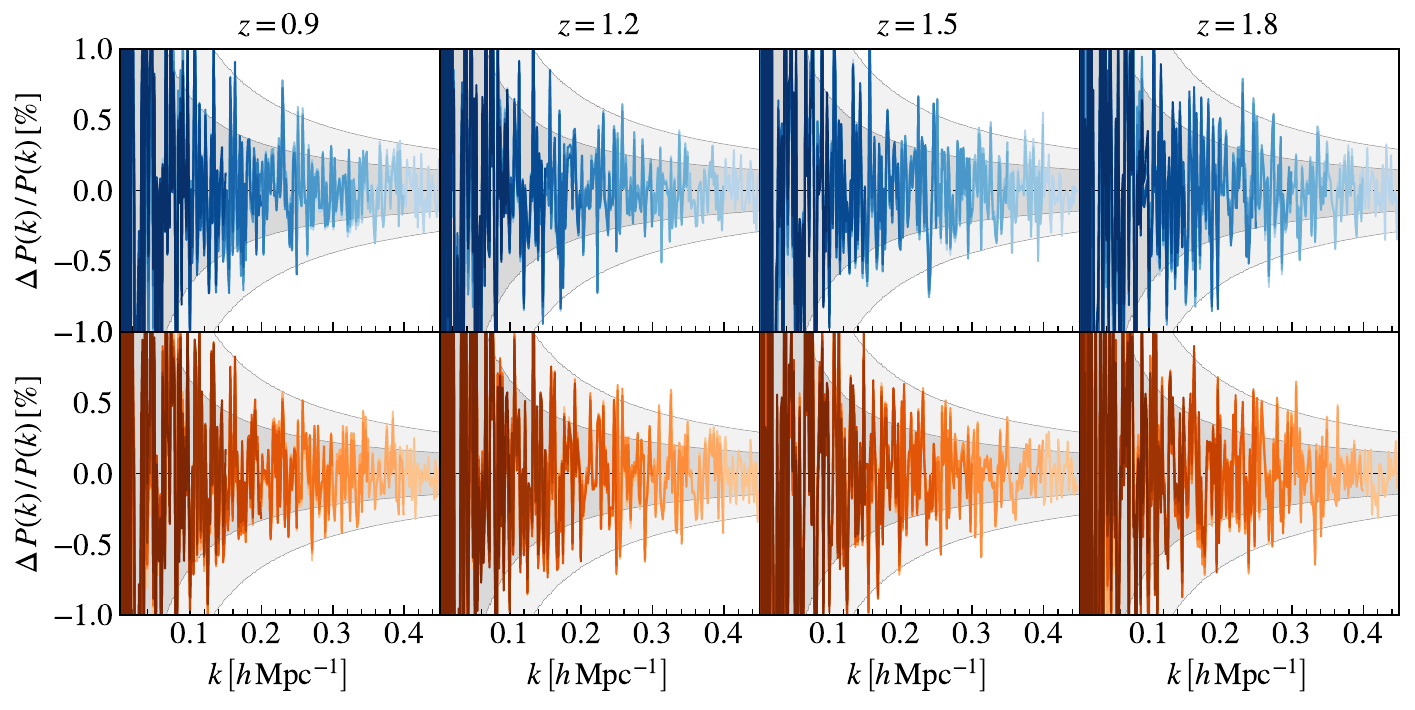}
    \caption{Residuals of the maximum-likelihood best fits against the input galaxy power spectrum data vectors, assuming the case with all bias parameters free to vary. Different columns correspond to different redshifts, as shown on top of the corresponding column, while different rows mark either the Model 1 (top row) or Model 3 (bottom row) \gls{hod} sample. Different colour shades mark the best fits obtained at different $\kmax$ values, from $0.1\kMpc$ up to $0.45\kMpc$.}
    \label{fig:bestfits_vs_data}
\end{figure*}

After having investigated the performance of the EFT model, we now turn our attention to the study of how cosmological constraints can be affected by different choices of model configuration. Specifically, we assume the same parameter space already used in Sect.~\ref{sec:eulerian_vs_lagrangian}, which also includes the Hubble parameter $h$ and the cold dark matter density parameter $\omegac$, while keeping the rest of the cosmological parameters fixed to their fiducial values, as listed in Table \ref{tab:flagship_cosmology}. A standard full-shape analysis of the redshift-space galaxy power spectrum would typically also include the scalar amplitude of the power spectrum, $\As$, since the anisotropies introduced by peculiar velocities make possible to break the strong $\As$-$\,b_1$ degeneracy that is otherwise present when considering real-space coordinates.\footnote{Specifically, the $\As$-$\,b_1$ degeneracy can be broken thanks to the different impact that these two parameters have on the amplitude of the leading-order power spectrum multipoles, $P^{\,(\ell)}\propto\As\,b_1^{2-\ell/2}$.} However, since this analysis revolves around the real-space galaxy power spectrum, we ought to choose a more conservative approach in order to obtain as least degenerate constraints as possible on the rest of the cosmological parameters. The sampling of $\As$ will be performed in the rest of the papers of this series, when considering the additional information content of the galaxy bispectrum (Euclid Collaboration: Eggemeier et al., in prep.) and \gls{rsd} (Euclid Collaboration: Camacho et al., in prep., Euclid Collaboration: Pardede et al., in prep).

Figures \ref{fig:pm_mod1} and \ref{fig:pm_mod3} show the three performance metrics defined in Sect.~\ref{sec:perf_metrics} for the Model 1 and Model 3 \gls{hod} samples respectively, assuming the full-box volume of the Flagship I simulation. In both cases, the \gls{fob} and \gls{fom} panels refer to the combination between the two cosmological parameters we are sampling over.

\subsubsection{Figure of bias}

In terms of \gls{fob}, we observe a consistent trend across each model configuration, indicating an unbiased combined measurement of the cosmological parameters even well within the mildly non-linear regime, at $\kmax\gtrsim0.3\kMpc$. The only exception is represented by the configuration in which both tidal bias parameters, $\bGtwo$ and $\bGthree$, are simultaneously kept fixed to the excursion-set relation (Eq. \ref{eq:bG2_exset}) and to the coevolution relation (Eq. \ref{eq:bGthree_coev}), respectively. This outcome is unsurprising, as we are reducing by two the total number of degrees of freedom of the model. Notably, the bias on the cosmological parameters gets larger at lower redshift, hinting to a departure from the coevolution relations as non-linear gravitational effects become more pronounced. Typically, this deviation occurs at scales of approximately $\kmax=0.3\kMpc$. However, we observe that this configuration behaves surprisingly well for high-redshift snapshots, exhibiting a \gls{fob} well within the $68\%$ confidence interval. 

In terms of overall stability of the results, we observe a deviation at low $\kmax$ values for the Model 3 sample. This is more strongly affecting high-redshift snapshots, for which the value of the \gls{fob} at $\kmax\sim0.1\kMpc$ already exceeds the corresponding $68\%$ confidence level, and only gets below the threshold when including additional signals from smaller scales. This effect is primarily attributed to the presence of projection effects, owing to the large dimensionality of the selected parameter space. Specifically, we find that all samples display a non-negligible correlation between the cold dark matter density parameter, $\omegac$, and the EFT counterterm $c_0$, resulting in a systematic shift of $\omegac$ for the lowest values of $\kmax$, where there is insufficient constraining power to accurately constrain $c_0$. Fixing the cubic tidal bias $\bGthree$ to the coevolution relation typically helps to restore the cosmological parameters to their fiducial positions. This happens due the further degeneracy between $\bGthree$ and $c_0$ over the mildly non-linear regime. The relative importance of the $\omegac$-$c_0$ degeneracy gets amplified only when considering the snapshots at the highest redshifts. As a partial confirmation of this trend, \cite{PezCroEgg2108} did not report either any low-$k$ systematic effect when analysing mock galaxies meant to reproduce the clustering properties of the BOSS -- CMASS and LOWZ -- and SDSS MGS samples, since, in that case, the considered redshift range was much lower ($0.1\lesssim z\lesssim 0.6$) than the one analysed in this work. As a further cross-check, the same effect is partially present when combining the full shape of the galaxy power spectrum and bispectrum in a joint analysis (Euclid Collaboration: Eggemeier et al., in prep.), albeit with a lower significance, due to the additional constraining power of higher-order statistics.

\subsubsection{Goodness of fit}

Similarly to the case at fixed cosmology, we find that the goodness of fit for the different models is consistent among the various model configurations, with only a small departure of the case where both tidal biases, $\bGtwo$ and $\bGthree$, are kept fixed to their corresponding relations and for the largest $\kmax$ values we consider in this analysis. This is visible in the case of the Model 1 sample at $z=0.9$, for which there is an increase in the averaged $\chi^2$ value at $\kmax\gtrsim 0.3\kMpc$, which also corresponds to the transition of the \gls{fob} to values above the $68\%$ percentile value. Otherwise, we find that the different model configurations provide a systematically consistent goodness of fit, with a reduced average $\chi^2$ value that is typically well within the $95\%$ percentile of the corresponding $\chi^2$ distribution.

When considering the two different \gls{hod} models, the $\chi^2$ for the Model 3 \gls{hod} samples is consistently larger than for the Model 1 case (see the middle panel of Fig.~\ref{fig:pm_mod3}). The most significant deviation is affecting the high-redshift snapshots, for which the average $\chi^2$ spuriously gets larger than the $95\%$ confidence interval for some of the selected $\kmax$ values. In practice, this deviation is still consistent to better than the $3\sigma$ confidence interval. In addition, we remind that here we are using a single noisy realization, meaning sample variance could partially be driving some of the constraints. Moreover, we are analysing the data vectors with an extremely high level of precision, due to choice of using the full volume of the simulation and to the high number density of the \gls{hod} samples. As a further evidence for the goodness of our fits, in Fig.~\ref{fig:bestfits_vs_data} we show the residuals between the maximum-likelihood theory vectors obtained at different values of $\kmax$ against the input data vectors, assuming the most relaxed model configuration. For all the samples that are under examination, we find that the broadband of the input galaxy power spectrum is perfectly recovered, and that the worst performance (in terms of goodness of fit) is only imputable to the scatter of the noisy data vector around the best fit -- with a significance that is larger for some of the samples, such as for the Model 3 sample at $z=1.8$.

Finally, in Figs.~\ref{fig:pm_mod1} and \ref{fig:pm_mod3} we show with a dashed black line the \gls{fob} and averaged $\chi^2$ of the case where the values of the tidal biases are set to 0. In this case, we observe a departure of the goodness of fit from the other configurations, in a redshift-dependent way, which is also accompanied by a breaking of the model in terms of \gls{fob}. This shows that the use of coevolution relations can drastically improve the performance of the model, with respect to simply set the non-local bias parameters to zero.

\subsubsection{Figure of merit}

As expected, the \gls{fom} monotonically increases when including additional information from more non-linear scales, with a relative gain with respect to the most relaxed configuration (all bias parameters free to vary, at $\kmax=0.1\kMpc$) that becomes larger moving towards lower redshifts. In fact, extending the fitting range to the maximal value of  $\kmax=0.45\kMpc$, we find that the trend for the \gls{fom} of the different bias configurations is approximately $10,15,20$ and $30$ times larger than the reference at $z=1.8, 1.5,1.2$ and $0.9$, respectively. The only exception is represented by the high-redshift snapshots of the Model 3 sample, for which the total galaxy power spectrum on mildly non-linear scales becomes dominated by the shot-noise correction earlier than for the rest of the samples, and for which we observe that the \gls{fom} reaches a plateau at $\kmax\gtrsim0.3\kMpc$. As expected, the case with two less degrees of freedom consistently gains more constraining power on the cosmological parameters, leading to much tighter constraints in particular at the largest $\kmax$ value we probe. However, we note that these gains are directly correlated with the breaking of the model in terms of \gls{fob} at $z=0.9$, \footnote{The breaking of this particular configuration, with both $\bGtwo$ and $\bGthree$ fixed to the bias relations, also exhibits a \gls{fom} at $\kmax=0.45\kMpc$ that is lower than the one obtained at lower $\kmax$. We do not explore this configuration, since the model cannot be used with this configuration, but we argue that a more careful investigation of this effect should be carried out using a larger set of simulations, to reduce the importance of cosmic variance, which might partially drive these effects.} and might therefore lead to biased cosmological constraints if used in a real-data analysis. Nevertheless, for most of the tested cases, combining the two relations still leads to acceptable results up to the maximal scale we are considering.

\subsubsection{Summary}

Overall, we find that fixing only the quadratic tidal bias $\bGtwo$ leads to the most stable results, with a \gls{fob} that is typically -- except for some spurious scale cut -- well within the $68\%$ percentile of the corresponding distribution, and with a \gls{fom} which is systematically larger than in the case where all the parameters are free to vary. The performance of the case with a fixed cubic tidal bias is also consistent, but with the caveat that the underlying bias parameters experience a strong degeneracy among themselves, as shown in Sect.~\ref{sec:results_fixed_cosmo}. However, we find that this case typically achieves a \gls{fom} larger than the one with fixed quadratic tidal bias, with the latter catching up only at large enough values of $\kmax$. Also, in a range of scales  up to $\kmax\sim0.3\kMpc$, the case with relations applied to both tidal biases matches almost identically the case with fixed cubic bias, highlighting again how this parameter has a much larger impact when constraining the cosmological parameters considered in this analysis.

Nevertheless, as already mentioned in Sec. \ref{sec:bias_rel}, we argue that the applicability of the excursion set and coevolution relations may partially fail when adopting them with real data, given the lack of any realistic dependency on other quantities different from the halo mass, such as assembly bias \citep{Croton2007, Barreira_2021, HadLiuSom2021, LazBarSch2023} or the scatter in the $\ave{N}(M_{\rm h})$ relation \citep{BehConWec2010, Zehavi2011}. For this reason, extended studies on more realistic simulated samples will be needed to properly benchmark these relations when applying them to real \Euclid observations.

As a final remark, in \citet{NicolaEtal2023arXiv} the authors carried out a comparison similar to the one presented in this paper, by forecasting the impact of different bias models in the analysis of the Rubin Observatory Legacy Survey of Space and Time (LSST) \citep{Ivezic2019} photometric observations. Differently from this work, their analysis showed that the use of minimal bias models with 1/2 less degrees of freedom could bias the recovery of cosmological parameters at more than $3\sigma$ also when restricting the analysis to relatively large scales ($\kmax\sim0.15\kMpc$). We argue that this apparent inconsistency is mainly imputable to the different observables selected in the two analyses. While in this work we make use of the real-space galaxy power spectrum at some fixed redshifts $z\gtrsim1$, their analysis is focused  on the 3$\times$2-point data combination in tomographic bins that are for the most part at lower redshifts ($0<z<1$), for which the assumption of a negligible tidal or higher-derivative bias is no longer valid. As a further evidence, the dashed black lines in Figs. \ref{fig:pm_mod1} and \ref{fig:pm_mod3} show how setting both tidal biases to 0 makes the Eulerian bias model break sooner when considering the snapshots at lower redshifts, especially the one at $z=0.9$.

\subsection{Dependence on sample volume}
\label{sec:results_diff_volumes}

\begin{figure*}
         \centering
    \includegraphics[width=2\columnwidth]{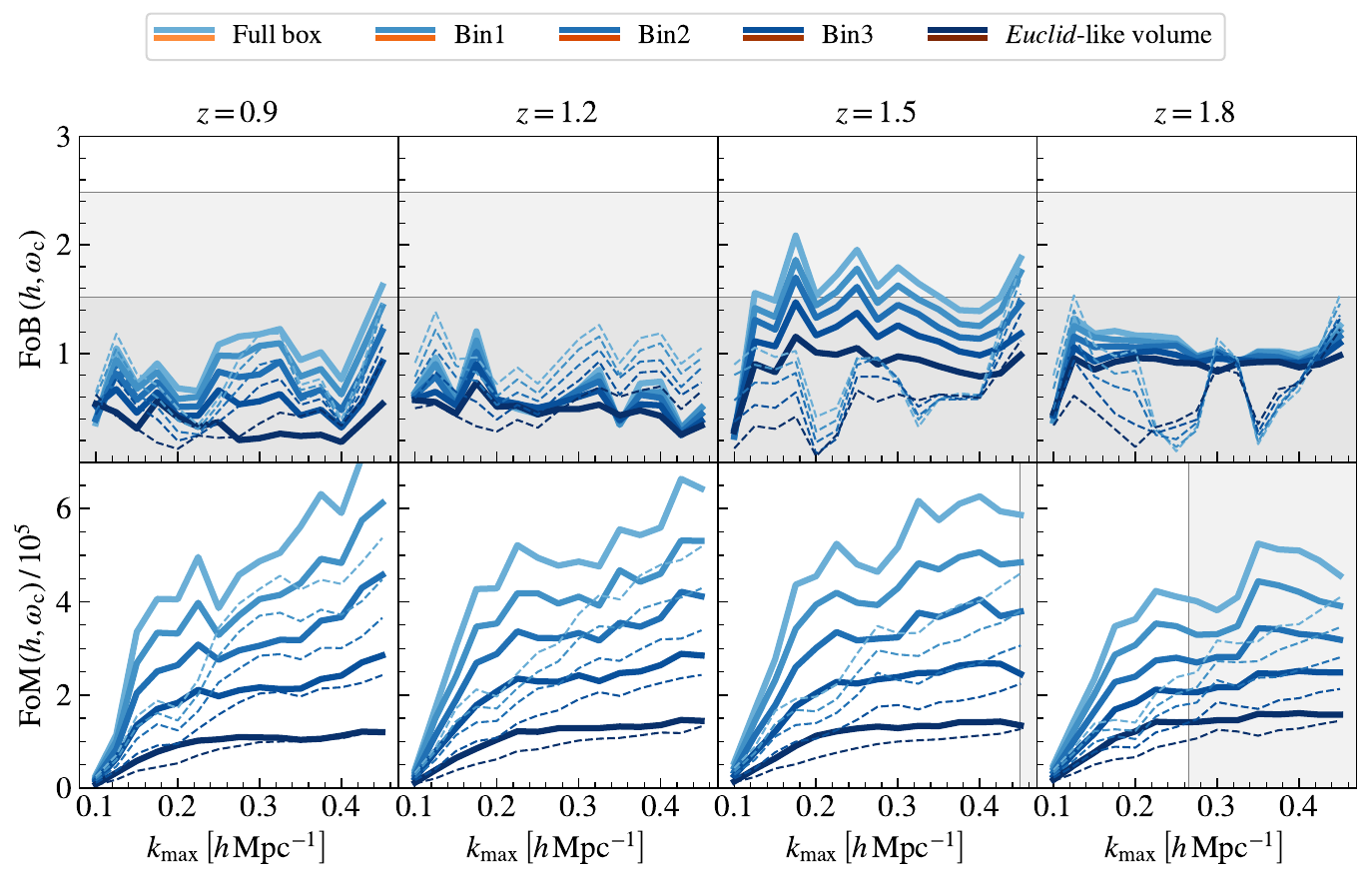}
    \includegraphics[width=2\columnwidth]{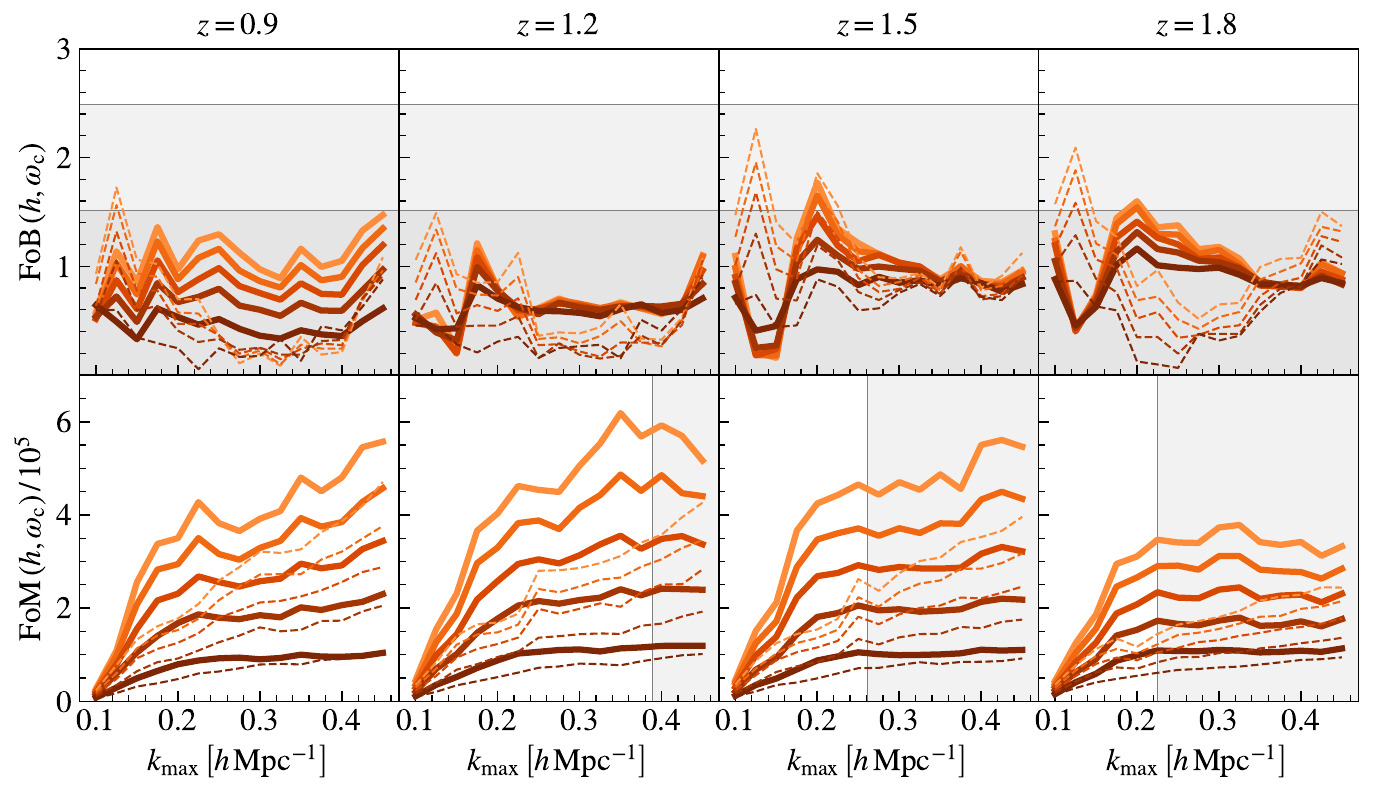}
    \caption{Evolution of the \gls{fob} and \gls{fom} as a function of the different choice for the rescaling of the reference volume, as shown in the legend. Solid and dashed curves correspond to the case with $\bGthree$ fixed to the coevolution relation and to the case with all the parameters free to vary, respectively. All configurations of the top two rows correspond to the fits of the Model 1 \gls{hod} samples, while the bottom two rows do the same for the Model 3 \gls{hod} samples. Grey bands in the \gls{fom} panels identify the scale at which the Poisson shot-noise contribution assumes the same height of the underlying clustering signal, marking the transition to the shot-noise-dominated scales.}
    \label{fig:pm_mod1_diff_vols}
\end{figure*}

As anticipated in Sect.~\ref{sec:vol_resc}, in addition to performing a model selection using extremely high-precision measurements -- with a Gaussian covariance matrix derived assuming the full box volume of the Flagship I simulation -- in this section we test the performance of the Eulerian bias expansion for smaller values of the galaxy sample volume, in order to provide more realistic forecasts for the analysis of the \Euclid spectroscopic data. Once again, we remind the reader that for these tests we consider \gls{hod} catalogs with a number density larger than the one expected for the real galaxy samples, and that more realistic mocks will be used in forthcoming analyses that will also consider observational systematic effects, such as target purity and incompleteness, and observational effects such as the radial and angular selection function.

In this section, we consider the same data vectors obtained from measurements of the Flagship simulation snapshots already used in the previous ones. The dependence on the volume is explored by rescaling the corresponding covariance. We consider four different volumes corresponding to possible Euclid-like shells, as explained in Sect.~\ref{sec:vol_resc}. For each comoving snapshot, at $z=(0.9,1.2,1.5,1.8)$, with reference volume $V_{\rm box}$, we define the volume $V_{\rm shell}$ of a spectroscopic bin corresponding to a total angular surface of $15\,000$ square degrees -- and with a depth of $\Delta z=(0.2,0.2,0.2,0.3)$ -- following the choices made in \cite{EuclidForecast2019}. The three additional volume rescalings are obtained by selecting the values that divide the interval $[\eta,1]$ into four equi-partitioned subintervals, where $\eta=V_{\rm box}/V_{\rm shell}$.

In Fig.~\ref{fig:pm_mod1_diff_vols} we show the trends of the \gls{fob} and \gls{fom} for the previously defined samples, with Model 1 and 3 in the top and bottom two rows, respectively. Thick solid lines correspond to the case with $\bGthree$ fixed to the coevolution relation, which we selected as one of the best performing model among the ones that we have tested in Sect.~\ref{sec:results_free_cosmo}. On the contrary, thin dashed lines correspond to the case with all the bias parameters free to vary. Different colours identify the different covariance matrices used in the fits, from the already shown full-box case -- in light blue/orange -- to the case corresponding to the \Euclid-like shells -- in dark blue/orange. As expected, we find that in both cases the \gls{fob} for the two cosmological parameters $\paren{h,\omegac}$ becomes progressively smaller, reflecting the increasing amplitude of the covariance matrices used in the fits of the input data vectors. At the same time, we observe  how this trend is tightly correlated to a decrease in the \gls{fom}, with the severity of the drop being almost proportional to the factor between the original and the rescaled volume, $\eta$.

It comes with no surprise that the reference bias model is well-performing up to the highest value of $\kmax$ we consider, even under realistic assumptions. Additionally, for the case with fixed $\bGthree$, we can observe how the precision on the cosmological constraints reaches a plateau above a typical threshold that corresponds to the transition scale between the regimes dominated by the signal and by the shot-noise contribution, respectively. For this reason, it is possible to gain additional constraining power by pushing the analysis to high values of $\kmax$ at $z=0.9$, for which the number density of the sample is significantly larger than the one of the high-redshift samples. In contrast, when all the parameters are free to vary, the trend for the \gls{fom} curves is to gain additional constraints from smaller scales, even above the scale of transition, possibly pointing to a further breaking of parameter degeneracies that are no longer present when fixing the value of $\bGthree$ to the coevolution relation.

As a final consideration, we note that the Eulerian bias expansion is performing significantly well for all the considered rescalings of the covariance matrix. The constraining power of the EFT model in terms of the combination $\curly{h,\omegac}$ can be enhanced employing one of the coevolution relations described in Sect.~\ref{sec:bias_rel} without the appearance of systematic errors, even when considering samples with a number density significantly larger than the one expected from the real \Euclid data. This analysis, limited to real space, motivates further tests including higher-order statistics, such as the galaxy bispectrum, and taking into account redshift-space distortions. These topics will be properly explored in the next entries of this series.


\section{Conclusions}
\label{sec:conclusions}

In this paper we carried out an analysis meant to assess the performance of state-of-art models for one-loop galaxy bias over a redshift range that is well representative of the spectroscopic galaxy sample that will be one of the main targets of \Euclid. We employed a set of four \gls{fof} halo catalogues from comoving snapshots of the Flagship I simulation at $z=(0.9,1.2,1.5,1.8)$, which were subsequently populated with H$\alpha$ galaxies using \gls{hod} prescriptions based on the Model 1 and 3 from \cite{Pozzetti2016}. Each snapshot features an outstanding volume of $(3780\Mpc)^3$ and a high comoving number density (from $\sim10^{-4}\kcMpc$ to $\sim10^{-3}\kcMpc$), which corresponds to a flux limit of $f_{{\rm H}\alpha} = 2\times 10^{-16}\,{\rm erg\,cm^{-2}\,s^{-1}}$. These snapshots can therefore be used to assess the accuracy of the perturbative bias expansion at a high level of precision.

We tested two galaxy bias models for the full shape of the real-space galaxy power spectrum. The first one adopts a Eulerian bias expansion, and is based on the recently developed \gls{eft} modelling, in which the impact of small-scale physics, as well as the integration of ultraviolet modes in SPT, can be captured by a set of counter terms, which reduce to a single one when considering real-space coordinates. The final parameter space consists of two cosmological parameters, the Hubble parameter $h,$ and the cold dark matter density parameter $\omegac$, plus a set of six nuisance parameters, consisting of the linear bias $b_1$, the quadratic bias $b_2$, the tidal quadratic and cubic biases, $\bGtwo$ and $\bGthree$, the matter counter term and higher-derivative bias $c_0$, and two extra parameters representing deviations from Poissonian shot noise: a constant offset, $\aPone$, and a scale-dependent term, $\aPtwo$. The second model adopts a similar one-loop expansion of the galaxy power spectrum, but this time using Lagrangian coordinates, and is based on the emulation of the individual terms of the expansion starting from a limited set of high-resolution \nbody simulations. This is achieved thanks to the cosmology-rescaling technique presented in \cite{Angulo2021}. In addition to the two cosmological parameters, this model
features the linear bias $b_1$, the quadratic bias $b_2$, the tidal quadratic bias $b_{s^2}$, the Laplacian bias $b_{\nabla^2\delta}$, and a stochastic parameter representing a constant deviation from Poissonian shot noise, $\aPone$. In this work, we made use of the two implementations available in the public codes \comet \citep{comet} for the EFT model ---with accuracy tests carried out against external benchmarks, as shown in Appendix \ref{app:modelling_challenge}--- and \bacco \citep{ZennaroEtal2021} for the hybrid model.

In the main section of this work, we present tests of the relative performance of these two galaxy bias models, while in the subsequent sections we show how we determined the range of validity of the EFT model and tested the impact of fixing one or more parameters of the Eulerian bias expansion to some physically motivated relations in a way that allows us to break strong parameter degeneracies and better constrain the cosmological parameters. In all cases, we determined the range of validity of a given bias relation and scale cut by means of three different performance metrics: the goodness of fit, the figure of bias, and the figure of merit. The latter two metrics are computed on the $\curly{h,\omegac}$ combination in order to quantify the accuracy and precision of the model in terms of these two parameters.

We compare the performance of the Eulerian and hybrid Lagrangian bias models using a rescaled covariance to match the size of \Euclid-like redshift shells (assuming a full-sky area of $15\,000\deg^2$). Our results highlight how both models are capable of providing unbiased measurements of the cosmological parameters up to $\kmax = 0.45\kMpc$ for all four redshift snapshots, consistently with the $1\sigma$ confidence interval for the \gls{fob} distribution. In terms of \gls{fom}, \bacco reaches the same amplitude of the maximally achievable \gls{fom} of the EFT case already at a lower $\kmax$ value, most likely due to the absence of the cubic tidal bias parameter that is instead free to vary in the EFT chains. As expected, when fixing this parameter, the EFT model performs similarly to \bacco, and for most of the configurations it achieves a slightly larger \gls{fom}. When considering the covariance matrix corresponding to the full-box of the comoving snapshots (from three to seven times larger than the \Euclid-like shells, depending on the considered redshift) we find that the EFT model manages to recover the cosmological parameters consistently with the $1\sigma$ confidence interval. However, there are some spurious cases at low values of $\kmax$ that are affected by projection effects, which can be alleviated by fixing the value of $\bGthree$. On the other hand, with this level of precision, we hit the intrinsic emulation error of \bacco, which leads to bias in the inferred parameters when considering high $\kmax$ values at low redshift. Including an extra component to the covariance matrix ---corresponding to $0.5\%$ of the amplitude of the galaxy power spectrum, based on the combined systematic error from the emulation and from the measurements used to train the emulator--- brings the \gls{fob} of \bacco back within the $1\sigma$ confidence interval. Also in this case, on scales for which the intrinsic error of \bacco is not the dominant contribution, that is, $\kmax\lesssim0.2\kMpc$, its \gls{fom} is consistent with that of the EFT model. Not including the $0.5\%$ extra error, we note that the \gls{fom} is consistent with ---and in some cases even larger than--- the corresponding EFT results, demonstrating the potential gain to be obtained by improving the accuracy of next-generation emulators of the full shape of the galaxy power spectrum.

We then focus exclusively on the EFT model: we first describe the fits we performed at fixed cosmology to check the self-consistency of the one-loop galaxy bias expansion in terms of the linear bias $b_1$ and the scale-independent shot-noise parameter $\aPone$. The fiducial values for these parameters were fitted from the measured galaxy-to-matter-power-spectrum ratio, assuming a leading-order recipe on scales of $k<0.08\kMpc$. The result is that, when leaving all parameters free to vary, it is possible to recover ---at better than $2\sigma$--- the value of both parameters for the majority of the samples ---two \gls{hod} models times four different redshifts--- and scale cuts up to $\kmax=0.45\kMpc$. The only significant deviation takes place at the lowest redshift we consider, $z=0.9$, for which we observe a departure from the fiducial values soon after $\kmax=0.2\kMpc$. The latter is however consistent with sample variance expectations, as observed in a set of ten different noisy realisations of a synthetic theory data vector (see Appendix \ref{app:sample_variance}). The systematic errors are partially alleviated when considering the position in parameter space corresponding to the maximum of the likelihood, showing how the deviations might be imputable to projection effects. Fixing one of the two tidal biases to either an excursion-set-derived relation or to the coevolution relation still results in constraints that are consistent with the $2\sigma$ confidence interval. In particular, the former is preferred in terms of constraints of the model parameters because of the simultaneous breaking of the strong degeneracies with both the quadratic bias $b_2$ and the cubic tidal bias $\bGthree$. We find that fixing both parameters at the same time still works extremely well (with a typical recovery of $b_1$ and $\aPone$ within the 68\% confidence interval) even for the largest $\kmax$ values when considering the high-redshift snapshots. However, at low redshift, this choice can lead to deviations of more than $3\sigma$ for some of the configurations we test, especially when the maximum scale included in the fit is above a typical scale of $\kmax\sim0.3\kMpc$.

We explicitly checked whether or not a next-to-leading-order correction to the shot-noise contribution ($\aPtwo$) can improve the model performance. In all cases considered, the marginalised posterior distribution for $\aPtwo$ is consistent with zero within $2\sigma$, suggesting this additional parameter is not needed, at least for the description of these galaxy samples. Additionally, the scales that can constrain this parameter soon become dominated by the underlying shot-noise correction, effectively breaking the perturbative description of the latter in a Taylor expansion. A more significant test should be carried out considering more realistic galaxy samples, in terms of galaxy number density, and also including \gls{rsd} and observational systematic effects.

When the parameter space is extended to also include the two cosmological parameters, we note a good recovery of the fiducial values across the whole range of separations we test. The \gls{fob} exhibits an increasing trend moving towards high redshifts and low values of $\kmax$, which is due to projection effects when marginalising over all the nuisance parameters. This trend can be partially corrected by fixing the cubic tidal bias $\bGthree$ to the coevolution relation, and indeed with this configuration it is possible to consistently recover unbiased ---within the $68\%$ confidence interval --- constraints on the $\paren{h,\omegac}$ pair. Also in this case, fixing both tidal biases at the same time can lead to biased cosmological constraints, with the amplitude of the systematic errors increasing towards lower redshifts, which suggests a premature breaking of the tested relations. In terms of goodness of fit, we do not observe a significant change in the average $\chi^2$ when fixing some of the model parameters to the relations presented in Sect.~\ref{sec:bias_rel}. Finally, the \gls{fom} of the cosmological parameters clearly increases when reducing the degrees of freedom of the model. However, the configuration with both tidal biases fixed results in biased constraints for the low-redshift samples, while the case with only $\bGthree$ fixed can be employed down to smaller scales. This configuration displays a relative gain in \gls{fom} that ranges from 1.5 to 2 times, with respect to the case with all the parameters free. Relative gains in the \gls{fom} are more concentrated at $\kmax\lesssim0.3\kMpc$, where the model with fixed $\bGthree$ experiences a steep increase that is then followed by a more modest growth. The configuration with $\bGtwo$ fixed also displays a \gls{fom} larger than the case with all parameters free to vary, but with a steady slope that manages to catch up with the other configuration only for the largest $\kmax$ values that we tested. Overall, we find that the one-loop galaxy bias expansion is sufficiently accurate on the redshift range that we are exploring, $1\lesssim z\lesssim2$, even deep within the mildly non-linear regime, at $\kmax\sim0.4\kMpc$, with a statistical significance on the cosmological parameters that can be enhanced by fixing some of the degrees of freedom of the model.

In order to understand the impact of a different statistical precision on the input data vectors, we rescaled the Gaussian covariance matrix used in the fitting procedure to match the volume of a Euclid-like spectroscopic bin, with three additional intermediate volume choices selected between the Euclid-like bin and the original volume of the comoving box.
A smaller volume therefore corresponds to a reduced amplitude in the covariance matrix, resulting in a decrease in both \gls{fob} and \gls{fom} in a way that is proportional to the fraction of lost volume. We therefore confirm that these models of galaxy bias can be eventually used to analyse the real spectroscopic data collected by \Euclid.

This paper stands as the first installment of a series of works meant to validate the theoretical framework that will be used to analyse the large-scale galaxy distribution as observed in the actual measurements of \Euclid. Here we focus on the modelling of the real-space galaxy power spectrum of the spectroscopic sample, which stands as an important test for the complementary photometric analysis that is going to be carried out by \Euclid, in the shape of the popular 3$\times$2-point data combination. In parallel, in Euclid Collaboration: Eggemeier et al. (in prep.) we consider the joint analysis of the real-space galaxy power spectrum and  bispectrum, exploring a consistent description of non-linear bias in both observables. Two additional papers in the series (Euclid Collabpration: Camacho et al., in prep., Euclid Collaboration: Pardede et al., in prep.) will extend the modelling tests to redshift space. In parallel, a different set of papers will be devoted to a similar analysis of configuration-space statistics (Euclid Collaboration: Guidi et al., in prep., Euclid Collaboration: K{\"a}rcher et al., in prep., Euclid Collaboration: Pugno et al., in prep.).

%
%

\begin{acknowledgements}

\AckEC \\

The majority of the analysis carried out in this manuscript has been produced by a joint effort among several Euclid members and work centers. A special acknowledgement goes to the Max Planck Computing and Data Facility (MPCDF) in Garching (Germany), where most of the results presented in this article have been obtained. We acknowledge the hospitality of the Institute for Fundamental Physics of the Universe (IFPU) of Trieste for the group meeting held there in November 2022. \\

C.M.'s research for this project was supported by a UK Research and Innovation Future Leaders Fellowship [grant MR/S016066/2].  C.M.'s work is supported by the Fondazione ICSC, Spoke 3 Astrophysics and Cosmos Observations, National Recovery and Resilience Plan (Piano Nazionale di Ripresa e Resilienza, PNRR) Project ID CN\_00000013 ``Italian Research Center on High-Performance Computing, Big Data and Quantum Computing'' funded by MUR Missione 4 Componente 2 Investimento 1.4: Potenziamento strutture di ricerca e creazione di "campioni nazionali di R\&S (M4C2-19 )" - Next Generation EU (NGEU). A.E. is supported at the Argelander Institut für Astronomie by an Argelander Fellowship. A.B. acknowledges support from the Excellence Cluster ORIGINS which is funded by the Deutsche Forschungsgemeinschaft (DFG, German Research Foundation) under Germany’s Excellence Strategy - EXC-2094-390783311. R.E.A. acknowledges the support of the ERC-StG number 716151 (BACCO), and the project PID2021-128338NB-I00 from the Spanish Ministry of Science. M.B. is supported by the Programma Nazionale della Ricerca (PNR) grant J95F21002830001 with title ``FAIR-by-design''. M.K. is funded by the Excellence Initiative of Aix-Marseille University - A*MIDEX, a French ``Investissements d'Aveni'' programme (AMX-19-IET-008 - IPhU).\\

This research made use of \texttt{matplotlib}, a Python library for publication quality graphics \citep{Hunter:2007}.

\end{acknowledgements}

%
%

\bibliography{main.bib}

%

\begin{appendix}

\section{Matter power spectrum fits and the fiducial cosmology}
\label{app:matter_power_spectrum}

In the main body of this article, we carried out tests meant to assess the level of accuracy of different models for the one-loop galaxy power spectrum, using measurements coming from the Flagship I simulation. In addition to this analysis, we also tested our model for the matter power spectrum on a set of measurements obtained on the fly while running the \texttt{PKDGRAV3} code in the redshift range $\brackets{0.7,2.4}$. Each measurement consists of the matter power spectrum, measured in the range of wave modes defined by the interval $\brackets{0.01, 4}\,\kMpc$, using 18 linearly spaced bins up to $k\sim0.03\kMpc$ and other 84 logarithmically spaced bins after.

We run two independent analyses, the first using the next-to-leading order matter power spectrum obtained in the \gls{eft} framework (Eq. \ref{eq:Pmm_final}) and the second using the non-linear matter power spectrum from the \bacco emulator. We use an analytical Gaussian covariance matrix to describe the error on the matter power spectrum measurements, assuming the full volume of the simulation box, that is, $(3780\Mpc)^3$. For both models, we limit the maximum mode of the fit to $\kmax=0.25\kMpc$, within the expected \gls{pt} range of validity for the relevant redshifts.

\begin{figure}[b!]
    \centering
    \includegraphics[width=\columnwidth]{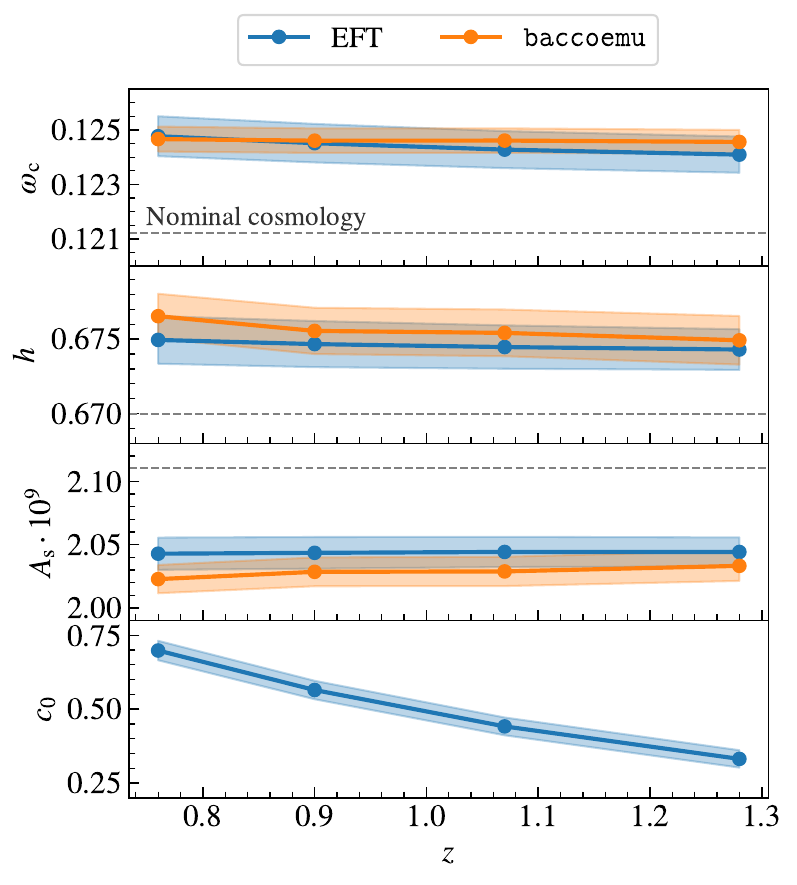}
    
    \caption{Marginalised one-dimensional constraints as a function of redshift obtained by fitting the measured matter power spectrum with the EFT model (blue) and the \bacco emulator (orange). Solid lines and shaded bands mark the mean and the standard deviation of the posterior distribution, respectively. For both models, the fit is carried out up to maximum wave mode $\kmax=0.25\kMpc$, using a Gaussian covariance matrix corresponding to the full box volume of the Flagship I simulation. Dashed lines denote the nominal fiducial values of the parameters $\curly{h,\omegac,\As}$. In the bottom panel we show the marginalised constraints on the $c_0$ EFT counterterm parameter.}
    \label{fig:marg_1d_Pmm}
\end{figure}

In Fig.~\ref{fig:marg_1d_Pmm} we show the marginalised 1d posterior distributions for $\curly{h,\omegac,\As}$ as a function of redshift. For this test we limit the redshift range to the first four snapshots ($z<1.3$), since the results are sufficient to draw conclusions on the agreement between data vector and theory models. Also shown, with dashed lines, are the nominal values of the parameters provided in \citet{Potter2017}. We find that all cosmological parameters are obtained with a bias of 3--4$\sigma$ from their fiducial values, in a way that is consistent across redshifts, as highlighted by the almost constant trends in each of the panels. In addition, the EFT model and \bacco are consistent at better than 1$\sigma$ with each other, pointing to a systematic effect that cannot be attributed to the particular model used to describe the data vectors.

\begin{figure}[b!]
    \centering
    \includegraphics[width=\columnwidth]{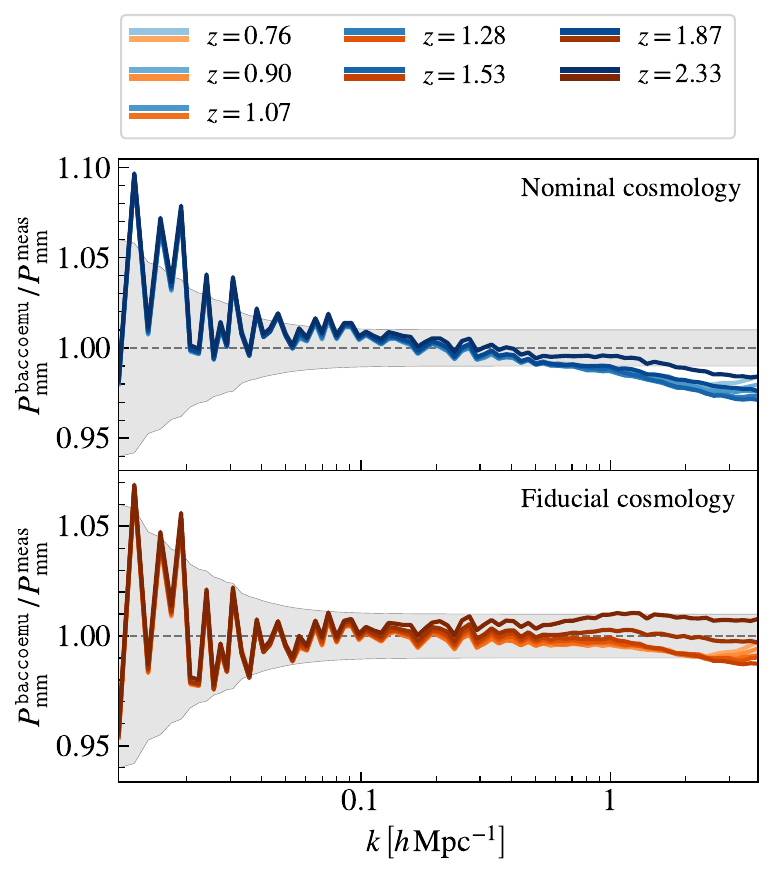}
    
    \caption{Ratio between the Flagship matter power spectrum measurements and predictions obtained using \bacco, considering both the nominal cosmology (upper panel) and the fiducial one (bottom panel). In both cases, different colours correspond to a different redshift, as shown in the legend. The grey shaded band mark the intrinsic error of the power spectrum measurements to which an additional $1\%$ contribution has been added to include the error contribution of the emulator, as explained in Sect.~\ref{sec:eulerian_vs_lagrangian}.}
    \label{fig:comp_Pmm_flagship}
\end{figure}

In the upper panel of Fig.~\ref{fig:comp_Pmm_flagship} we show the ratio of the non-linear matter power spectrum from the \bacco emulator assuming the nominal cosmology to the measured one for the different redshifts. We note that the discrepancy among the two sets of curves is apparently due to a different tilt in the full-shape of the matter power spectrum, here corresponding to the nominal value of $\ns=0.96$. Indeed we note that the disagreement between nominal and recovered cosmology can be alleviated for all parameters by assuming a different value of $\ns$, while keeping all the remaining parameters fixed to their nominal value (with the exception of the scalar amplitude $A_s$, since the latter has to be modified to recover the nominal value of $\sigma_8$). This is also supported by the matter transfer function files that were generated to set-up the initial conditions for the Flagship I simulation, which confirmed how all the parameters affecting the shape of the transfer function $\curly{\omegac,\,\omegab,\,M_\nu}$ are consistent with the nominal values listed in Table \ref{tab:flagship_cosmology}.

We therefore performed new fits to the Flagship I matter power spectra keeping all parameters fixed (including the Hubble parameter $h$) to the nominal values, and only sampling those controlling the primordial matter power spectrum, that is, the scalar amplitude $\As$ and index $\ns$. The best-fit value that we found for the spectral index is close to the value of $\ns=0.97$.  This leads to consistent results across the redshift range we are considering, as well as for a broad range of  scales, up to $k\sim4\kMpc$, as shown in the bottom panel of Fig.~\ref{fig:comp_Pmm_flagship}.

The new values of the spectral index and scalar amplitude, along with nominal values for the other parameters constitute what we refer to as the fiducial cosmology. This is the one given in Table~\ref{tab:flagship_cosmology} and adopted for all predictions throughout this paper.

\section{Standard perturbation theory}
\label{app:formulas}

In this section we report the full expressions for several quantities that are defined in Sect.~\ref{sec:theory} and used throughout the rest of the paper. For a more detailed and exhaustive description of the framework on which cosmological \gls{pt} is based, we refer the reader to the comprehensive review by \cite{Bernardeau2002}.

The main idea behind cosmological \gls{pt} is that the generic solution to the growth of non-linear density and velocity fluctuations -- $\delta(\kv)$ and $\theta(\kv)$ -- in an expanding universe can be expressed in terms of linear theory solutions, $\deltainit(\kv)$. Assuming an \gls{eds} universe, it is possible to perfectly separate the time- and space-dependence of $\delta(\kv)$ and $\theta(\kv)$ \citep{GoroffEtal1986, Jain1994}. In an arbitrary $\Lambda$CDM cosmology, it can be shown that approximate solutions can be found with the same separation \citep{Donath2020}, especially in a range of redshift for which the \gls{eds} cosmology is still a valid approximation. In this case we can write the density and velocity divergence fields using an expansion of the form
\begin{align}
    & \delta\,(\kv,\tau)=\sum_{n=1}^{\infty}D_n(\tau)\,\delta^{\,(n)}(\kv)\,,
    \label{eq:app_delta_expansion}\\
    &\theta\,(\kv,\tau)=-\,\mathcal{H}(\tau)f(\tau)\sum_{n=1}^{\infty}D_n(\tau)\,\theta^{\,(n)}(\kv)\,,
    \label{eq:app_theta_expansion}
\end{align}
where $\tau$ is the conformal time defined via $\de t=a(\tau)\,\de \tau$, $a(\tau)$ is the cosmic scale factor, $\mathcal{H}(\tau)\equiv\de\ln{a(\tau)}/\de\tau$ is the conformal Hubble expansion factor, and $f(\tau)\equiv \de\ln D_1(\tau)/\de\ln a(\tau)$ is the growth rate. The $n$-th order growth factor $D_n$ characterises the time-dependence of the density and velocity field, and reduces to $D_n=a^{\,n}$ in the \gls{eds} limit. Assuming the conservation of mass, momentum, and the Poisson equation, the individual $n$-th order corrections to the density and velocity fields can be written as
\begin{align}
    \begin{split}
        \delta^{\,(n)}(\kv)=&\int_{\qv_1}\ldots\int_{\qv_n}\dirac(\kv-\qv_{1\ldots n})\,F_n(\qv_1,\ldots,\qv_n) \\
        &\times \deltainit(\qv_1)\ldots\deltainit(\qv_n)\,,
    \end{split}
    \label{eq:app_delta_n}\\
    \begin{split}
        \theta^{\,(n)}(\kv)=&\int_{\qv_1}\ldots\int_{\qv_n}\dirac(\kv-\qv_{1\ldots n})\,G_n(\qv_1,\ldots,\qv_n) \\
    &\times\deltainit(\qv_1)\ldots\deltainit(\qv_n)\,,
    \end{split}
    \label{eq:app_theta_n}
\end{align}
where the $n$-th order \gls{pt} kernels $F_n$ and $G_n$ are homogeneous functions of the wave vectors $\left(\qv_1,\ldots,\qv_n\right)$, and are built starting from the fundamental mode-coupling functions,
\begin{align}
    & \alpha\,(\kv_1,\kv_2)\equiv\frac{\kv_{12}\cdot\kv_1}{k_1^{\,2}}\,,
    \label{eq:app_alpha}\\
    & \beta\,(\kv_1,\kv_2)\equiv\frac{k_{12}^{\,2}(\kv_1\cdot\kv_2)}{2\,k_1^{\,2}k_2^{\,2}}\,,
    \label{eq:app_beta}
\end{align}
with $\kv_{12}=\kv_1+\kv_2$.  At linear order these quantities clearly become unity, $F_1=G_1=1$, so to recover linear theory predictions, that is, $\delta^{\,(1)}(\kv)=\theta^{\,(1)}(\kv)=\deltainit(\kv)$. At higher order, these kernels can be derived using recursive relations, which read
\begin{align}
    F_n\,\paren{\qv_1,\ldots,\qv_n}=&\sum_{m=1}^{n-1}\frac{G_m(\qv_1,\ldots,\qv_m)}{(2n+3)\,(n-1)} \nonumber\\
    &\times \Big[(2n+1)\,\alpha(\kv_1,\kv_2)\,F_{n-m}\,(\qv_{m+1},\ldots,\qv_n) \nonumber\\
    &\hspace{11pt} +2\,\beta(\kv_1,\kv_2)\,G_{n-m}(\qv_{m+1},\ldots,\qv_n)\Big]\,,\\
    G_n\,\paren{\qv_1,\ldots,\qv_n}=&\sum_{m=1}^{n-1}\frac{G_m(\qv_1,\ldots,\qv_m)}{(2n+3)\,(n-1)} \nonumber\\
    & \times \Big[3\,\alpha(\kv_1,\kv_2)\,F_{n-m}(\qv_{m+1},\ldots,\qv_n) \nonumber\\
    &\hspace{11pt}+2n\,\beta(\kv_1,\kv_2)\,G_{n-m}(\qv_{m+1},\ldots,\qv_n)\Big]\,,
\end{align}
where $\kv_1=\qv_1+\ldots+\qv_m$ and $\kv_2=\qv_{m+1}+\ldots+\qv_n$.
As a classical example, needed for the calculation of the one-loop galaxy power spectrum, the second-order \gls{pt} kernels for the non-linear evolution of the matter density and velocity fields are defined as
\begin{align}
    &F_2(\kv_1,\kv_2)=\frac{5}{7}+\frac{1}{2}\frac{\kv_1\cdot\kv_2}{k_1k_2}\left(\frac{k_1}{k_2}+\frac{k_2}{k_1}\right)+\frac{2}{7}\frac{\left(\kv_1\cdot\kv_2\right)^2}{k_1^{\,2}\,k_2^{\,2}}\,,
    \label{eq:F2_kernel}\\
    &G_2(\kv_1,\kv_2)=\frac{3}{7}+\frac{1}{2}\frac{\kv_1\cdot\kv_2}{k_1k_2}\left(\frac{k_1}{k_2}+\frac{k_2}{k_1}\right)+\frac{4}{7}\frac{\left(\kv_1\cdot\kv_2\right)^2}{k_1^{\,2}\,k_2^{\,2}}\,,
    \label{eq:G2_kernel}
\end{align}
while the explicit expression for the third-order kernel of the matter density field, $F_3(\kv_1,\kv_2,\kv_3)$, can be found in \cite{GoroffEtal1986}, and it is not reported here for practical purposes.

When considering the galaxy power spectrum $\Pgg$, we need to evaluate the usual two-point statistics defined by the ensemble average of the galaxy density field times itself,
\be
        \ave{\deltag(\kv)\,\deltag(\kv')}=(2\pi)^3\,\dirac(\kv+\kv')\,\Pgg(k)\,.
\ee
At third order in the perturbations of $\delta$, we obtain the one-loop expression for the power spectrum presented in Eq. (\ref{eq:Pgg_oneloop}), where all the next-to-leading order corrections are grouped into mode-coupling and propagator-like contributions. In the former, the loop integrand is proportional to $\Plin(\abs{\kv-\qv}) \, \Plin(q)$, reflecting the mixing of modes due to non-linear evolution, while in the latter the integrals are carried out on the factor $\Plin(k) \, \Plin(q)$, corresponding to a time propagation of the initial density field. Separating these two groups into individual corrections, each one multiplied by a given combination of bias parameters, we end up with the following scheme,
\be
    \begin{split}
        \Pgg^{\,\rm 1\mbox{-}loop}(k) = \: & b_1^{\,2}\,\Ponel(k) \\
        & + b_1 b_2 \, P_{b_1 b_2}(k) + b_1 \bGtwo \, P_{b_1 \bGtwo} (k) \\
        & + b_1 \bGthree \, P_{b_1 \bGthree} (k) + b_2^{\,2} \, P_{b_2 b_2} (k) \\
        & + b_2 \bGtwo \, P_{b_2 \bGtwo} (k) + \bGtwo^{\,2} \, P_{\bGtwo \bGtwo} (k) \,,
    \end{split}
    \label{eq:Pgg_indiv_corr}
\ee
where all the previous terms can be represented as loop integrals,
\be
    \begin{split}
        \Ponel(k) = \: & P^{\,\rm 1\mbox{-}loop,\,MC}(k) + P^{\,\rm 1\mbox{-}loop,\,Prop}(k) \\
        = \:& 2 \int_{\qv} F_2^{\;2}(\kv-\qv,\qv) \, \Plin(\abs{\kv-\qv}) \, \Plin(q)  \\
        & + 6 \, \Plin(k) \int_{\qv} F_3(\qv,-\qv,\kv) \, \Plin(q)\,,
    \end{split}
\ee
\be
    P_{b_1b_2}(k) = 2 \int_{\qv} F_2(\kv-\qv,\qv) \, \Plin(\abs{\kv-\qv}) \, \Plin(q)\,,
    \label{eq:pb1b2}
\ee
\be
    \begin{split}
        P_{b_1\bGtwo}(k) = \: & P_{b_1\bGtwo}^{\,\rm MC}(k) + P_{b_1\bGtwo}^{\,\rm Prop}(k) \\
        = \; & 4 \int_{\qv} F_2(\kv-\qv,\qv) \, S(\kv-\qv, \qv) \, \Plin(\abs{\kv-\qv}) \, \Plin(q) \\
        & + 8\, \Plin(k) \int_{\qv} F_2(\kv,-\qv) \, S(\kv-\qv,\qv) \, \Plin(q)\,,
    \end{split}
\ee
\be
    P_{b_1\bGthree}(k) = -\frac{16}{7} \Plin(k) \int_{\qv} S(\kv-\qv,\qv) \, S(\kv,\qv) \, \Plin(q)\,,
\ee
\be
    P_{b_2b_2}(k) = \frac{1}{2} \int_{\qv} \Plin(\abs{\kv-\qv}) \, \Plin(q)\,,
\ee
\be
    P_{b_2\bGtwo}(k) = 2 \int_{\qv} S(\kv-\qv,\qv) \, \Plin(\kv-\qv,\qv) \, \Plin(q)\,,
\ee
\be
    P_{\bGtwo\bGtwo}(k) = 2 \int_{\qv} S^{2}(\kv-\qv,\qv) \, \Plin(\kv-\qv,\qv) \, \Plin(q)\,.
    \label{eq:pg2g2}
\ee
In the previous set of equations, the $P_{b_1\bGthree}$ contribution is characterised by a single propagator-like term, which is perfectly degenerate with the second addend contributing to $P_{b_1\bGtwo}$. Following the expansion that has been adopted in this paper, the degeneracy results from the equality 
\be
    P_{b_1\bGthree}(k)=\frac{2}{5}P_{b_1\bGtwo}^{\,\rm MC}(k)\,.
    \label{eq:g2-g3_degeneracy}
\ee
As shown in the main body of this article, breaking this degeneracy with the information coming from the galaxy power spectrum alone is not possible, and for this reason we often keep one of the two tidal bias parameters fixed to some physically motivated relation, such as the excursion-set-based relation (Eq. \ref{eq:bG2_exset}) for $\bGtwo$, or the coevolution relations for both $\bGtwo$ and $\bGthree$ (Eqs. \ref{eq:bGtwo_coev}\,--\,\ref{eq:bGthree_coev}).
 
The behaviour of all the loop integrals listed above is to consistently converge to zero at infrared modes, since the non-linear kernel $F_n(\kv_1,\ldots,\kv_n)$ scales as $k^2$ when $\kv\equiv\kv_1+\ldots+\kv_n$ goes to zero, reflecting the range of validity of linear theory predictions. The only exception is represented by the $P_{b_2b_2}$ term, which features a non-zero asymptote for $k\rightarrow 0$. This limit can be manually set to zero via a redefinition of the loop integral, such that
 \be
    P_{b_2b_2}(k) = \frac{1}{2} \int_{\qv} \Plin(q) \, \brackets{\Plin(\abs{\kv-\qv}) - \Plin(q)}\,.
    \label{eq:Pb2b2_resc}
\ee
In turn, the extra contribution
\be
    P_{b_2b_2}^{\,\rm noise}(k) = \int_{\qv} \Plin^{\,2}(q)
\ee
can be absorbed by the constant shot-noise parameter, $\aPone$, which we have defined in Sect.~\ref{sec:modelling_of_the_non-linear_galaxy_power_spectrum}, as they both correspond to constant shift in the amplitude of the galaxy power spectrum.


\section{Implementation of the wiggle vs no-wiggle split}

In this section, we investigate the prescriptions used to obtain a smooth template $\Pnw$ starting from the linear matter power spectrum. This is an important aspect of the theoretical recipe that we adopt, as the wiggle-no wiggle split is essential for the correct implementation of \gls{ir}-resummation, as shown in Sect.~\ref{sec:theory_eft}. While several different algorithms can be found in the literature, here we test three different methods.

The first one is based on a one-dimensional Gaussian smoothing (GS1D), and consists of a rescaling of the original formula for the featureless matter power spectrum $\Peh$, originally presented in \cite{EisHu9804}, to match the broadband amplitude of the linear matter power spectrum. In practice, we follow the approach of \cite{VlahEtal2016}, who defines the smooth component of the linear matter power spectrum as
\be
    \Pnw(k)=\Peh(k)\,{\cal F}\brackets{\frac{\Plin(k)}{\Peh(k)}}\,,
    \label{eq:Pnw_GS1D}
\ee
where ${\cal F}$ is meant to filter out the broadband difference between $\Plin$ and $\Peh$. We choose a functional form for ${\cal F}$ corresponding to a Gaussian filter, that is,
\be
    {\cal F}\brackets{f(k)} = \; \frac{\logten({\rm e})}{\sqrt{2\pi}\lambda}\int \de q\,\frac{f(q)}{q} \exp\brackets{-\frac{1}{2\lambda^2}\log^{\,2}_{10}\paren{\frac{k}{q}}}\,,
\ee

where $\lambda$ determines the variance of the Gaussian filter used to rescale the ratio of the linear to the featureless power spectrum. In this analysis, we fix its value to $\lambda=0.25$.

\begin{figure}
    \centering
    \includegraphics[width=\columnwidth]{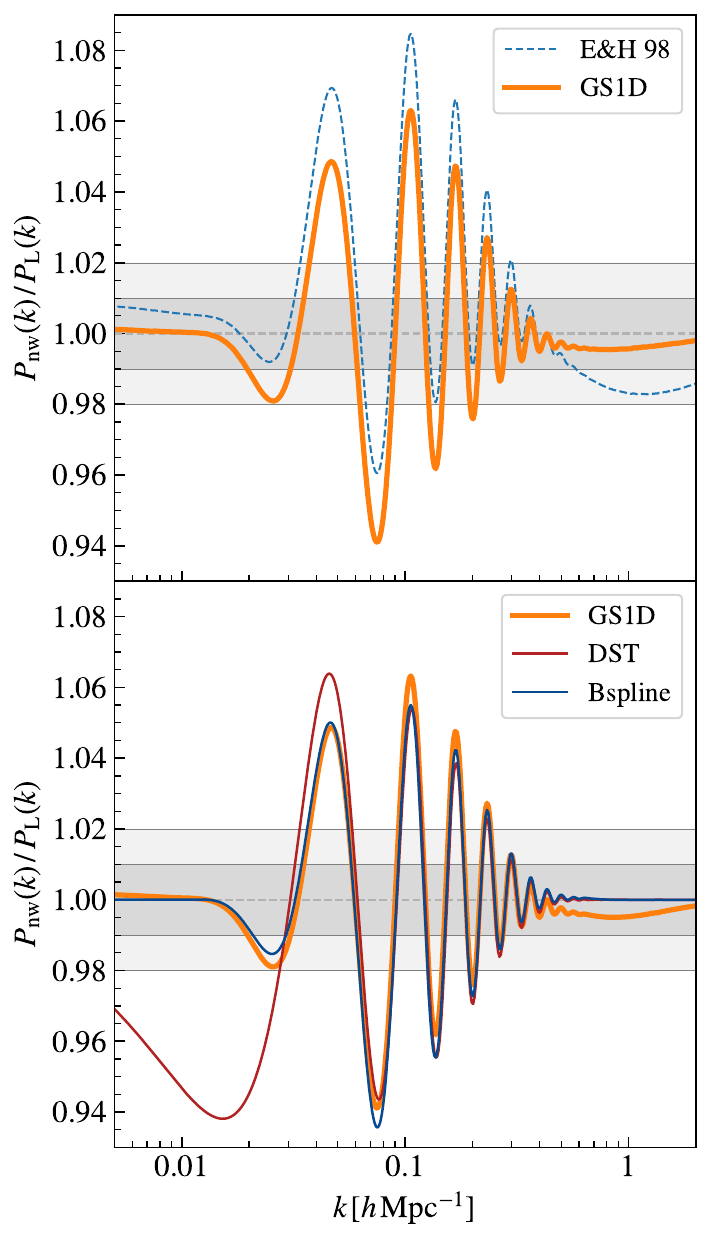}
    \caption{Ratio between the no-wiggle and the linear matter power spectrum. \emph{Top}: Comparison between the raw featureless power spectrum $\Peh$ and the no-wiggle power spectrum $\Pnw$ computed using Eq. \eqref{eq:Pnw_GS1D}. In both cases, the power spectra are computed at $z=0$ using the fiducial cosmology from Table \ref{tab:flagship_cosmology}. \emph{Bottom}: Comparison between the three different methods we tested to obtain the no-wiggle power spectrum $\Pnw$. The thick orange line corresponds to the method we selected, \ie the convolution with a Gaussian smoothing function, while the other two lines represent a Discrete Sine Transform (red) and a basis spline (blue). In both panels, the grey shaded bands represent the $1\%$ (dark grey) and $2\%$ (light grey) thresholds.}
    \label{fig:Pnw_vs_Peh}
\end{figure}

The result of this approach is presented in the top panel of Fig.~\ref{fig:Pnw_vs_Peh}, where we show a comparison between the shape of the original \cite{EisHu9804} smooth function and the one presented in Eq. \eqref{eq:Pnw_GS1D} using the fiducial cosmology from Table \ref{tab:flagship_cosmology} at $z=0$. From the comparison, it is clear that the broadband of $\Peh$ features a non-negligible tilt with respect to the one of the linear matter power spectrum, reaching deviations of up to $2\%$ across the whole \gls{bao} wave mode interval. This difference can be corrected using the filtering strategy, whose output is completely consistent with the broadband shape of $\Plin$, with the only exception of a small residual of ${\sim}\,0.5\%$ peaking at $0.5\kMpc\lesssim k\lesssim 1\kMpc$.

The second approach makes use of a discrete sine transform (DST), and has been originally proposed in \cite{Hamann2010} (see \citealp{classpt} and \citealp{IvanovEtal2020} for an application to real data). This is based on a fast Fourier transformation of the input matter power spectrum, and in the removal of the bump corresponding to the \gls{bao} peak. This step is carried out based on the value of the second derivative of the sine transform, and the generated gap is subsequently filled using a cubic-spline interpolation. The new function is finally transformed back into Fourier space to deliver a power spectrum shape deprived from \gls{bao} oscillations.

The third and final approach is based on the approximation of the \gls{bao} wiggles with a basis spline (B-spline) curve, starting from a set of knots $\{k_i,P_i\}$, and subsequently finding the spline coefficients that maximise the likelihood with the original power spectrum \citep[see \eg][for a similar implementation]{VlahEtal2016}.

The bottom panel of Fig.~\ref{fig:Pnw_vs_Peh} shows a comparison between the three methods summarised above. We note a non-negligible difference, with discrepancies on the \gls{bao} scales that can reach a fraction of percent, depending on the considered approach \citep[see][for similar conclusions]{MoradinezhadEtal2020}. Nevertheless, at linear order these differences are diluted by the recombination of the smooth and wiggling component at a later stage, so that the net result on the \gls{ir}-resummed non-linear galaxy power spectrum is going to be much smaller than the values exhibited in this plot. We highlight a major discrepancy between the DST method and the other two that reaches its maximum at a scale $k\sim0.01\kMpc$, which roughly coincides with the position of the turnaround in the matter power spectrum. Because of this major discrepancy, and since all three methods behave in a slightly different way on the \gls{bao} scales, we decided to adopt the GS1D method throughout the analysis presented in this paper. We observe a marginal residuals of this method on mildly non-linear scales, at $0.5\kMpc\lesssim k\lesssim1\kMpc$, which, once again, is not likely to significantly bias our results

At the same time we highlight how this approach is most likely going to perform worse when considering cosmologies beyond the vanilla $\Lambda$CDM model, such as those including massive neutrinos. In these cases, depending on the magnitude of the deviation from $\Lambda$CDM, the broadband shape of $\Peh$ can deviate from that of $\Plin$ by up to $10\%$ (as quoted in \citealp{EisHu9804}). We therefore adopt one of the other two methods for future analyses, such as for the upcoming analysis of the Flagship II simulation.


\section{Comparison between model implementation}
\label{app:modelling_challenge}

\begin{figure*}
    \centering
    \includegraphics[width=2\columnwidth]{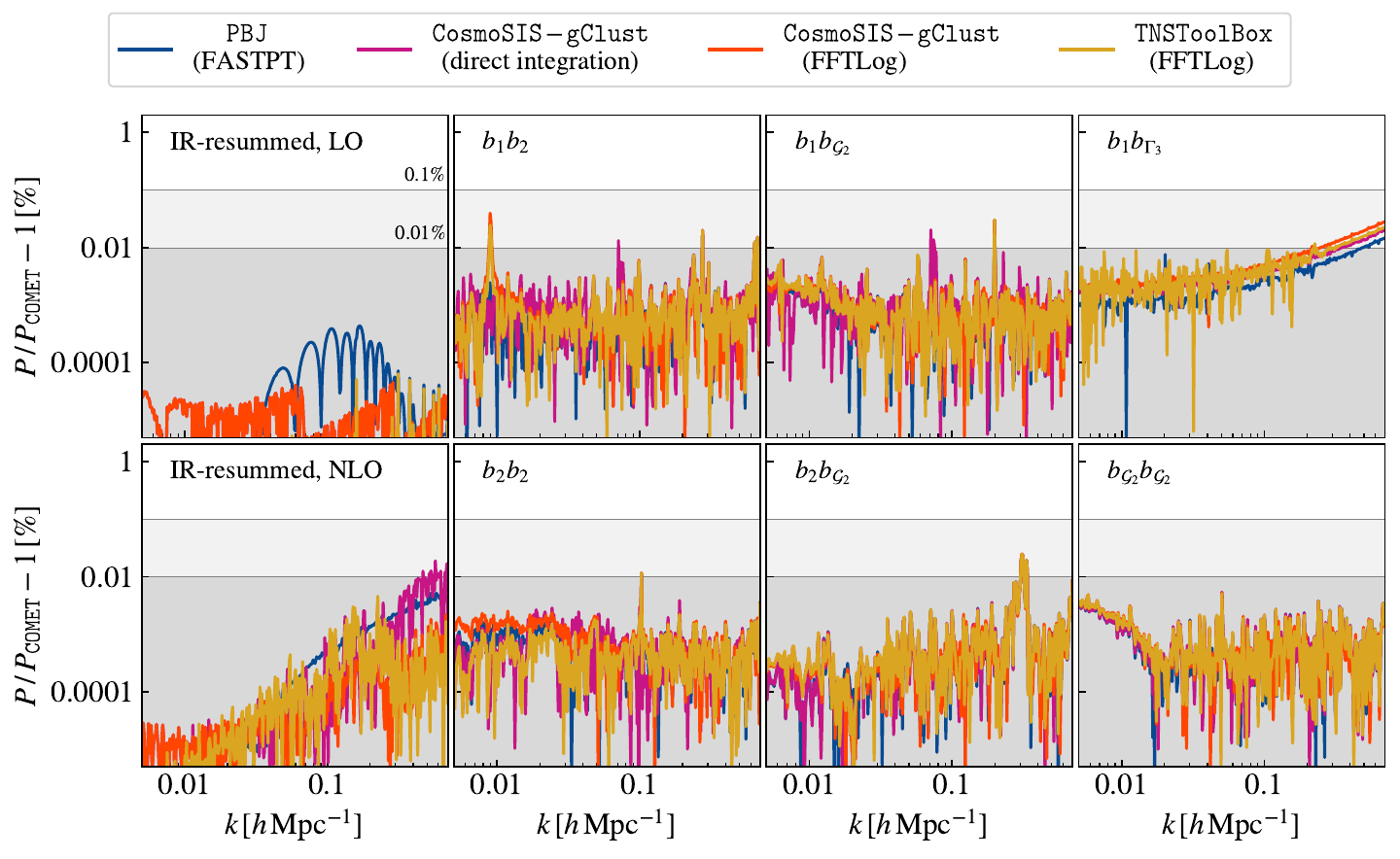}
    \caption{Comparison among the different methods to obtain the one-loop bias integrals, colour-coded according to the individual code/method, as listed in the legend. The reference code for these plots is the \comet emulator. Different panels corresponds to different diagrams: the first column shows the leading order \gls{ir}-resummed matter power spectrum -- for which no actual integration is required -- and the next-to-leading order correction. The remaining columns show the bias diagrams from Eqs. (\ref{eq:pb1b2}\,--\,\ref{eq:pg2g2}).}
    \label{fig:model_comparison}
\end{figure*}

In order to estimate the systematic error budget due to the implementation of the algorithm for the one-loop model presented in Sect.~\ref{sec:theory_eft} and Appendix \ref{app:formulas}, we make use of four individual implementations, based on four independent codes provided by several group members. Each code features a different way to compute the loop corrections presented in Sect.~\ref{app:formulas}.

These include a two-dimensional integration implemented within the \texttt{Cuba} library\,\footnote{\url{http://www.feynarts.de/cuba/}} \citep{Hahn2005} used to generate the training set of the \comet emulator \citep{comet}, and implemented in the \texttt{CosmoSIS-gClust}\,\footnote{To be made publicly available soon.} code (courtesy of A. Moradinezhad Dizgah). Alternative methods take advantage of a Fast Fourier Transform approach, such as \texttt{FastPT}\,\footnote{\url{https://github.com/JoeMcEwen/FAST-PT/}} \citep{McEwen_2016} -- implemented in the \texttt{PBJ} code \citep{Oddo2019, Oddo2021, Rizzo2022, Moretti2023} -- and \texttt{FFTLog}\,\footnote{\url{https://jila.colorado.edu/~ajsh/FFTLog/}} \citep{Hamilton2000, Simonovic2018} -- used in the \texttt{TNSToolBox}\,\footnote{\url{https://github.com/sdlt/TNS_ToolBox/}} code (courtesy of S. de la Torre) and again in \texttt{CosmoSIS-gClust}.

In Fig.~\ref{fig:model_comparison} we show the systematic deviation between different computations of the same terms, including the leading- and next-to-leading order correction to the \gls{ir}-resummed matter power spectrum, and the six one-loop bias corrections defined in Eqs. (\ref{eq:pb1b2}\,--\,\ref{eq:pg2g2}). For this exercise, we compute a common data vector using the fiducial Flagship cosmology from Table \ref{tab:flagship_cosmology} at $z=0$. We assume as a reference measurement the one performed using a direct integration of the wave number $q$ within the range of scales $\brackets{0.00001,100}\kMpc$, which proved to be a sensible choice to achieve the convergence of the different loop integrals. Finally, we compare the different terms evaluated using the four codes described in the previous paragraph. We find an overall optimal consistency among the different theory implementations, with all terms being in agreement at better than $0.01\%$ on the overall range of scales shown in Fig.~\ref{fig:model_comparison}. We observe a slightly worse concordance between different computations of the propagator-like terms, that is, $\Pmmnlo$, which contains the contribution $P_{13}$, and more importantly $P_{b_1b_{\Gamma_3}}$, for which the discrepancy can become as large as ${\sim}\,0.05\%$ on scales of $k\sim1\kMpc$. However, we consider this difference completely negligible, since this value is much smaller than the statistical precision of the data vectors, and since this scale is completely outside of the range of scales considered in this work. 

\section{Comparison between posterior averaged and maximum likelihood position}
\label{app:chi2}

In the main body of this article, we determine the goodness of fit of a given model configuration in terms of the posterior-averaged $\chi^2$ statistics. For this purpose, we simply iterate through the sampled positions of the posterior distribution and define the weighted mean as
\be
    \ave{\,\chi^2} = \sum_i w_i \,\chi^2_i\,,
\ee
where $\chi^2_i$ and $w_i$ correspond to the $\chi^2$ and corresponding weight for the $i$-th sampled position, respectively. As already mentioned in Sec. \ref{sec:chi2}, this quantity can be estimated with less uncertainty from a sampled posterior distribution with respect to the maximum-likelihood position. However, the conclusions drawn in Sec. \ref{sec:results} are independent from which definition is used to quantify the goodness-of-fit statistics.

\begin{figure}[h!]
    \centering
    \includegraphics[width=\columnwidth]{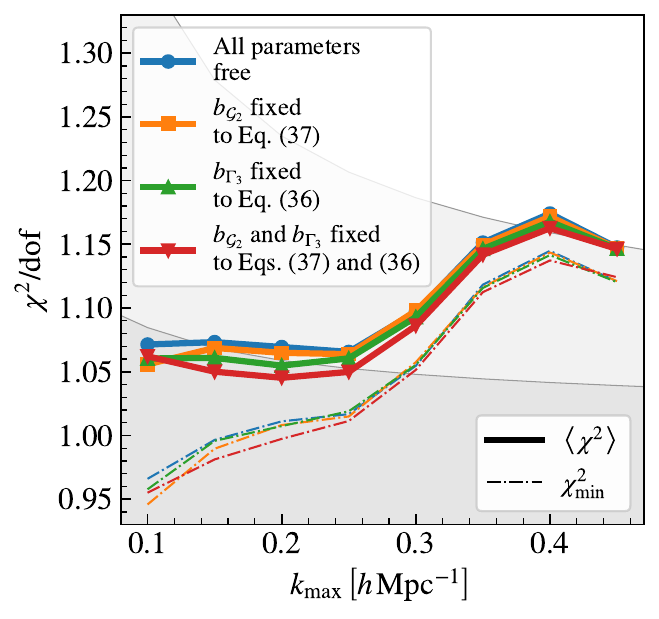}
    \caption{Normalised $\chi^2$ values for the Model 3 sample at $z=1.2$. Solid and dot-dashed lines corresponds to the posterior-averaged and minimum $\chi^2$, respectively. Different colours represent different bias modelling assumtpions, as listed in the legend. The two grey shaded areas represent the 68\% and 95\% of the $\chi^2$ distribution with the same number of degrees of freedom of the considered model.}
    \label{fig:comp_ave_vs_min_chi2}
\end{figure}

In Fig. \ref{fig:comp_ave_vs_min_chi2} we show the goodness-of-fit performance metric for the case of the Model 3 sample at $z=1.2$, which is one of the few cases with a posterior-averaged $\chi^2$ getting above the $2\sigma$ in Fig. \ref{fig:pm_mod1} and \ref{fig:pm_mod3}. Differently from those two plots, in this case we compare the trend when using the posterior-averaged $\chi^2$ (solid lines) and maximum-likelihood value (dash-dotted lines). As expected, the latter is constantly outperforming the former, with a relative improvement that becomes larger at low $\kmax$ values. Moving to larger $\kmax$, we note that for this specific case the maximum-likelihood position makes all the points consistent within $2\sigma$ of the considered $\chi^2$ distribution. However, the marginal gains are not significantly changing the interpretations that can be made when using the posterior-averaged values.

\begin{figure}[h!]
    \centering
    \includegraphics[width=\columnwidth]{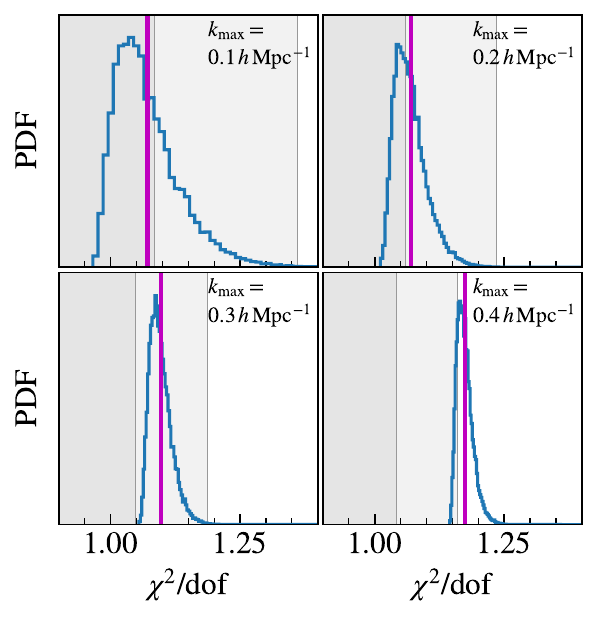}
    \caption{Normalised $\chi^2$ distribution for the Model 3 sample at $z=1.2$ and four different value of $\kmax$, as shown in the upper right corner of the individual panels. The shaded grey areas mark the same confidence intervals as in Fig. \ref{fig:comp_ave_vs_min_chi2}. The magenta vertical line marks the posterior-averaged value.}
    \label{fig:chi2_dist_diff_kmax}
\end{figure}

As a complement, in Fig. \ref{fig:chi2_dist_diff_kmax} we show the complete distribution of $\chi^2$ values, and we compare it with the 68th and 95th prercentile of the reference $\chi^2$ distribution with the same numbers of degrees of freedom. Different panels correspond to different $\kmax$ values, while the solid magenta line marks the position of the posterior-averaged value. The only configurations for which there is a partially significant gain in picking the maximum-likelihood value is clearly the one at $\kmax=0.4\kMpc$. Otherwise, our conclusions are not affected by the choice of which $\chi^2$ value is used to estimate the goodness-of-fit performance metric. For this reason, in Sec. \ref{sec:results} we show results employing the posterior-averaged $\chi^2$ statistics.

\section{Dependency on the selected sampler}
\label{sec:samplers}

As anticipated in Sect.~\ref{sec:fitting_procedure}, in order to obtain a posterior distribution for each of the model configurations that we test, we need to select a robust algorithm to sample the large multi-dimensional parameter space of the models described in Sect.~\ref{sec:theory}. Indeed, the choice of the sampler is critical in presence of multi-modal distributions. As already thoroughly discussed in the main body of this paper, this situation is prevalent when we focus solely on the full shape of the galaxy power spectrum, using only two-point statistics, due to the strong degeneracies between model parameters. In these situations traditional algorithms, which are meant to explore smaller and more well-behaved -- Gaussian-like -- random variables, may be under-performing, and therefore affecting the efficiency of the sampling.

In this analysis we compare three different samplers, testing them against a reference subset of the galaxy power spectrum data vectors already used to validate the theoretical model for one-loop galaxy bias. The properties of these sampling algorithms are listed hereafter.

\subsection*{Metropolis--Hastings sampler}

This approach \citep{Metropolis1953, Hastings1970} is a Markov-Chain Monte Carlo (MCMC) method based on the construction of a random walk inside the parameter space. Subsequent points in the Markov chain are determined based on a proposal function that has to be specified as a free parameter of the model. For each new point, the likelihood of the model determines whether the candidate state is accepted or discarded, that is,  whether to move to the candidate state or stay in the current state. For this standard algorithm we make use of a non-public code,\footnote{\texttt{COMPASS}, courtesy of Ariel G. S\'{a}nchez, Mart\'{i}n Crocce, and Rom\'{a}n Scoccimarro.} which features an implementation of a Metropolis-Hastings algorithm coupled with the likelihood for the galaxy power spectrum described in Sect.~\ref{sec:theory}.

\begin{figure*}
    \includegraphics[width=\columnwidth]{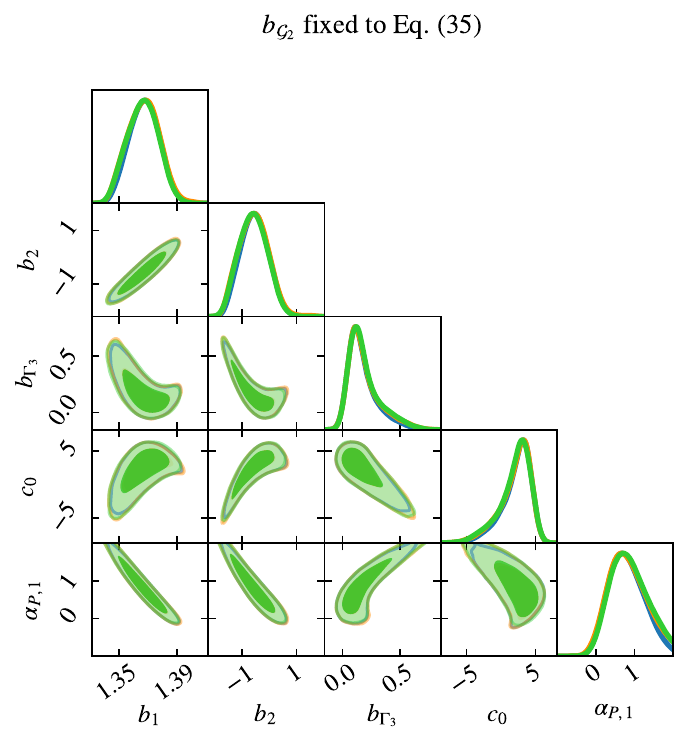}
    \includegraphics[width=\columnwidth]{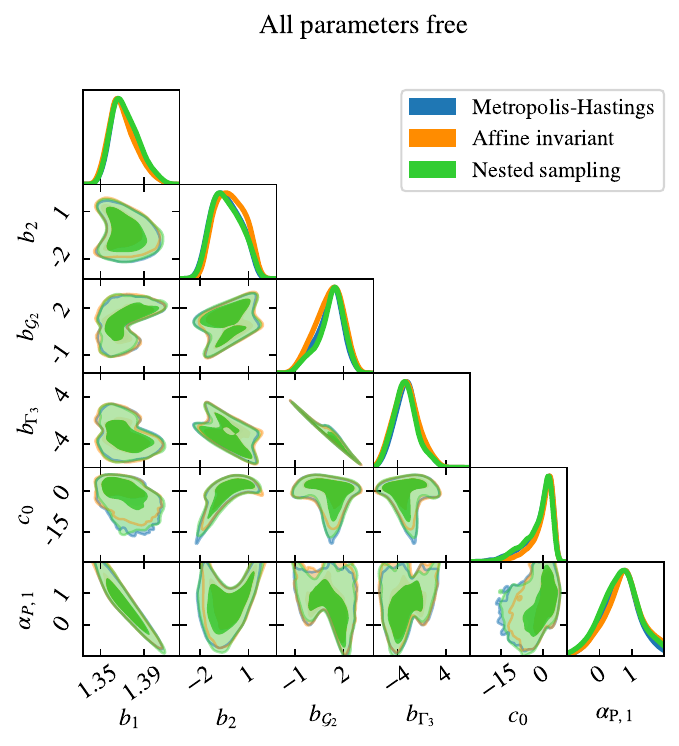}
    \caption{Comparison between runs at fixed cosmology obtained using different samplers, carried out using a reference galaxy power spectrum (Model 3 \gls{hod} sample at $z=0.9$). The left panel shows the posterior distribution of the nuisance parameters with $\bGtwo$ fixed to the local Lagrangian relation (Eq. \ref{eq:bGtwo_coev}), while in the right panel all the parameters are left free to vary. Different colours correspond to different samplers, as listed in the legend.}
    \label{fig:comp_samplers}
\end{figure*}

\subsection*{Affine-invariant sampler} 

This MCMC approach \citep{Goodman2010} is still based on a random walk across the parameter space, but with an improved ensemble sampler, in which a large number of walkers interact with each other in a way that reduces the dependence of the sampling on the aspect ratio of the particular posterior distribution under consideration. We use the affine-invariant sampler implemented in the public Python package \texttt{emcee} \citep{ForMac1303}.\footnote{\url{https://emcee.readthedocs.io/en/stable/}}

\subsection*{Nested sampling} 

This approach \citep{Skilling2006} relies on iteratively refining a set of live points, initially drawn from a prior distribution, to explore the parameter space. At each iteration, the point with the lowest likelihood is selected from the set and replaced with a new point. The latter is drawn from a constrained region of the prior distribution, where the likelihood must be higher than that of the replaced point. The selected point is then assigned a weight, based on its likelihood and the proportion of the prior volume it represents: several iterations will allow us to thus construct the posterior distribution.
 These algorithms are particularly suited for multi-modal distributions, given they are not subject to getting stuck in local minima of the loglikelihood, as it commonly happens with standard algorithms based on Markov chains. We use the public package \texttt{Multinest} \citep{Fer0910, Feroz2019}, which can be interfaced with Python using a dedicated wrapper module.\footnote{\url{https://johannesbuchner.github.io/PyMultiNest/}}

\subsection*{Comparison}

The comparison between the marginalised two-dimensional distributions obtained with the different samplers is shown in Fig.~\ref{fig:comp_samplers}, using as data the galaxy power spectrum of the Model 3 \gls{hod} sample at $z=0.9$ with $\kmax=0.2\kMpc$. We test two different cases, one where we vary the nuisance parameters $\paren{b_1, b_2, \bGthree, c_0, \aPone}$ while keeping the quadratic tidal bias $\bGtwo$ fixed to its local Lagrangian relation (\ref{eq:bGtwo_coev}), and one where the latter is also allowed to freely vary with the rest of the parameters.\footnote{As shown in the main body of this article, fixing $\bGtwo$ is the only case for which it is possible to completely break the strong degeneracies of the parameter space.}

In the first case, we note an almost perfect match among the three different set of contours, with a statistically negligible difference only appreciable at the tails of the posterior distributions, which can be anyhow partially explained by the intrinsic variance of the individual realizations of the posterior distributions. The second case shows a slightly larger discrepancy, mostly due to the presence of a multi-modal profile driven by the strong degeneracy between $\bGtwo$ and $\bGthree$. In this case, the peak of the posterior distribution is less consistent when using different samplers, and we find that the size of the discrepancy is a very strong function of $\kmax$. In details, the main differences arise because of the non-homogeneous sampling of different peaks, for which traditional Markov chain algorithms may get stuck in a particular minimum of the distribution, therefore affecting the overall convergence of the chain. Given the purpose that nested sampling algorithms were initially developed for, we decided to employ the latter as our baseline sampler, and used it to sample the parameter space for all the results presented in this work.


\section{Sample variance effects}
\label{app:sample_variance}

\begin{figure*}
    \centering
    \includegraphics[width=2\columnwidth]{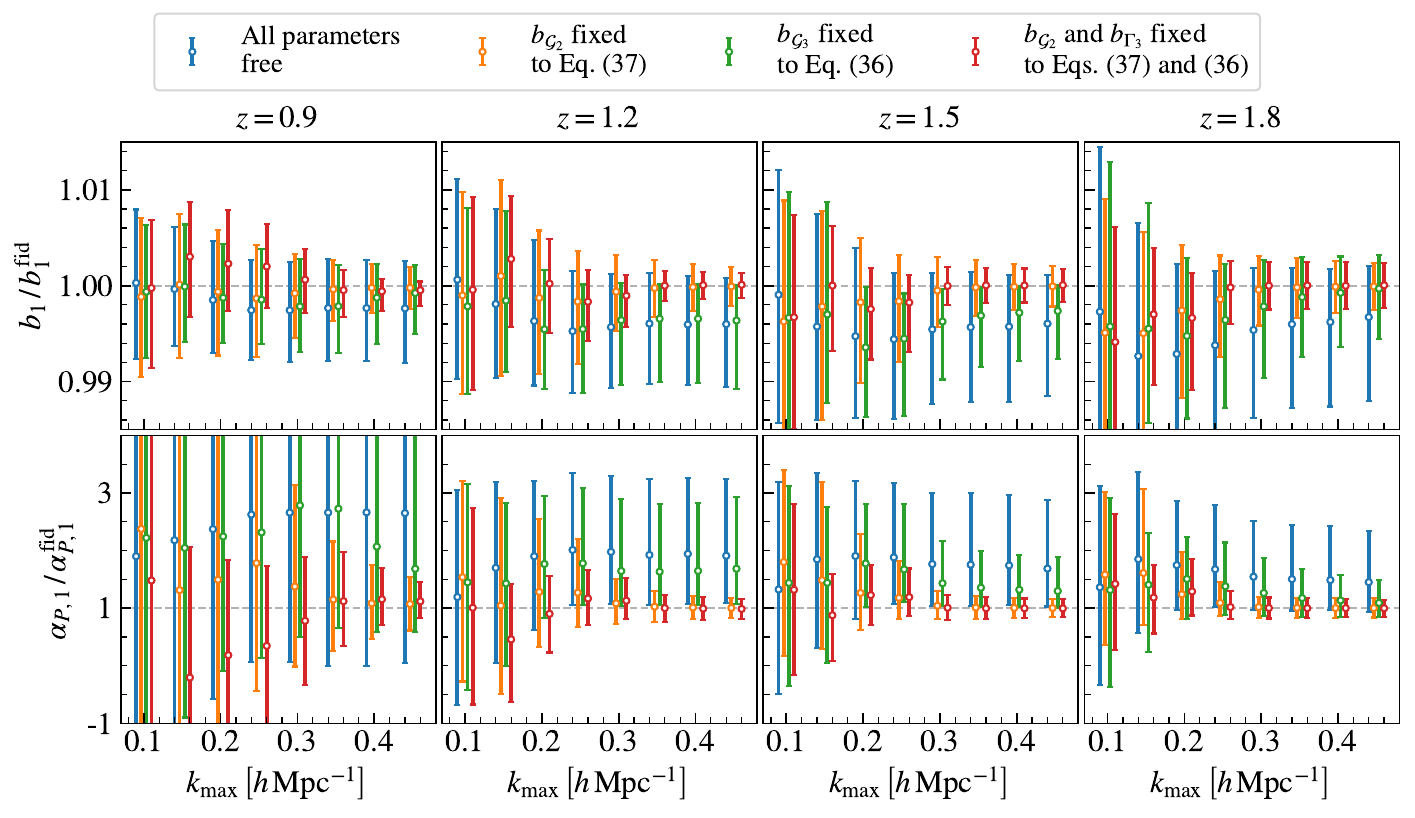}
    \caption{Marginalised 1d constraints on the linear bias $b_1$ and the shot-noise parameter $\aPone$ obtained from a set of synthetic theory vectors created using the same recipe from Eq. \eqref{eq:Pggnlo}, for the four different redshifts already explored with the Flagship data vectors. Different colours correspond to different assumptions on the total number of degrees of freedom of the model, as listed in the legend.}
    \label{fig:marg_1d_datavec_withoutnoise}
\end{figure*}

\begin{figure*}
    \centering
    \includegraphics[width=2\columnwidth]{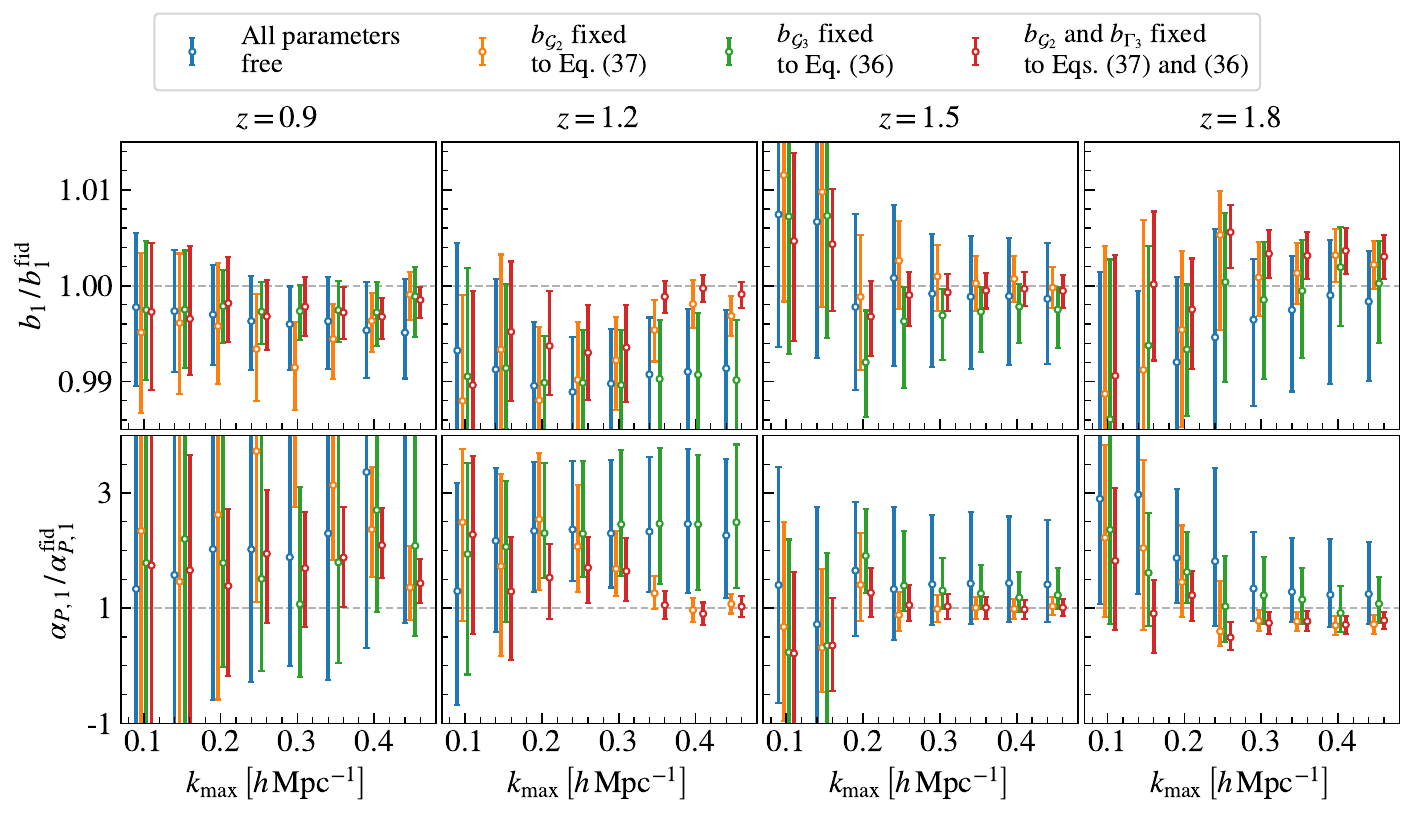}
    \caption{Same as in Fig.~\ref{fig:marg_1d_datavec_withoutnoise}, but using theory vectors displaced by Gaussian noise realisation from a box of size $3780\Mpc$.}
    \label{fig:marg_1d_datavec_withnoise}
\end{figure*}

 Even though the volume of the Flagship I simulation is quite large, most of our results are, to some extent, affected by the sample variance of the single \nbody realization available. 
 
 In this section we quantify this effect reproducing some of the parameters fits starting from noiseless, synthetic data vectors obtained as the real-space galaxy power spectrum predictions from the \comet emulator. These are generated adopting the reference cosmological parameters from Table \ref{tab:flagship_cosmology} at the four different redshifts of the comoving snapshots, $z=\curly{0.9,1.2,1.5,1.8}$. For each of the four samples, the nuisance parameters are derived as follows: $b_1$ and $\aPone$ are fixed to the best-fit value listed in Table \ref{tab:fiducial_b1_aP1} for the Model 1 \gls{hod} samples, $b_2$ is computed using a $b_2(b_1)$ relation derived from the corresponding \gls{hod} model, $\bGtwo$ and $\bGthree$ are obtained from the excursion-set relation (\ref{eq:bG2_exset}) and the coevolution relation (\ref{eq:bGthree_coev}), respectively, $c_0$ is set to unity, and $\aPtwo$ is set to zero. The covariance matrix for each sample is computed assuming only the Gaussian component from the full Flagship I box, and therefore corresponds to a volume of about $58\cGpc$.

Figure~\ref{fig:marg_1d_datavec_withoutnoise} shows the marginalised one-dimensional constraints of the $b_1$ and $\aPone$, similarly to Fig.~\ref{fig:marg_1d_fixed_cosmo}, for fits at fixed cosmology of the synthetic data vectors. Even though the data vectors are smooth -- no noise component has been added to any of the $\Pgg(k_i)$ bins -- and also perfectly consistent with the theory model used to fit them, we observe some discrepancy of the constraints with the fiducial values, both with all parameters free as for the case with $\bGthree$ fixed to the coevolution relation. In both cases, the statistical constraints are less tight than the ones obtained with other bias relations, with a partial break of the degeneracies between parameters at higher redshift then the coevolution relation is applied. Such deviations are induced by projection -- or prior volume -- effects in the marginalised posteriors, due to the strong degeneracies between parameters. In particular, these affect $b_2$, $\bGtwo$ and $\bGthree$, and cannot be broken completely even when reducing the dimensionality of the parameter space by fixing $\bGthree$, as the $b_2$-$\bGtwo$ degeneracy is still present.
Consistently among the different redshifts, we note that there is a trend for $b_1$ and $\aPone$ to be under- and overestimated, respectively, similarly to what we observe for the real Flagship I data vectors.

On the contrary, the case where $\bGtwo$ is fixed as a function of $b_1$ is systematically better in terms of amplitude of the error bars and accuracy in the recovery of the parameters, because the posterior distribution of this case is closer to a Gaussian distribution. This is due to the simultaneous breaking of both the $b_2$-$\bGtwo$ and $\bGtwo$-$\bGthree$ degeneracies, for which a clear example can be found in the left plot of Fig.~\ref{fig:comp_samplers}.

As a follow-up test, we add to the synthetic data vectors a Gaussian noise consistent with the covariance assumed for the fits to the simulation. Figure~\ref{fig:marg_1d_datavec_withnoise} shows the new marginalised constraints on $b_1$ and $\aPone$. In this case it is possible to observe a much larger discrepancy from the fiducial values, present also for the configurations in which $\bGtwo$ is fixed as a function of $b_1$. We remark how the case at $z=1.2$ features a large fluctuation at intermediate scales, $\kmax\sim0.2\kMpc$, which is symptomatic of the particular realization of the Gaussian noise. We repeated this exercise, on the sample at $z=1.2$, with ten different noise realizations, and in all cases we observe a slightly different trend as a function of $\kmax$ for the marginalised constraints of the configurations we are considering, that is, with all the parameters free to vary, or with either $\bGtwo$ or $\bGthree$ fixed in terms of $b_1$. For some of the configurations we observe trends as the one seen in the $z=0.9$ Flagship I data vectors, for which we find a sharp running in the $(b_1,\aPone)$ plane at $\kmax\gtrsim0.2\kMpc$. We therefore conclude that a more realistic analysis, also including observational effects and a proper survey window function should be conducted using a wider set of simulations, in order to reduce the overall impact of cosmic variance.

\end{appendix}

\end{document}